\documentclass[twocolumn]{aastex63}

\received{April 11, 2022}
\revised{July 5, 2022}
\accepted{July 27, 2022}
\submitjournal{AJ}

\shorttitle{Radio Structure of RQQs}
\shortauthors{McCaffrey et al.}

\usepackage{pgfplots}
\usepackage{appendix}
\usepackage{array}
\usepackage{booktabs} 
\usepackage{multirow}
\usepackage[flushmargin]{footmisc}
\usepackage{xcolor}
\pgfplotsset{compat=1.16}  

\begin{document}

\title{Kpc-scale Radio Structure in $z\sim0.25$ Radio-Quiet QSOs}

\author[0000-0002-9321-9559]{Trevor V McCaffrey}
\affiliation{Department of Physics, Drexel University\\
32 S.\ 32nd Street\\ 
Philadelphia, PA 19104, USA}
\affiliation{National Radio Astronomy Observatory \\ 1003 Lopezville Rd \\ Socorro, NM 87801, USA}

\author[0000-0001-9324-6787]{Amy E Kimball}
\affiliation{National Radio Astronomy Observatory \\ 1003 Lopezville Rd \\ Socorro, NM 87801, USA}

\author[0000-0003-3168-5922]{Emmanuel Momjian}
\affiliation{National Radio Astronomy Observatory \\ 1003 Lopezville Rd \\ Socorro, NM 87801, USA}

\author[0000-0002-1061-1804]{Gordon T Richards}
\affiliation{Department of Physics, Drexel University\\
32 S.\ 32nd Street\\ 
Philadelphia, PA 19104, USA}

\begin{abstract}

We present analysis of a homogeneous, optically selected, volume-limited ($0.2<z<0.3$) sample of 128 radio-quiet quasi-stellar objects (QSOs) recently observed at 6\,GHz with the Very Large Array (VLA) in A-configuration ($\sim0\farcs33$ resolution). 
We compare these new results to earlier (2010--2011) 6-GHz observations with the VLA in C-configuration ($\sim3\farcs5$).
While all of these radio-quiet QSOs (RQQs) were unresolved on a $3\farcs5$ scale ($\sim$14\,kpc at $z=0.25$), we resolve notable complex sub-galactic structures in about half of the RQQs at $0\farcs33$ resolution ($\sim$1.3\,kpc at $z=0.25$).  By comparison of flux density measurements between the two sets of observations, we demonstrate that significant sub-galactic-scale radio structure is present in at least 70\%\ of the RQQ population, and that the central component accounts for an average of $\approx$65\% of the total detected radio power.  One RQQ, J0935+4819, shows striking symmetric, double-lobed morphology, and appears to be the first identified example of a radio-{\it quiet} QSO with FR~II type morphology on $\sim$arcsec scale (projected size of $\gtrsim6\mathrm{kpc}$).
In addition to revealing RQQ sub-galactic morphology, we employ counterparts from legacy (FIRST at 1.4\,GHz) and recent (VLA Sky Survey at 3\,GHz) VLA surveys to investigate radio spectral indices and potential variability over decades-long timescales for a subset of the RQQs, and for the cores of radio-intermediate and -loud sources in the parent sample of 178 QSOs.
These results support the growing notion that the RQQ population is not a monolithic phenomenon, but instead consists of a mixture of mainly starburst-powered and jet-powered galaxies.

\vspace{28pt}    
\end{abstract}

\section{Introduction}

While the first quasi-stellar objects (QSOs) were discovered upon the basis of their radio properties \citep{Burbidge67,BB67_book,Schmidt69}, 
our understanding of their radio emission remains incomplete.  Emission from radio-loud QSOs (RLQs)
is known to be powered by the jets of their active galactic nuclei (AGNs) because of, among other lines of reasoning: extended morphologies (large-scale radio components extending beyond the host galaxy), observations of apparent superluminal motion \citep{Chi+13,Maini+16,HR16,HR17}, or the presence of high brightness temperature component(s) \citep[non-thermal emission with $T_\mathrm{B} > 4\times10^4$\,K at 6 GHz;][]{condon1992}.
The dominant source of radio emission in radio-\textit{quiet}\footnote{For the purposes of this paper, we define radio-quiet QSOs to be QSOs with $L_\mathrm{6\,GHz}<10^{23}\;\mathrm{W\,Hz}^{-1}$.} QSOs (RQQs), however, remains a subject of much controversy.  Some arguments assert that star formation processes are the likely dominant contributor \citep{kimball2011,Condon+2013,Padovani+2015,kellerman16}, while others have argued that star formation simply is not powerful enough, and instead mechanisms such as AGN-driven winds \citep{ZG14, Zakamska+16} or scaled down jets \citep[e.g.,][]{ulvestad2005} are primarily responsible. 

The advent of the Karl G.\ Jansky Very Large Array \citep[VLA;][]{EVLA} allowed \citet{kimball2011}, followed by \citet{kellerman16}, to determine the faint end of the QSO radio luminosity function (RLF) for the first time, using a volume-complete, homogeneous QSO sample, albeit at relatively low redshift ($0.2<z<0.3$).
Their sample is optically selected and thus is not subject to selection bias towards any radio properties.  The authors concluded that the RLF is consistent with two independent components contributing to the radio emission in QSOs: a likely AGN-driven component dominating at $L_\mathrm{6\,GHz}\gtrsim10^{24}$\,W\,Hz$^{-1}$ (the RLQs), whereas a bump in the RLF at $\sim10^{23}$\,W\,Hz$^{-1}$ \citep[similar in radio luminosity to that of nearby starburst galaxies;][]{Condon+2002,Condon+2013} could be due to an additional, independent, star-formation component dominating in the lower-luminosity RQQs.

An obstacle to answering the question ``what is the dominant source of radio emission in RQQs?” is that the most sensitive wide-area imaging campaigns have not been undertaken at a sufficient spatial resolution to reveal the sub-galaxy-scale radio morphologies across a large fraction of the RQQ population.  Indeed, the conclusions of \citet{kimball2011} were based on the shape of the RLF, as their observations were designed as a detection experiment to investigate radio luminosity rather than morphology.  Moreover, while the images of their targets show no sign of large-scale jets, they \textit{do} reveal that the radio emission in RQQs is confined to the dimensions of the host galaxy [$\sim$14\,kpc at $z=0.25$ \citep[e.g.,][]{Lerner2018}].  Having radio emission constrained to that scale is also consistent with AGN-driven winds \citep{ZG14,Zakamska+16} or small-scale jets \citep{ulvestad2005,HR16,HR17}.  To date, the largest and most sensitive radio survey capable of resolving such galaxy-confined emission is that of \citet{hodge2011}; however, its area of sky (Stripe 82, $\sim300\,\mathrm{deg}^2$) is too small to probe much of the RQQ population.   We note that none of the 178 QSOs from \citet{kimball2011} were observed by \citet{hodge2011}.  Progress is being made on this front; \citet{Morabito+22} presented a calibration strategy to overcome lasting ionospheric and computational difficulties of sub-arcsecond ($\sim$0\farcs25) imaging with the international LOw-Frequency Array \citep[LOFAR;][]{LOFAR}.  While their methods so far have been applied to small fields \citep[e.g., the Lockman Hole;][]{Sweijen+22}, such high-resolution imaging will soon be applied across the entire northern sky as a complement to the LOFAR Two-metre Sky Survey \citep{LoTSS1,Shimwell+22_LoTSS2}, reaching 1$\sigma$ sensitivities as low as $\approx$18\,$\mu$Jy\,beam$^{-1}$ at 1.4\,GHz for sources with a radio spectral index $\alpha=-0.7$.


In this paper, we take the investigations of \citet{kimball2011} and \citet{kellerman16} a step further, obtaining high-resolution ($\sim0\farcs33$; $\sim1.3$\,kpc at $z=0.25$) follow-up observations of the 128 RQ sources from their volume-limited sample of 178 QSOs. All 128 were unresolved in the original observations with resolution $\sim3\farcs5$ ($\sim14$\,kpc at $z=0.25$); we now aim to resolve emission structures on size scales 1.3--14\,kpc within the host galaxies across the sample of 128 RQQs.  Following \citet{kimball2011}, we define ``radio-quiet" as having radio luminosity $L_\mathrm{6 GHz}<10^{23}$\,W\,Hz$^{-1}$, where the purported star-formation component dominates the RLF of QSOs.

While we use a specific limit on radio luminosity to define radio-quiet/radio-loud categories for our analysis, it is important to acknowledge that various definitions and diagnostics are used throughout the literature, potentially leading to different conclusions as to the properties of RQ/RL sources.  \citet{Kellerman1989} first reported a potential bimodality in the Bright Quasar Survey \citep{SG83} based on $R_\mathrm{O}$, the ratio between 5-GHz radio luminosity and 4400-Å optical luminosity.  This criterion remains a useful radio-loudness diagnostic \citep[e.g.,][]{Balokovic2012}, along with $q$---the ratio between radio and far-infrared luminosities---as well as $R_X$---a ratio of radio to X-ray luminosities \citep[e.g.,][]{TW03_Rx}.  Such ratios are particularly useful for comparing sources across large redshift ranges, because QSOs' radio and optical luminosities show evolution with redshift \citep{padovani11}; defining RQ/RL classes based on a luminosity ratio serves somewhat to mitigate these evolutionary effects.  \citet{Condon+2013} probed evolution of the 1.4-GHz RLF using optically selected QSOs, and found that the luminosity-based demarcation between RQ/RL populations increases in luminosity from low redshift ($0.2<z<0.45$) to high redshift ($1.8<z<2.5$).  For the sample presented here, drawn from \citet{kimball2011} and \citet{kellerman16}, the narrow redshift range ($0.2<z<0.3$) mitigates effects from evolution; furthermore, for the typical $i$-band magnitude of 17.0 in this population (optical luminosity of $10^{23}$\,W\,Hz$^{-1}$ for $z=0.25$), the radio luminosity of $10^{23}$\,W\,Hz$^{-1}$ corresponds to a radio-to-optical ratio of~1.  Therefore, our definition of radio-quiet/radio-loud is similar to a definition based on radio-to-optical ratio over this redshift range. 

As an additional step in our investigation, we use counterparts from the literature [the Faint Images of the Radio Sky at Twenty-cm (FIRST) survey \citep{first,Helfand15} and the VLA Sky Survey \citep[VLASS;][]{vlass}] to investigate spectral indices and potential variability of the RQ targets, as well as radio-loud (RL) and radio-intermediate\footnote{We define ``radio-intermediate" to be the luminosity range $10^{23}$--$10^{24}$\,W\,Hz$^{-1}$ between RQ and RL QSOs.} (RI) sources from the parent sample of 178 QSOs.
Cores of RLQs are known to exhibit multi-wavelength variability on timescales of years, months, and even days \citep{Hovatta+2007, Thygarajan+2011}.  \citet{barvainis2005} further showed that RQQs are capable of mirroring this behavior in the radio, 
suggesting that the underlying mechanisms that drive radio emission in RQQs may ultimately be similar to those of their radio-loud counterparts. 
In a recent study, \citet{Nyland+2020} examined a sample of 26 quasars that were determined to be radio-quiet based on upper limits in the FIRST survey \citep[1993--2003;][]{first}, but 
recent detections in Epoch~1 of the VLA Sky Survey \citep[observed 2017--2019;][]{vlass} and further VLA follow-up observations indicate that they have now transitioned into a clear radio-loud state.  Such results indicate that RQ and RL populations overlap, and that at least in some cases the difference between the two classes may simply be related to the duty cycle of jet activity.

In summary, we present new targeted VLA observations---combined with previous targeted VLA observations, legacy VLA survey data, and new VLA survey data---to probe the radio emission in RQQs with unprecedented (for this sample) angular resolution and years-long timescales, using an established volume-complete RQQ population. The paper is organized as follows.  Section~\ref{section:targets} introduces the target sample.  In Section~\ref{section:observations}, we describe our observation and data reduction techniques.  We discuss the morphologies present across our sample in Section~\ref{section:morphologies}, with images provided in the Appendix.  Section~\ref{section:variability} presents our spectral index and variability analysis, and we summarize our findings in Section~\ref{section:summary}.  Throughout, we assume a concordance cosmology with $H_0=70\ \mathrm{km\ s}^{-1}\ \mathrm{Mpc}^{-1}$ and $\Omega_\Lambda = 0.7$.

\section{Target selection}
\label{section:targets}

\begin{deluxetable*}{cccccccccc}
\tablecaption{Target Quasar Radio Properties}
\label{tab:radprops_ex}
\tablewidth{0pt}
\tablehead{
\colhead{Name} & \colhead{$S_\mathrm{p}$} & \colhead{$S_\mathrm{core}$} &  \colhead{$S_\mathrm{tot}$} & \colhead{$\sigma_\mathrm{rms}$} & \colhead{$S_\mathrm{core}/S_\mathrm{tot}$} & \colhead{$S_\mathrm{A}/S_\mathrm{C}$} & \colhead{$\log L_\mathrm{6GHz}$} & \colhead{Morph.} & \colhead{$\theta_\mathrm{maj}\times\theta_\mathrm{min}$} \\
\colhead{(SDSSJ)} & \colhead{($\mu$Jy beam$^{-1}$)} & \colhead{($\mu$Jy)} & \colhead{($\mu$Jy)} & \colhead{($\mu$Jy beam$^{-1}$)} & \colhead{} & \colhead{} & \colhead{($\log$[W\,Hz$^{-1}$])} & \colhead{} & \colhead{(mas $\times$ mas)}
}
\decimalcolnumbers
\startdata
080829.17+440754.1 & 36.3$\pm$8.3 & 36.3$\pm$8.3 & 36.3$\pm$8.3 & 7.0 & 1.00 & 0.60 & 21.90 & U & $<$472$\times$237 \\
081652.24+425829.4 & 178.5$\pm$7.4 & 189.2$\pm$7.6 & 199.0$\pm$14.0 & 7.6 & 0.90 & 0.90 & 22.49 & SR & 153$\times$105 \\
082205.24+455349.1 & 425.0$\pm$10.0 & 425.0$\pm$10.0 & 531.0$\pm$33.0 & 6.8 & 0.80 & 2.40 & 23.15 & M & 150$\times$89 \\
083443.80+382632.8 & 116.8$\pm$7.1 & 116.8$\pm$7.1 & 116.8$\pm$7.1 & 6.7 & 1.00 & 0.90 & 22.45 & U & $<$238$\times$39 \\
084313.41+535718.8 & 122.8$\pm$8.2 & 238.0$\pm$11.0 & 349.0$\pm$31.0 & 6.8 & 0.35 & 0.90 & 22.66 & E & 838$\times$168 \\
084755.63+263147.6 & 27.0$\pm$2.7 & 27.0$\pm$2.7 & 41.0$\pm$6.3 & 3.6 & 0.66 & 2.40 & 21.98 & SR & 312$\times$271 \\
085640.78+105755.8 & 305.0$\pm$7.1 & 467.7$\pm$8.9 & 509.0$\pm$101.0 & 6.9 & 0.60 & 1.82 & 23.04 & SR & 510$\times$237 \\
085828.69+342343.8 & 103.8$\pm$8.8 & 103.8$\pm$8.8 & 189.0$\pm$24.1 & 7.0 & 0.55 & 0.90 & 22.55 & M & 370$\times$140 \\
090151.13+103020.4 & 46.5$\pm$6.2 & 46.5$\pm$6.2 & 46.5$\pm$6.2 & 6.5 & 1.00 & 0.69 & 21.71 & U & \nodata \\
090454.99+511444.5 & 18.4$\pm$6.0 & 18.4$\pm$6.0 & 18.4$\pm$6.0 & 7.1 & 1.00 & 0.40 & 21.41 & U & $<$1500$\times$180 \\
\enddata
\tablecomments{
    Column~1: Full name of QSO in SDSSJ format;
    Columns~2, 3, and~4: Peak, core, and total 6-GHz flux densities from the VLA at $\sim0\farcs33$ resolution with uncertainties computed by the {\tt imfit} task in CASA; for Unresolved sources and Resolved sources whose core component is Unresolved and clearly separated from 
    its extended emission, columns 2 and 3 are equal.
    Column~5: RMS noise of radio image.
    Column~6: Ratio between the core flux density measured in column~3 and the total measured in column~4.
    Column~7: Ratio between our measured total 6-GHz flux density at $0\farcs33$ and that measured at $3\farcs5$ in \citet{kellerman16}.
    Column~8: Base-10 logarithm of 6-GHz luminosity, computed with total flux densities given in column~3.
    Column~9: Targets' radio morphology at $0\farcs33$ determined by eye (see Section \ref{section:morphologies} for a description of morphologies).
    Column~10: Targets' angular size deconvolved with the beam using the {\tt imfit} task in CASA; upper limits are given for unresolved sources and no estimate is available for sources too faint to derive an accurate size.  A subset of the table is shown here for brevity; a machine-readable version of the full table is available in the electronic journal.}
\end{deluxetable*}

Our radio-quiet sources (defined by their 6-GHz radio luminosity, as discussed further on in this section) are from the volume-complete QSO sample defined in \citet{kimball2011} and \citet{kellerman16}.  Their population was drawn from the \citet{schneider10} quasar catalog in the seventh data release of the Sloan Digital Sky Survey \citep{sdss}, with a narrow redshift range so as to minimize evolutionary effects.  Their selection criteria, repeated below, ensure that the parent sample is volume-complete and homogeneously selected; the selection purely on optical properties avoids any bias toward radio-loud objects.  Selection criteria for the parent sample were:
\begin{enumerate}
    \item object targeted for SDSS spectroscopy according to ``low $z$" color criteria (see below);
    \item $M_i<-23$, the historical definition for QSOs \citep{SG83};
    \item SDSS $i$-band magnitude $14 < i < 19$, after correction for extinction by Galactic dust \citep{schlegel98};
    \item $0.2 < z < 0.3$;
    \item Galactic latitude $b > 30$\degr.
\end{enumerate}
For SDSS spectroscopic targeting, the ``low $z$" criterion refers to photometric sources whose colors are non-stellar (i.e., inconsistent with the stellar locus) in a color cube defined by the SDSS $ugri$ bands; this selection ensures that sources in this sample were targeted for spectroscopy only by their optical properties. From the SDSS spectroscopic data, QSOs are identified by having at least one emission line with $\mathrm{FWHM}>1000\ \mathrm{km\ s}^{-1}$ \citep{Richards+2002a}.

Of 179 sources satisfying the above criteria, one QSO was later removed from the sample by \citet{kellerman16} 
because it happens to lie along the line of sight to a nearby strong radio galaxy, which is likely to confuse any radio follow-up.

From \citet{kellerman16}'s final sample of 178 QSOs, we selected 128 sources qualifying as ``radio-quiet" for the high-resolution follow-up observations that are presented in this paper.  The criterion we used to classify sources as ``radio-quiet" is $L_\mathrm{6GHz}<10^{23} \mathrm{\,W\,Hz}^{-1}$, where $L_\mathrm{6GHz}$ is the 6-GHz radio luminosity calculated from the \citet{kimball2011} observations. 

\begin{deluxetable}{cccccr}
\tablewidth{7in}
\tablecaption{Target Quasar Optical Properties}
\label{tab:optprops_ex}
\tablehead{
\colhead{Name} & \colhead{$z$} &  \colhead{$\Delta$} & \colhead{$M_i$} & \colhead{$m_i$} & \colhead{log\,$R_i$} \\
\colhead{(SDSSJ)} & \colhead{}  & \colhead{$('')$} & \colhead{} & \colhead{} & \colhead{}
}
\decimalcolnumbers
\startdata
080829.17+440754.1 & 0.275 & 0.150 & $-23.61$ & 17.13 & $-1.15$ \\
081652.24+425829.4 & 0.234 & 0.150 & $-23.98$ & 16.36 & $-0.72$ \\
082205.24+455349.1 & 0.300 & 0.067 & $-23.69$ & 17.27 & $0.07$ \\
083443.80+382632.8 & 0.288 & 0.078 & $-23.80$ & 17.05 & $-0.67$ \\
084313.41+535718.8 & 0.218 & 0.153 & $-23.91$ & 16.25 & $-0.52$ \\
084755.63+263147.6 & 0.282 & 0.100 & $-23.61$ & 17.20 & $-1.07$ \\
085640.78+105755.8 & 0.274 & 0.030 & $-24.32$ & 16.41 & $-0.29$ \\
085828.69+342343.8 & 0.257 & 0.222 & $-24.82$ & 15.75 & $-0.98$ \\
090151.13+103020.4 & 0.201 & 0.170 & $-24.20$ & 15.77 & $-1.59$ \\
090454.99+511444.5 & 0.225 & 0.071 & $-23.82$ & 16.42 & $-1.73$ \\
\enddata
\tablecomments{Column~1: Full name of QSO in SDSSJ format;
    Column~2: Redshift from the SDSS-DR7 catalog \citep{schneider10};
    Column~3: Angular offset of the QSO's closest detected radio component from its optical position; typical uncertainties in the optical positions of SDSS quasars are $\approx0\farcs1$;
    Column~4: Absolute $i$-band magnitude from the SDSS-DR7 catalog;
    Column~5: Apparent $i$-band magnitude from the SDSS-DR7 catalog;
    Column~6: Base-10 logarithm of $R_i$, the ratio between the measured total 6-GHz flux density at $0\farcs33$ and the optical flux density in the $i$ band\footref{footnote:iband}.  A subset of the table is shown here for brevity; a machine-readable version of the full table is available in the electronic journal.}
\end{deluxetable}

We note that in the 2010--2011 C-configuration observations---with angular resolution of 3\farcs5 ($\sim$14\ kpc)---127 of the 128 RQQs showed no sign of extended radio morphology.  Just one target (J1458+4555) showed any possibility of having extended structure, with a second radio counterpart separated by a distance of 5\arcsec; however, inspection of the optical SDSS image reveals a neighboring galaxy that coincides with the second radio component.  This galaxy may in fact be a physical companion to the QSO; its photometric redshift reported from SDSS is $0.310\pm0.0751$ \citep{photoZ_sdss} whereas J1458+4555 has a spectroscopic redshift of 0.285738.

\section{Observations and Data Reduction}
\label{section:observations}

Analyses presented in this paper are based primarily on the 128 RQQs that were observed in VLA's A-configuration from 2019 August 3 to 2019 September 20 via observing program VLA/19A-343.  We compare these results to earlier observations of the same RQQs that were observed in C-configuration as part of program VLA/10B-104.  The C-configuration observations targeted the parent sample of 178 QSOs, whereas the new A-configuration observations targeted just the 128 RQQs from the parent sample.  The C-configuration observations, performed from 2010 October 17 to 2011 January 17, have been presented in detail in \citet{kimball2011} and \citet{kellerman16}.  In this section, we discuss the 2019 follow-up observations (presented here for the first time) and where relevant we provide additional details of the earlier observations.

\subsection{Observations}

Both sets of observations were performed in the radio astronomy ``C-band" which comprises the frequency range 4--8~GHz.  The 2010--2011 observations were performed during the commissioning and verification phase of the upgrade from the legacy VLA hardware and software to the new Karl G.\ Jansky VLA.  As only 2~GHz of bandwidth was available for C-band observations at that time, the original set of observations covered the frequency range 5--7~GHz.  The 2019 observations used the full 4--8~GHz range of the C-band receiver by utilizing two pairs of the VLA's 3-bit samplers in each antenna.  Each sampler pair delivered 2 GHz of bandwidth in both right and left circular polarizations.  The correlator was set up to split each 2 GHz of bandwidth into 16 subbands, each 128 MHz wide with $64\times2$\,MHz spectral channels.

The 2010--2011 observations were performed in the VLA's C-configuration, whereas the 2019 observations used the more extended A-configuration.  At a reference frequency of 6~GHz, the typical angular resolution for C-configuration is $\sim$3\farcs5 ($\sim$14~kpc for $z=0.25$), and the typical angular resolution for A-configuration is $\sim$0\farcs33 ($\sim$1.3~kpc for $z=0.25$).  Baseline lengths in the C-configuration range from a minimum of 350~m to a maximum of 3.4~km; baseline lengths in the A-configuration range from a minimum of 0.68~km to a maximum of 36.4~km.  The largest angular scale of emission detectable at 6~GHz is 4\arcmin\ in C-configuration and $\sim4\farcs5$ in A-configuration.

The A-configuration observations were designed to reach signal-to-noise (S/N) levels of at least five, for peak flux densities equal to those calculated from the original 2010--2011 observations (including two RQQs with marginal detections).  Required on-source observing times were determined using the VLA Exposure Calculator Tool\footnote{\url{https://obs.vla.nrao.edu/ect/}};
we assumed 15\% loss of bandwidth due to known radio frequency interference (RFI), leaving an expected useful bandwidth of 3.4~GHz. The faintest detection in the 2010--2011 observations was J0847+2631 with peak flux density of 16.8~$\mu$Jy~beam$^{-1}$, which for the new observations required 40 minutes on-source to reach a target 1-$\sigma_\mathrm{rms}$ sensitivity of 3.36~$\mu$Jy\,beam$^{-1}$ (where $\sigma_\mathrm{rms}$ refers to the rms noise of the image). 
The next 12 most-faint ($\lesssim33.5\mu$Jy~beam$^{-1}$) targets required on-source times of 15--35 minutes to reach target sensitivities in the range of 3.6--5.5~$\mu$Jy\,beam$^{-1}$.  The remaining 116 targets ($>33.5\mu$Jy\,beam$^{-1}$) were observed for 10~minutes each, split into two 5-minute scans separated by about an hour in order to obtain improved $uv$-coverage. 

All of the 2019 observations used the radio source 3C~286 as both the primary flux density scale calibrator and bandpass calibrator, with complex gain calibrators selected individually for each science target.  We incorporate a systematic uncertainty of 3\% in the flux density values we determine from these observations, following NRAO guidelines\footnote{\url{https://science.nrao.edu/facilities/vla/docs/manuals/oss/performance/fdscale}} based on the fundamental accuracy of the flux density scale for 3C~286 \citep{perleyButlerFlux}.  The error values reported in Table~\ref{tab:radprops_ex} comprise this 3\% uncertainty added in quadrature with the statistical errors reported for the flux density measurements from CASA (see below).

\subsection{Data Reduction}

We used the Common Astronomy Software Applications package \citep[CASA;][]{casa} (versions 5.6.2 and 5.6.3) to reduce our VLA data and perform all imaging in this analysis\footnote{Images and flux density measurements in \citet{kellerman16} were performed in AIPS.}.  We used the standard VLA calibration pipeline to automatically flag and calibrate each data set.  We experimented with further flagging manually, but deemed such manual intervention to be unnecessary (possibly even counterproductive), because the data directly output from the the VLA pipeline had been efficiently flagged and weighted; further manual flagging had a tendency to actually increase the image noise.  We therefore chose to proceed with imaging analysis directly after the successful pipeline run.

While the imaging process is largely tunable and several small details may differ from target to target, we outline our general process for imaging here.  The CASA task {\tt tclean} was used with the natural weighting scheme to deconvolve all images.  We used a pixel size of $0\farcs06$ in order to adequately sample the synthesized beam (FWHM $\sim0\farcs33$).  Image cutouts for sources showing any structure are presented in Appendix~\ref{sec:app_images}, along with optical cutouts from SDSS and the corresponding image cutouts from the 2010--2011 observations \citep{kellerman16}.

\begin{figure}[h!]
    \epsscale{1.1}
    \plotone{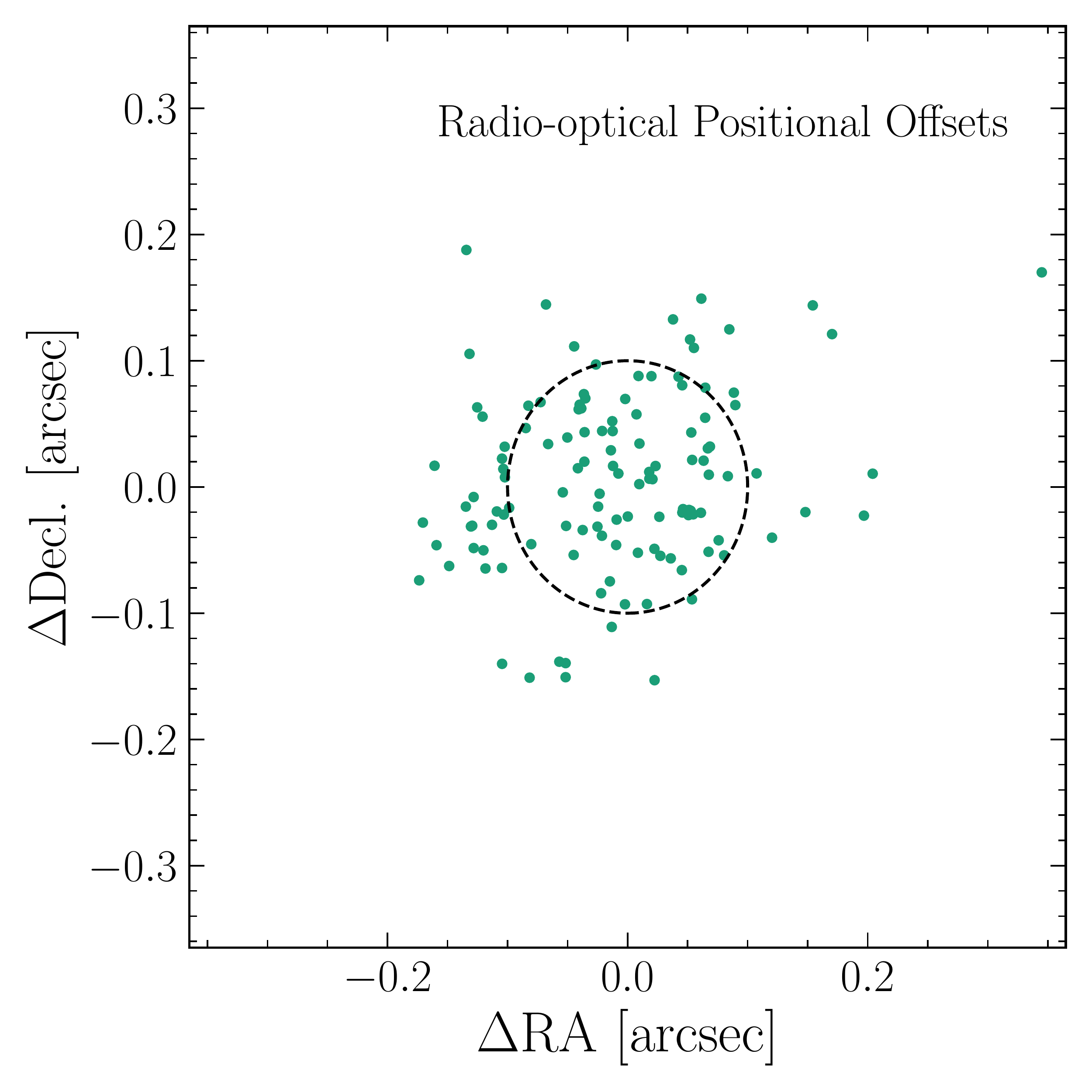}
    \caption{Angular offsets between radio and optical positions of our targets.  Typical SDSS positional uncertainties are $\approx0\farcs1$, indicated by the circle centered at the origin.  Not pictured is J1102$+$0844, which has an angular offset of 0\farcs81.}
    \label{fig:fig1_position_offsets}
\end{figure}

For each target, we have imaged the full field-of-view, i.e., the FWHM of the antenna primary beam ($\sim$7\farcm3$\times$7\farcm3 at 6 GHz), in order to include any potential bright sources that---if not properly deconvolved---could result in imaging artifacts and increase the image noise at the location of the science target; when such cases appeared we employed phase-only self-calibration to improve the image.  If remaining noise or artifacts appeared to affect the target location, we would try one or more of the following with CASA in an attempt to achieve the sensitivity goal (typically $\approx7$\,$\mu$Jy~beam$^{-1}$): a conservative round of amplitude self-calibration, multi-term multi-frequency synthesis \citep[MT-MFS;][]{mtmfs}, or aw-projection \citep{wproj, aproj}.  Nevertheless, some images containing nearby bright sources remain dynamic-range limited with sensitivities as poor as $\sim30~\mu\mathrm{Jy~beam}^{-1}$.  Almost all (122 of 128) targets have a detected radio component associated with their known SDSS optical position; radio-optical offsets are shown in Figure~\ref{fig:fig1_position_offsets}.

\begin{figure}[bh!]
    \epsscale{1.25}
    \plotone{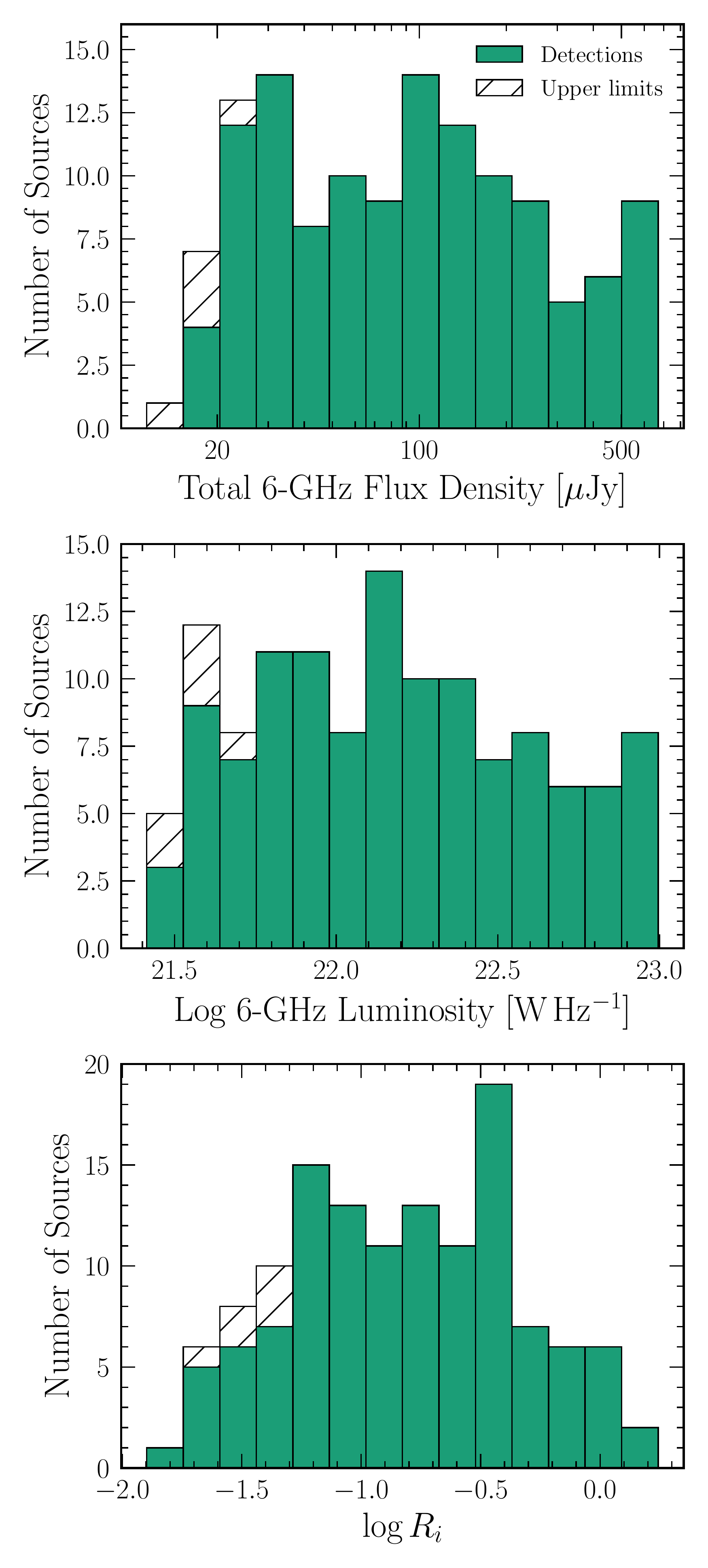}
    \caption{Distributions of the radio properties of the RQQ targets observed at $\sim0\farcs33$ resolution.  {\em Top}: total 6-GHz flux density measured with the VLA.  {\em Middle}: 6-GHz luminosity.  {\em Bottom}: base-10 logarithm of $R_i$, where $R_i$ represents the ratio between our measured 6-GHz flux density and the source's optical $i$-band flux density \citep{schneider10}.
    \label{fig:fig2_radio_distributions}}
\end{figure}

Compilations of radio and optical properties of the RQQ sample are provided in Tables \ref{tab:radprops_ex} and \ref{tab:optprops_ex}, respectively.  Notes on individual sources can be found in Appendix~\ref{section:notes}.  Distributions of integrated 6-GHz flux density and luminosity, as well as the ratio between 6-GHz and optical $i$-band flux density\footnote{The $i$-band flux densities are computed from SDSS apparent magnitude as: $S_i = 10^{(9.56-i/2.5)}\mu$Jy \citep{ABmag,sdss}\label{footnote:iband}.}, $R_i$, are shown in Figure~\ref{fig:fig2_radio_distributions}.

\section{Radio Morphology of RQQs}
\label{section:morphologies}
In this section, we introduce morphology classifications of RQQs based on our new observations, then compare morphologies and flux densities to the lower-resolution observations from 2010--2011.  Corresponding VLA images for the new observations can be found in Appendix~\ref{sec:app_images} for all sources showing structure (``Resolved" classification; see next sub-section)\footnote{A tar.gz file containing image cutouts of all sources is available at \url{https://github.com/trevormccaffrey/RQQs/blob/main/RQQs_15arcsec_cutouts.tar.gz}.}.

\subsection{Morphology Classification}
\label{subsec:morphology}

These new A-configuration observations allow us to investigate kpc-scale radio structure in the population of $z\sim0.25$ RQQs.  Upon visual inspection, we find that RQQs present a range of radio morphology on sub-galactic scales (1.3\,kpc $\lesssim\Delta\lesssim$ 14\,kpc); for the purposes of analysis, we introduce the following classes and subclasses to categorize the range of observed morphology:

\begin{figure*}[th!]
    \plotone{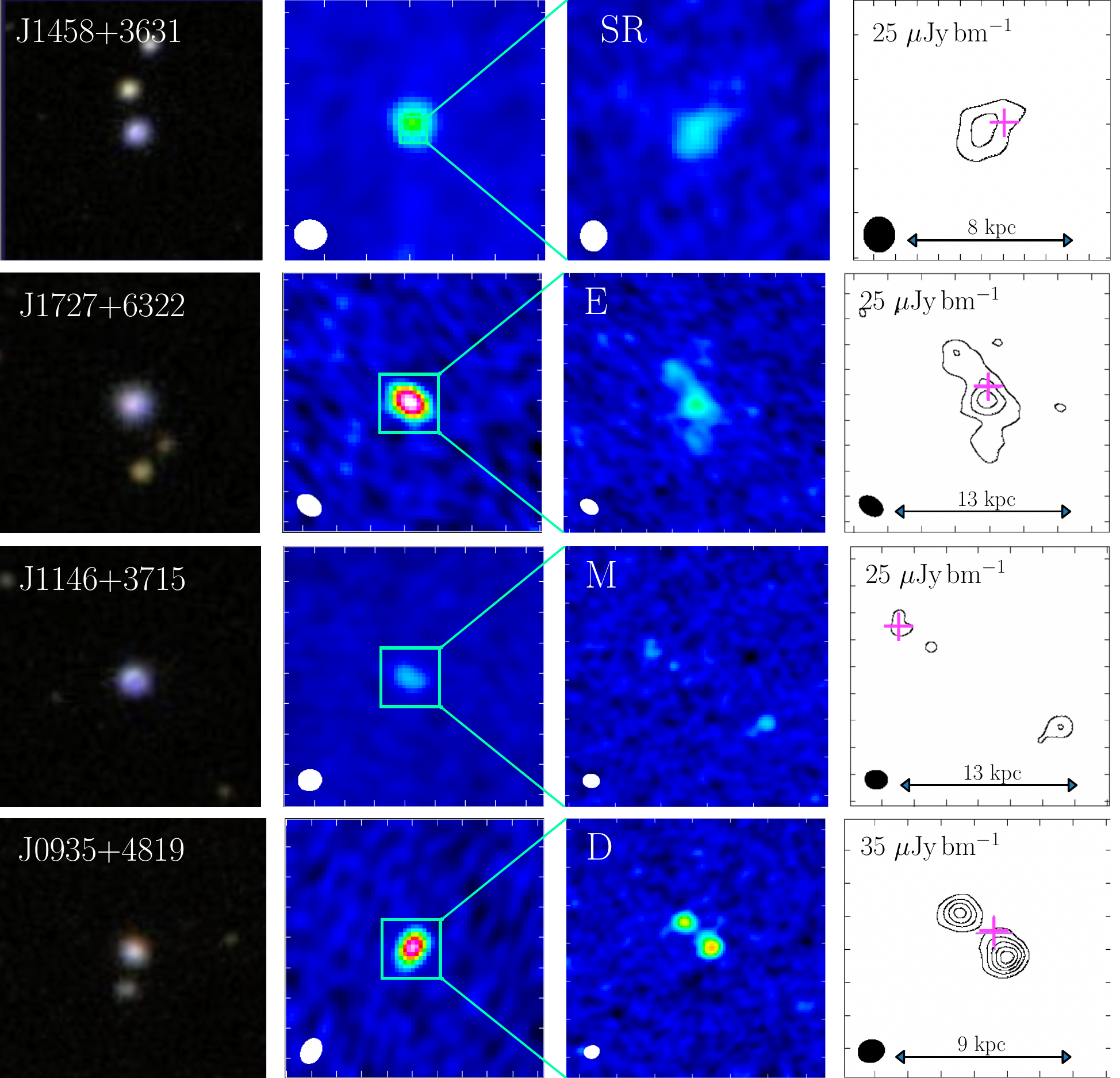}
    \caption{Example images of sources with ``Resolved" morphology; one source from each sub-class is shown.  \textbf{Column~1} shows each object's arcminute-wide SDSS color image.  \textbf{Column~2} shows the corresponding $\sim3\farcs5$-resolution 6-GHz image from \citet{kellerman16}, where all sources appeared unresolved.  \textbf{Column 3} shows the source's higher-resolution ($\sim0\farcs33$) A-configuration image, where the size of the image relative to the lower-resolution is indicated by the green inset region from Column~2; each source's A-configuration morphology sub-classification is noted in the upper left corner of this panel.  \textbf{Column~4} further displays contour maps of the A-configuration images to highlight extended features; the number in the upper left of each contour map indicates the base contour value (typically equivalent to a detection with signal-to-noise $\approx$3.5--4; consecutive contours are spaced by multiples of $\sqrt{2}$; see Column~4 of Table~\ref{tab:radprops_ex} for exact $\sigma_\mathrm{rms}$ values).  An approximate linear size scale is given in each contour map.  The SDSS location of each QSO is indicated by the magenta cross in the contour maps.  The approximate FWHM of the synthesized beam (resolution element) for each radio observation is indicated by the ellipse in the lower left corner of each image.  Image cutouts and individual notes for \emph{all} Resolved sources in the sample are included in Appendix~\ref{sec:app_images}.}
    \label{fig:fig3_image_inset_example}
\end{figure*}

\begin{itemize}

\item \textbf{U}---\textit{Unresolved}: the source remains unresolved on A-configuration scales ($\sim0\farcs33$, corresponding to $1.3$\,kpc at $z=0.25$).  A total of 67/128 of our sources are Unresolved.

\item \textbf{R}---\textit{Resolved}: we observe resolved structure and/or extended radio emission in 55/128 RQQs, revealing a variety of appearances.  We adopt the sub-classes below to describe this diversity.  A representative image from each of the different sub-classes is presented in Figure~\ref{fig:fig3_image_inset_example}; images of all Resolved sources appear in Appendix~\ref{sec:app_images}. 

\begin{itemize}

\item \textbf{R-SR}---\textit{Slightly Resolved}: a formal Gaussian fit to the emission component in the image\footnote{Source fitting was performed using the {\tt imfit} task in CASA.} indicates that the source is resolved (i.e., the integrated source flux density is significantly greater than the peak flux density) and the source appears visually to have slight extended structure beyond the $\sim1.3$\,kpc synthesized beam.  There are 31/128 sources in this category.

\item \textbf{R-E}---\textit{Extended}: the source has a dominant radio
component coinciding with the QSO's optical position, and also has obvious widespread emission extending well beyond the resolution of the synthesized beam.  We identify 7/128 of our sources with clear extended morphology.

\item \textbf{R-M}---\textit{Multi-component}: the source has a radio component coincident with the optical SDSS position of the QSO, and also has one or more emission components---well separated from the core---contributing to the source's total radio emission.  We utilize the source-finding tool PyBDSF \citep{PyBDSF} with a S/N cutoff of 3 to distinguish faint emission components from unrelated noise peaks.  There are 16/128 sources in this category.  

\item \textbf{R-D}---\textit{Double}: the source shows two apparent radio ``lobes" symmetric about the central QSO position.  There is only one RQQ in our sample that has a Double morphology: J0935+4819 (shown in the fourth panel of Figure~\ref{fig:fig3_image_inset_example}).

\end{itemize}

\item \textbf{ND}---\textit{Non-detection}: no potential radio component from the source is detected.  Typically, 3-$\sigma_\mathrm{rms}$ point-source upper limits for Non-detections in our sample are $\sim$21\,$\mu\mathrm{Jy}$ (corresponding to $\sim10^{21.6}$\,W\,Hz$^{-1}$ at $z=0.25$), but vary slightly depending on sensitivity reached in each image, with our most sensitive non-detection (J1703+1910) achieving a 3-$\sigma_\mathrm{rms}$ point-source upper limit of 11.4\,$\mu\mathrm{Jy}$ ($\sim10^{21.4}$\,W\,Hz$^{-1}$).  Only 6/128 total sources are Non-detections in our A-configuration observations; all were robust detections in the original C-configuration observations (with peak flux densities in the range from 19.7--81~$\mu$Jy\,beam$^{-1}$).

\end{itemize}

\begin{figure}[t!]
    \epsscale{1.17}
    \plotone{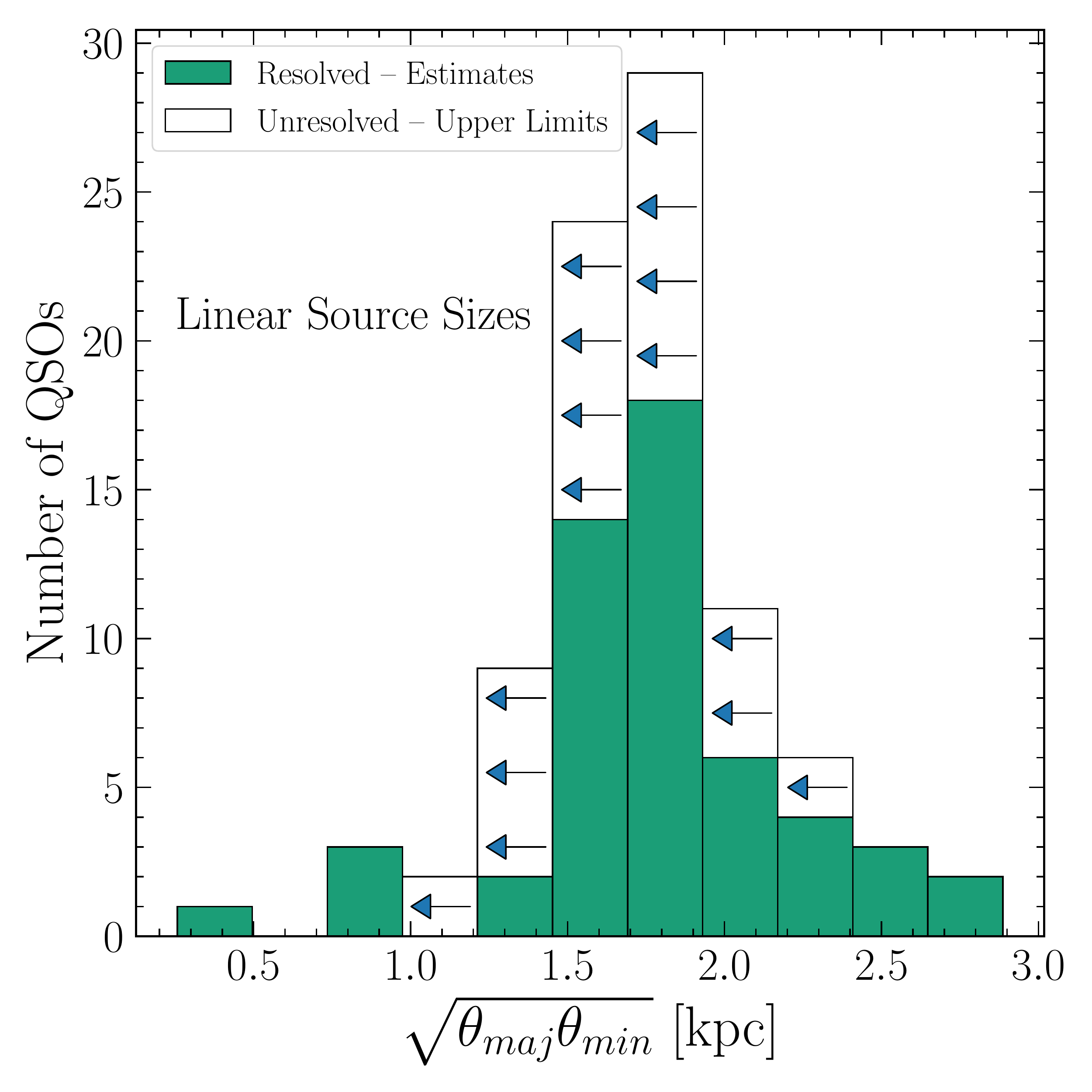}
    \caption{Histogram of projected physical size of RQQ radio emission, computed as the geometric mean of the major and minor axes (deconvolved).  For Multiple-Component sources, the size corresponds to the component that is co-located with the optical QSO.  For Extended sources (i.e., with emission connected to the core component), the size corresponds to the extent of the extended emission.  Upper limits are shown for Unresolved sources.  A total of 90/128 sources are included in this figure; the remaining 38 are too faint to derive accurate size estimates (see Table~\ref{tab:radprops_ex}).
    \label{fig:fig4_sourcesizes_geommean}}
\end{figure}

\begin{figure}[t!]
    \epsscale{1.17}
    \plotone{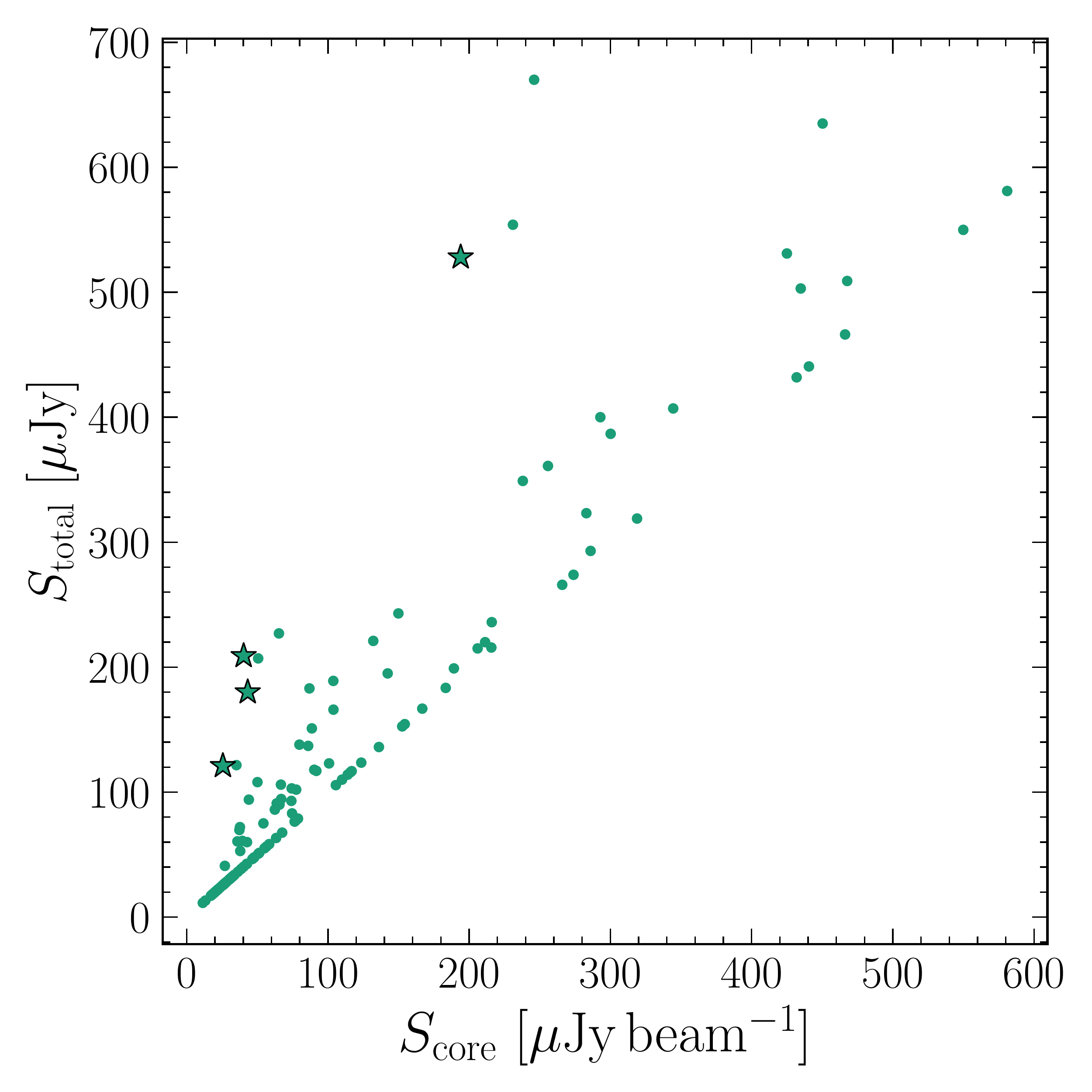}
    \caption{Total 6-GHz flux density in A-configuration versus that associated with the SDSS optical counterpart for all detections.  The core is the dominant radio component for all RQQs except for J0935$+$4819, J1118$+$3103, J1146$+$3715, and J1233$+$6443 (marked by star symbols)---all of which exhibit edge-brightened, extended morphology.}
    \label{fig:fig5_Stot_vs_Score}
\end{figure}

\begin{figure}[ht!]
    \epsscale{1.17}
    \plotone{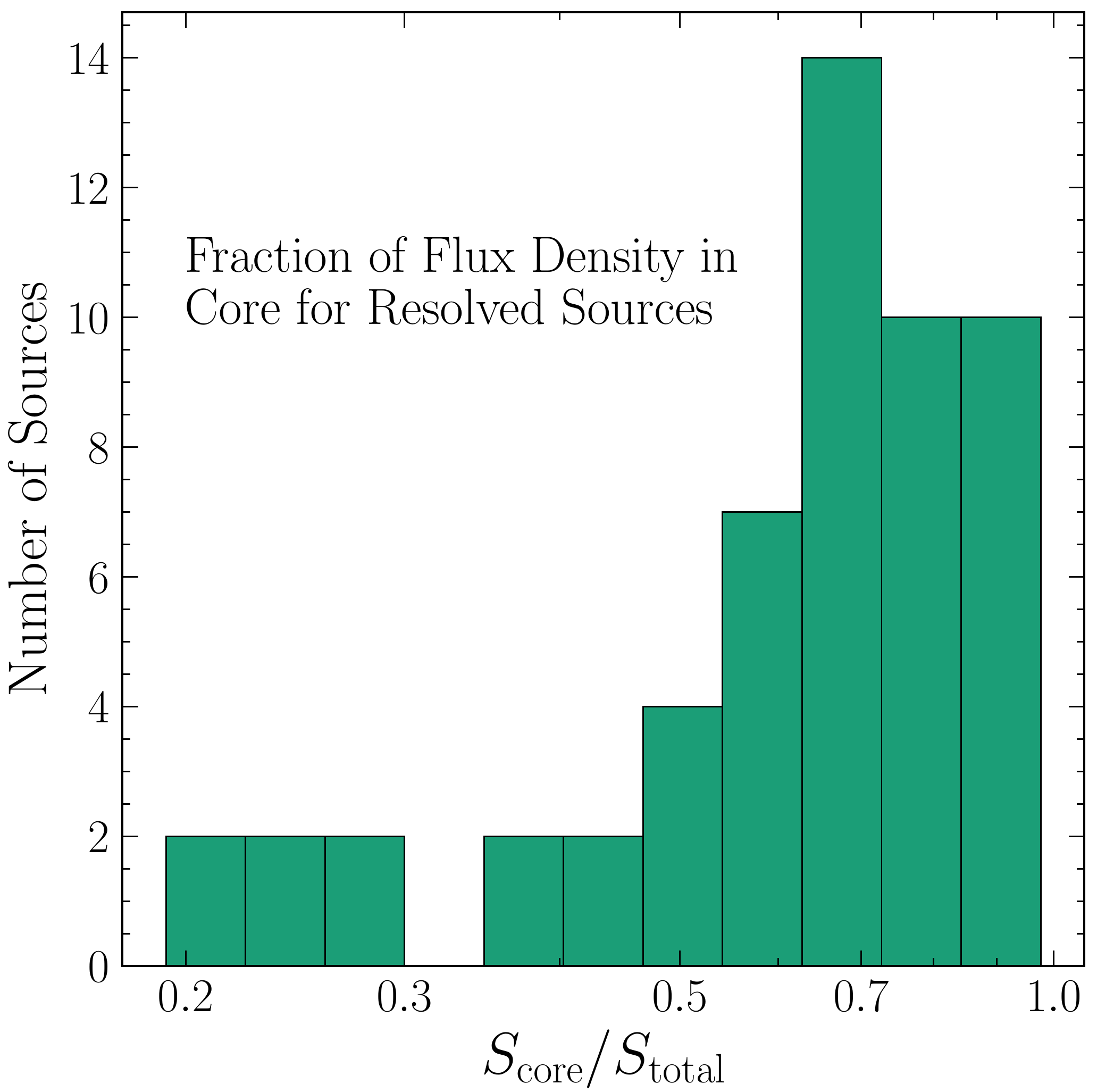}
    \caption{Histogram showing ratio of core- to total-6-GHz emission detected in A configuration for the 55 Resolved sources.  For sources without a core clearly separated from more extended structure, $S_\mathrm{core}$ is the peak flux density (determined by Gaussian fit) of the source.  The 67 Unresolved sources are not included in this figure; core-to-total ratio for unresolved sources is by definition equal to unity.}
    \label{fig:fig6_Score_over_Stot}
\end{figure}

Figures~\ref{fig:fig4_sourcesizes_geommean}, \ref{fig:fig5_Stot_vs_Score}, and~\ref{fig:fig6_Score_over_Stot} characterize the RQQs in terms of the physical size and---for the Resolved sources---central concentration of their radio emission.  Figure~\ref{fig:fig4_sourcesizes_geommean} shows the projected physical size for all sources (based on the geometric mean of the axis sizes after deconvolution).
The bulk of the detected radio emission for roughly 80\% of the sample is constrained within a $\sim2$~kpc region.
Figure~\ref{fig:fig5_Stot_vs_Score} shows total 6-GHz flux density detected in 2019, $S_\mathrm{total}$, plotted against flux density $S_\mathrm{core}$ associated with the core component (i.e., the radio component coincident with the QSO optical position)---both taken from the $0\farcs33$ observations---for all detections.  For sources without a core clearly separated from the source's extended structure, we take $S_\mathrm{core}$ to be the peak of the Gaussian fit to the radio source; for the single source with Double morphology, this corresponds to the peak of the southwest lobe partially coincident with the QSO optical position. Figure~\ref{fig:fig6_Score_over_Stot} shows the distribution of $S_\mathrm{core}/S_\mathrm{total}$, effectively displaying what fraction of overall radio emission in these 55 resolved sources is confined within a single $\sim0\farcs33$ resolution element of these observations.

\begin{figure}[ht!]
    \epsscale{1.17}
    \plotone{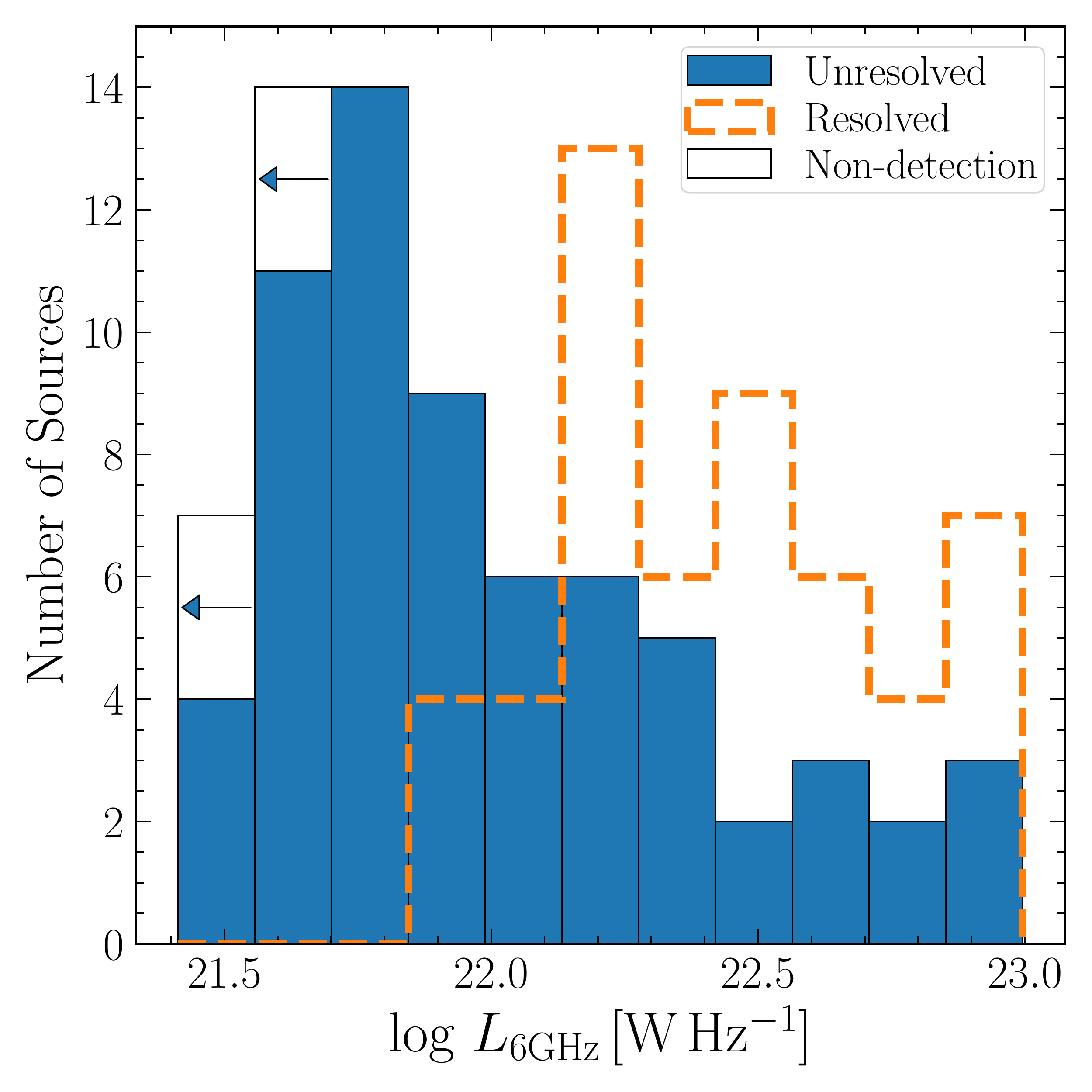}
    \caption{Luminosity distributions at 6 GHz for the different morphology classes of our sample.  The open histogram with arrows emphasizes that the Non-detections are plotted as 3-$\sigma_\mathrm{rms}$ upper limits.  See Section~\ref{section:morphologies} for descriptions of morphology classes. 
    \label{fig:fig7_morphologies_L6_v2}}
\end{figure}

Luminosity distributions from the three main morphology classes are shown in Figure~\ref{fig:fig7_morphologies_L6_v2}.  Resolved sources are typically found among the higher-luminosity range of our sample ($L_\mathrm{6GHz}\simeq10^{22.2}$--$10^{23.0}$ W\,Hz$^{-1}$), whereas the less-luminous range ($L_\mathrm{6GHz}\simeq10^{21.4}$--$10^{22.2}$ W\,Hz$^{-1}$) is populated primarily by Unresolved sources.  Implications of this result, in the context of how a visual morphology classification depends on an interferometer's sensitivity as a function of angular scale, are discussed in Section~\ref{subsec:angular_resolution}.

\subsection{Image constraints on the origin of radio emission in RQQs}
\label{subsec:imageconstraints}

These new VLA observations are the first to probe kpc-scale radio emission across a volume-complete sample of homogeneously selected RQQs.  We see that extended (but sub-galactic) radio emission is visible in a large fraction of the population: 55 of 128 RQQs are Resolved.  Unresolved central components (i.e., coincident with the optical position) could be dominated by AGN-related radio emission---small-scale jets, shocks, or winds \citep{ZG14,Zakamska+16}---or nuclear starburst emission \citep[e.g.,][]{condon1992b,colina1993,anantharamaiah2000,lonsdale2006}, whereas emission outside the central component may arise from diffuse or clumpy star-formation, or could indicate AGN-driven features such as jet or wind interactions with the intragalactic medium \citep[e.g.,][]{Klindt+2019,Rosario+21}. Note that some sources may have faint diffuse emission that we fail to detect due to sensitivity limits of these high-resolution observations (discussed further in \S\ref{subsec:angular_resolution}).

Among the 16 Multi-component sources, a separate component is typically found within a projected distance of about 10~kpc (3\arcsec\ at $z=0.2$), potentially indicating a star-forming region near the outskirts of the host galaxy.  However, it is possible that one or more of these additional components correspond to an unrelated background source.  To estimate this likelihood, we look to \citet{Smolcic2017_radioCOSMOS}, who performed a sensitive high-resolution ($\sim0\farcs75$) survey of the $\sim$2.6\,deg$^{2}$ COSMOS field; the authors found an overall source density of $\sim4200$\,deg$^{-2}$ above about $11.5$\,$\mu\mathrm{Jy}$\,beam$^{-1}$.  Using their result, the expectation value of the number of background interlopers with flux density $\gtrsim11.5$\,$\mu\mathrm{Jy}$\,beam$^{-1}$ within 3\arcsec\ of any of the 128 RQQs is about equal to~1.  We therefore expect that it is likely that one among the Multi-Component class may be mis-classified due to an unassociated background source, but it is safe to assume that nearly all of the detected components are physically associated with the RQQs.

Convincing visual evidence for an AGN as the dominant origin of RQQ radio emission might manifest as a morphology that is distinctly associated with powerful QSOs and radio galaxies, albeit scaled down, such as small-scale jets or lobes analogous to the classical picture of powerful RLQs as double-lobed or triple sources \citep[e.g., see Figure 1 of][]{Kimball+11b}.  In fact, J0935+4819---the source classified as a ``Double" and shown in Figure~\ref{fig:fig3_image_inset_example}---does have a morphology that is strikingly similar to that classical picture: two round, similarly bright lobes appear on either side of the central black hole (whose position is determined by the optical QSO position).  The southwest lobe of J0935+4819 source has a slight extension toward the northeast that clearly coincides with the optical position, suggesting possible radio emission from a core in addition to the two lobes.  This source's appearance is similar to that of Gigahertz-Peaked Spectrum (GPS) and Compact Steep Spectrum (CSS) sources with symmetric-double morphology.  The GPS and CSS labels indicate two classes of powerful radio objects, with the main differences being that GPS sources are more compact ($\lesssim1$~kpc) whereas CSS sources are larger ($\lesssim15$~kpc) and with steep spectra.  As discussed in reviews by \citet{Odea1998}, \citet{sadlerReview}, and \citet{Odea+Saikia21}, there are two main theories explaining GPS and CSS sources: (1) in the ``frustrated" scenario, GPS/CSS sources consist of jets that are confined by dense gas in the host galaxy, preventing them from evolving into large radio galaxies; (2) in the ``evolutionary" scenario, newly formed jets manifest as GPS sources, later becoming CSS sources as they grow and age, eventually evolving into large classical radio galaxies.  The projected physical extent of J0935+4819's lobes is $\gtrsim6$~kpc and its spectral index is $\alpha\sim-0.7$ (for $f_\nu\propto\nu^\alpha$) between 1.4 and 6~GHz.\footnote{Spectral indices of the entire sample are discussed in \S\ref{section:variability}.}  The visual appearance of J0935+4819 is therefore consistent with classification as a low-power CSS.  
RQQ J0935+4819 appears to be the first example of a radio-\textit{quiet} QSO that shows kpc-scale FR~II-like morphology \citep{Fanaroff1974} at a radio luminosity as low as 10$^{22.9}$\,W\,Hz$^{-1}$.  
Other radio AGNs of this size and luminosity tend to manifest as FR~I sources \citep[viz.\ Figure~1 of][]{anBaan}, which may indicate that they are ``frustrated" sources with a jet interrupted by the dense interstellar-medium\footnote{There also exists a large population of compact ``FR~0" sources \citep{Baldi+2015,Baldi+2018}, some of which may host radio jets on much smaller size scales.  \citet{Odea+Saikia21} suggest that many of these are in fact low-power CSS or GPS sources.}.  In contrast, J0935+4819 may be an example of an RQQ that could properly evolve into FR~II radio galaxy as it grows.  Regardless, the symmetric-double morphology of J0935+4819 is strongly suggestive that AGN activity dominates its radio emission, given the analogous appearance with classical double radio galaxies and QSOs.

Other intriguing images that may indicate the presence of radio jets are those with apparent linear or widespread flocculent emission extending outward from the core in just one direction, or symmetrically in two directions.   Among the Extended RQQs: J0843+5357, J1235+4104, J1304+0205, and J1727+6322 appear to have widespread two-sided emission; J1408+4303 and J1444+0633 appear to have linear emission extending out in one direction.  Among the Multi-component RQQs: J1146+3715 has two additional components along the same (projected) direction, while J1118+3103 has two symmetrically located components (along with a third), and J1004+1510 has an extension of the central component that extends directly toward an additional emission component.  Such morphology does not prove the presence of a radio jet, but---similar to J0935+4819---is the type of morphology that small-scale jets may manifest if present.  Images of these sources are all presented 
and described in detail in Appendix~\ref{notes}.

Further investigation of scaled-down jets as a possible origin of radio emission in the RQQs likely requires deeper follow-up observations with the VLA or observations with Very Long Baseline Interferometry (VLBI), which detects only high brightness temperatures \citep[although deep VLBI observations may still be sensitive to extreme starburst activity, e.g., Arp 220;][]{lonsdale2006}.  
\citet{ulvestad2005}, \citet{Chi+13}, and \citet{Maini+16} have used VLBI to detect jets in small numbers of purported RQQs, but many detections in their samples qualify as either radio-intermediate or -loud by the definition used in this paper---so we would expect a significant jet contribution in those objects---or fail to meet the historical $M<-23$ optical criterion of bona fide QSOs \citep{SG83}.  Similarly, \citet{HR16,HR17} detect a jet in 3/18 $M<-23$ QSOs in the COSMOS field, but these targets are not RQ by our 6-GHz luminosity criterion\footnote{These three sources are all brighter than $10^{24}$~W/Hz at 1.4 GHz; assuming a spectral index of $\alpha=-0.7$, the faintest would have 6-GHz luminosity of $10^{23.9}$~W/Hz.}.

\subsection{Comparison of A-configuration and C-configuration Flux Densities}
\label{subsec:angular_resolution}

To further investigate the intrinsic radio structure of RQQs and their host galaxies, we employ the advantage that this sample has been observed at the same frequency with different angular-resolution scales.  Comparison of flux density measured from observations made at low-resolution (VLA's C-configuration) versus high-resolution (VLA's A-configuration) allows us to probe their radio morphology using a quantitative method.  Such an approach has previously been used successfully for automated morphology classification of large catalogs made from wide-field sky surveys performed with the VLA \citep{kimballIvezic}.

Recall that these follow-up observations were designed to achieve $\geq5\sigma$ {\it point-source} sensitivity based on the Kellermann et al.\ flux densities measured in C-configuration.  However, if the radio emission is not centrally concentrated but instead is spread throughout the host galaxy, then emission in the higher-resolution image will be distributed across multiple resolution elements, yielding lower surface-brightness per beam, i.e., a lower point-source flux density.  Therefore, at higher resolution the emission from extended sub-galactic structures may be hidden within the image noise, leaving us sensitive only to the brightest emission, such as a bright core component.\footnote{We note that the largest angular scale of these A-configuration observations is 4\farcs5 or larger (see \url{http://go.nrao.edu/vla-res}) whereas \citet{kellerman16} determined that the radio emission of each of these RQQs is unresolved on a 3\farcs5 scale.  Therefore, we should not miss any emission due to interferometric spatial filtering.}
If we had observed for a longer amount of time to reach lower image noise (equivalent to better surface-brightness sensitivity), such extended emission may have become apparent above the noise level \citep[e.g.,][]{BR01}.  Therefore, a decrease in flux density between the two epochs cannot be attributed simply to variability, because detection in the A-configuration images depends strongly on how radio emission is distributed throughout the host galaxy.

\begin{figure}[!ht]
    \epsscale{1.15}
    \plotone{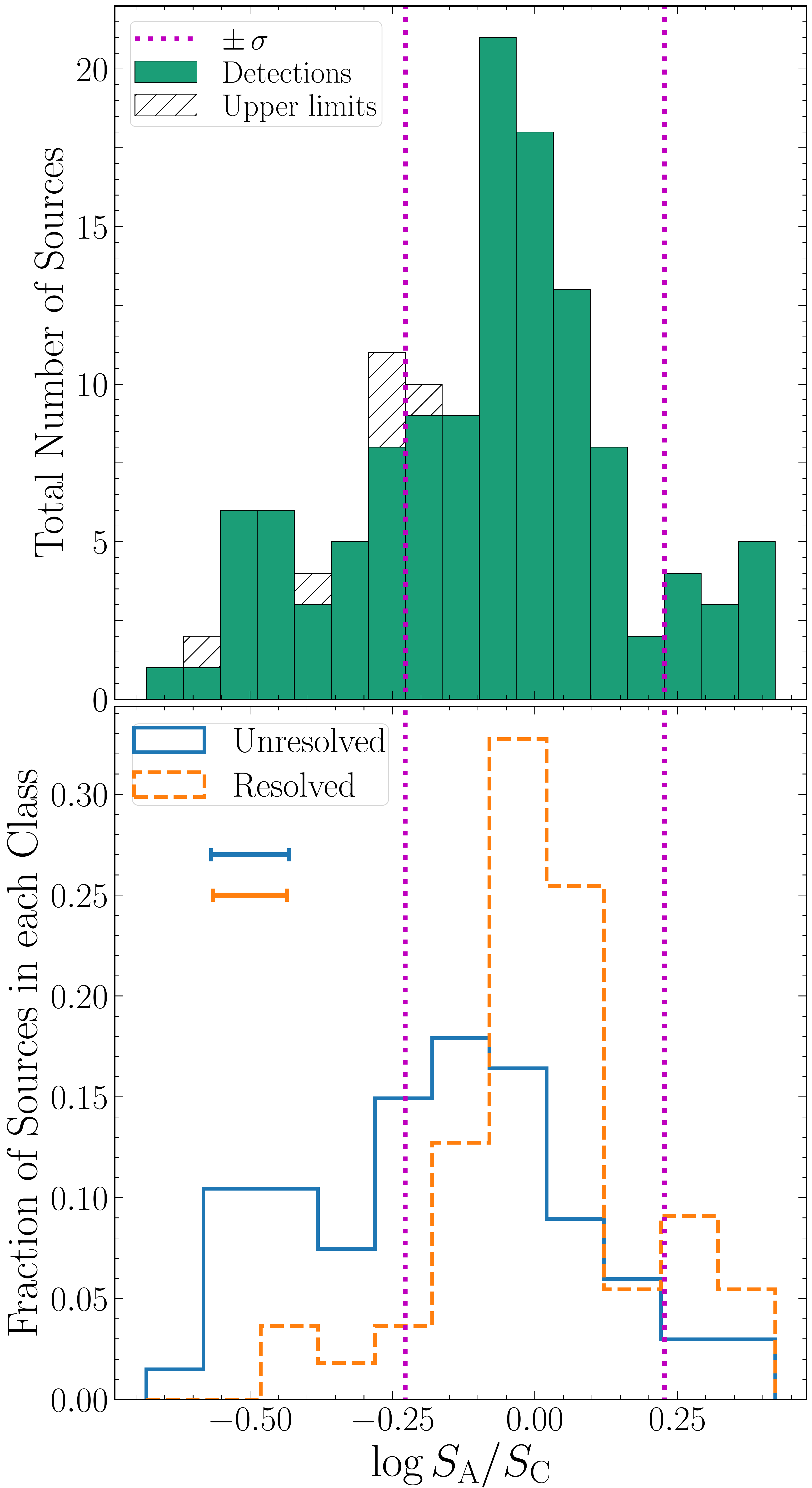}
    \caption{\textit{Top:} Distribution of $\log(S_\mathrm{A}/S_\mathrm{C}$) (see text) for the RQQ sample. Magenta dotted lines mark $\pm$1-$\sigma$ offsets from 0 (where $\sigma$ is the standard deviation of the sigma-clipped main distribution).  As discussed in \S\ref{subsec:angular_resolution}, the excess of sources $<-\sigma$ compared to those $>+\sigma$ suggests the presence of radio emission below the surface-brightness sensitivity of our follow-up observations.  \textit{Bottom:} Marginal distributions of $\log(S_\mathrm{A}/S_\mathrm{C})$ for sources that appear unresolved in our  0\farcs33-scale images [solid blue; median $\log(S_\mathrm{A}/S_\mathrm{C})\approx-0.15$)] and those that have been resolved [dash orange; median $\log(S_\mathrm{A}/S_\mathrm{C})\approx0.0043$)], plotted as the fractional total from each class.  Median error bars of $\log(S_\mathrm{A}/S_\mathrm{C})$ for each class of object are shown for reference.
    \label{fig:fig8_fluxratio_2panel}}
\end{figure}

To compare the flux density values measured in the high- and low-resolution observations, we use the ratio $S_\mathrm{A}/S_\mathrm{C}$, where $S_\mathrm{A}$ is the source's total A-configuration 6-GHz flux density\footnote{The variable $S_\mathrm{A}$ is equal to $S_\mathrm{total}$ from \S\ref{subsec:morphology}, but for clarity is labeled here with subscript ``A" to distinguish from the measurements made in C-configuration.} ($0\farcs33$ scale) measured from our 2019 observation, and $S_\mathrm{C}$ is the C-configuration 6-GHz flux density ($3\farcs5$ scale) from 2010--2011.  The upper panel of Figure~\ref{fig:fig8_fluxratio_2panel} shows the distribution of $\log(S_\mathrm{A}/S_\mathrm{C})$ for our complete RQQ population. If we were sensitive to all radio emission at all angular scales, then our measured flux densities for non-variable sources would be equivalent to those measured in 2010--2011, and the distribution should cluster (within measurement errors) around $\log(S_\mathrm{A}/S_\mathrm{C})\sim0$.  However, the sample instead manifests a range of $\log(S_\mathrm{A}/S_\mathrm{C})$, with a median value $\approx-0.07\pm0.03$ 
and a skew with negative values of $\log(S_\mathrm{A}/S_\mathrm{C})$.  The lower panel of  Figure~\ref{fig:fig8_fluxratio_2panel} shows the marginal distributions for the Unresolved and Resolved classes, and demonstrates that the skew is due entirely to the Unresolved class (median $\log(S_\mathrm{A}/S_\mathrm{C})\approx-0.15\pm0.04$), whereas the distribution for Resolved sources is roughly symmetric around $\log(S_\mathrm{A}/S_\mathrm{C})=0$, with a median $\log(S_\mathrm{A}/S_\mathrm{C})\approx0.01\pm0.03$; implications are discussed in \S\ref{subsubsec:morphology}.

For any given source, we consider that flux density values $S_\mathrm{A}$ and $S_\mathrm{C}$ are consistent if $\log(S_\mathrm{A}/S_\mathrm{C})$ lies within $3\sigma$ of 0 (where $\sigma$ has been calculated for each source individually via propagation of errors).  By that definition, out of 55 Resolved (67 Unresolved) sources, there are 46 (37) with consistent values of $S_\mathrm{A}$ and $S_\mathrm{C}$.  The remaining 9 Resolved (30 Unresolved) sources have inconsistent flux density measurements, due either to variability and/or to the difference in angular resolution discussed earlier.  The 39 sources with a significant change between $S_\mathrm{A}$ and $S_\mathrm{C}$ are listed in Table~\ref{tab:var_sample}.  Of the 9 Resolved sources listed, 6 have $\log(S_\mathrm{A}/S_\mathrm{C})>0$ and 3 have $\log(S_\mathrm{A}/S_\mathrm{C})<0$.  Among the Unresolved sources, only 4 have $\log(S_\mathrm{A}/S_\mathrm{C})>0$, while the remaining 26 have $\log(S_\mathrm{A}/S_\mathrm{C})<0$.  Additionally, all six Non-detections have $\log(S_\mathrm{A}/S_\mathrm{C})<0$; they have 3$\sigma$ upper limits in the range from 11.4--22.2~$\mu$Jy\,beam$^{-1}$, whereas all were detected in the original C-configuration observations with flux density values in the range 19.7--81~$\mu$Jy.  We calculate upper limits for their $\log(S_\mathrm{A}/S_\mathrm{C})$ ratios to be about $-0.2$ to $-0.3$ for five of the six sources, and $-0.6$ for the sixth.  

In the absence of variability or resolution effects, we would expect the distribution of flux density ratios to be centered on unity with a spread determined by measurement errors.  Variability could yield a flux density change in either direction, whereas the difference in angular resolution would result in $S_\mathrm{A}<S_\mathrm{C}$ due to missing emission in the high-resolution A-configuration images.

\subsubsection{Implications for Morphology}
\label{subsubsec:morphology}

\begin{figure}[t!]
    \epsscale{1.17}
    \plotone{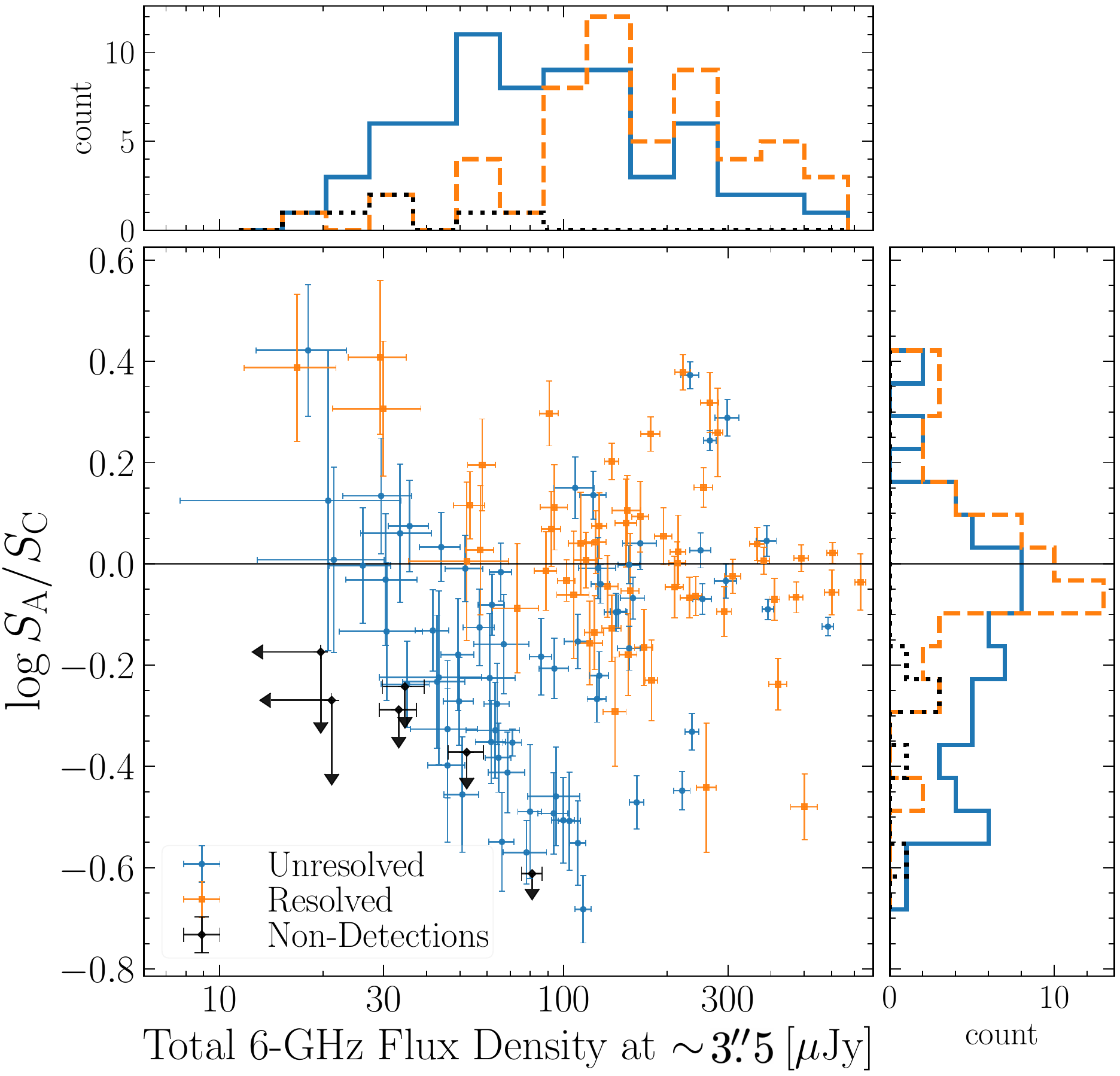}
    \caption{$\mathrm{Log}(S_\mathrm{A}/S_\mathrm{C})$ as a function of $S_\mathrm{C}$.  The lower-left quadrant exclusively contains Unresolved RQQs, consistent with an interpretation that these sources have sub-galactic radio emission that is too faint to be detected in our A-config observations (see text of \S\ref{subsec:angular_resolution}).  Marginal distributions for Unresolved and Resolved sources are shown on the top and right axes (blue solid: Unresolved; orange dashed: Resolved; black dotted: Non-detections).
    \label{fig:fig9_fluxratioscat_whistos}}
\end{figure}

For RQQs whose radio emission is truly compact (i.e. $<0\farcs33$), they would appear as Unresolved regardless of observation depth.  In contrast, the visual classification of RQQs with \textit{extended} emission depends strongly on the sensitivity of the observations in comparison to the surface brightness of the extended emission.  Therefore, we expect to find more Resolved RQQs at the bright end of the population.  
That expectation is confirmed in Figure~\ref{fig:fig9_fluxratioscat_whistos}, which shows $\log(S_\mathrm{A}/S_\mathrm{C})$ as a function of $S_\mathrm{C}$.  In this figure, we see that not only do the majority of Resolved sources populate the brighter end of the $S_\mathrm{C}$ distribution, but also that the Resolved sources with $S_\mathrm{C}>100\mu$Jy are distributed fairly symmetrically around $\log(S_\mathrm{A}/S_\mathrm{C})=0$.  Similarly, Unresolved sources with $S_\mathrm{C}>100\mu$Jy are also distributed somewhat symmetrically around $\log(S_\mathrm{A}/S_\mathrm{C})=0$.  Based on these observations, we conclude that the A-configuration images of RQQs with $S_\mathrm{C}\gtrsim100\mu$Jy are not systematically missing emission due to angular resolution effects.  Thus the morphology we observe for these RQQs is a good representation of their actual radio emission.

In contrast, the RQQs with $S_\mathrm{C}<100\mu$Jy show clearly different behavior in Figure~\ref{fig:fig9_fluxratioscat_whistos}.  Virtually no Resolved RQQs populate the lower-left quadrant of the main panel.  Furthermore, of the Unresolved RQQs with  $S_\mathrm{C}<100\mu$Jy, nearly all have ratios  $\log(S_\mathrm{A}/S_\mathrm{C})\leq0$.  This behavior is precisely the pattern we expect to see due to the change in angular resolution between the two sets of observations; when present, the effect will always yield a decrease of $S_\mathrm{A}$ compared to $S_\mathrm{C}$.  We therefore conclude that in the A-configuration images, we have ``missed" the presence of extended sub-galactic emission among many of the $S_\mathrm{C}<100\mu$Jy RQQs.  

While it is possible that some of the discrepancy between $S_\mathrm{A}$ and $S_\mathrm{C}$ may be due to variability (see discussion in \S\ref{subsubsec:variability}), for the moment we apply the assumption that all Unresolved sources with $S_\mathrm{C}<100\mu$Jy and $\log(S_\mathrm{A}/S_\mathrm{C})<0$ ($>3\sigma$ significance) have additional extended sub-galactic emission that we did not detect in the A-configuration images.  We can then estimate the fraction of missing emission based on the values of $\log(S_\mathrm{A}/S_\mathrm{C})$.  There are 15 such sources meeting these criteria, and their median value of $\log(S_\mathrm{A}/S_\mathrm{C})$ is $-0.44$, suggesting that on average we fail to detect almost two-thirds of the total radio emission in these sources, using the A-configuration images.  In comparison, the sources we identified as Resolved have a typical core-to-total flux density ratio of 0.65.  If we had performed deeper, much more sensitive observations, these faint RQQs with $S_\mathrm{C}<100\,\mu$Jy and $\log(S_\mathrm{A}/S_\mathrm{C})<0$ at $3\sigma$ significance that are currently classified as \emph{Unresolved} would likely have been classified as \emph{Resolved} instead.


Because all 6 Non-detections have upper limits of $\log(S_\mathrm{A}/S_\mathrm{C})<0$, we interpret them similarly to the detected Unresolved sources that also have $\log(S_\mathrm{A}/S_\mathrm{C})<0$: it is likely that their emission detected in C-configuration is spread throughout the host galaxy with a lower A-configuration peak flux density that falls below the noise limit of these new observations. 

Conversely, we assume that the Unresolved RQQs with $\log(S_\mathrm{A}/S_\mathrm{C})\gtrsim0$
(i.e., for which we do not have evidence for undetected extended emission) are those which are truly intrinsically compact.  There are only 41 such sources, indicating that up to $\sim70$\% (the other 87 of 128) of the RQQs in our target population have significant sub-galactic-scale radio emission (i.e., that contributed to the C-configuration detections); these complex sources comprise the 55 correctly identified Resolved sources, as well as the 26 Unresolved sources and 6 Non-detections with $\log(S_\mathrm{A}/S_\mathrm{C})<0$.
We emphasize that it would be difficult to reach such a conclusion without both low- and high-resolution VLA observations, yielding the ability to compare flux densities measured at different angular resolution.


\subsubsection{Implications for Variability} 
\label{subsubsec:variability}

A significant decrease in detected radio emission may be due to either variability or to resolution effects, as just discussed, but a significant \textit{increase} in detected radio emission 
can only be attributed to source variability.  We use the count of RQQs whose radio emission \emph{increased} to roughly estimate how many variable RQQs with \emph{decreased} radio emission exist within both the Resolved and Unresolved groups.

If all sources have the same variability duty cycle (i.e., are just as likely to increase as to decrease during the same time range), then to first order we expect the same number of sources to \emph{decrease} due to variability as to \emph{increase}.  Intermittent AGN jets that turn on and off would be an example of such a variability cycle \citep{Odea2008,Blandford+2019}.
Because 6 Resolved sources have $S_\mathrm{A}>S_\mathrm{C}$ ($>3\sigma$), which must be due to variability, we would similarly expect $\sim6$ sources with $S_\mathrm{A}<S_\mathrm{C}$ ($>3\sigma$) due to variability.  We find only 3 with this decrease in flux density, and the discrepancy could be attributed to resolution/sensitivity effects discussed above, combined with \citet{EddingtonBias} bias: an
RQQ with extended structure near the sensitivity limit will appear to be Unresolved rather than Resolved if negative fluctuations in the extended emission cause it to appear too faint to rise above the image noise; fluctuations in the opposite direction will simply cause the source to remain Resolved.  Therefore there may be $\sim3$ additional variable sources with $S_\mathrm{A}<S_\mathrm{C}$ that have undetected sub-galactic structure, and were thus classified as Unresolved rather than Resolved.  This adjustment amounts to $\sim12$ variable RQQs with extended subgalactic structure.

The Unresolved class contains 4 RQQs with increased radio emission at $>3\sigma$ significance, while 26 decreased.  As discussed earlier, many of these sources which decreased are due to resolution differences between observations, but applying the same assumption regarding RQQ variability duty cycles as before, we expect the decrease for $\sim4$ of these to be due to intrinsic variability.  Combining the estimate of 8 variable Unresolved sources with the above estimate of 12 variable Resolved sources, we conclude that there are $\sim20$ RQQs (in the sample of 128) with a variable component as a dominant source of radio emission.

\begin{table}[h!]
    \begin{center}
    \caption{RQQs with $\log(S_\mathrm{A}/S_\mathrm{C})\neq0$ ($>3\sigma$ significance)}
    \label{tab:table1}
    \begin{tabular}{l|c|c|c|c|c}
       &  & \textbf{Name} & \textbf{$\log{(S_A/S_C)}$} & $S_C$ & \textbf{Morph.}\\
       &  & (J2000) &  & ($\mu$Jy) & (see \S\ref{subsec:morphology})\\
      \hline
      \multirow{9}{*}{\textbf{R}} & \multirow{6}{*}{\textbf{Inc.}} & J1146+3715 & $0.30\pm0.06$ &  90 & R-M \\
        & & J1425+0803 & $0.20\pm0.04$ & 138 & R-SR \\
        & & J1045+2933 & $0.26\pm0.03$ & 179 & R-M \\
        & & J0822+4553 & $0.38\pm0.03$ & 222 & R-M \\
        & & J1444+0633 & $0.15\pm0.04$ & 255 & R-E \\
        & & J1235+4104 & $0.32\pm0.06$ & 266 & R-E \\
       \cline{2-6} 
       & \multirow{3}{*}{\textbf{Dec.}} & J0948+4335 & $-0.44\pm0.13$ & 260 & R-SR \\
        & & J1004+1510 & $-0.24\pm0.05$ & 420 & R-M \\
        & & J1129+5120 & $-0.48\pm0.07$ & 501 & R-SR \\
       \hline
       \multirow{30}{*}{\textbf{U}} & \multirow{4}{*}{\textbf{Inc.}} & J0952+2051 & $0.42\pm0.13$ &  18 & U \\
        & & J1000+1047 & $0.37\pm0.03$ & 233 & U \\
        & & J1617+0638 & $0.24\pm0.02$ & 266 & U \\
        & & J1627+4736 & $0.29\pm0.04$ & 299 & U \\
        \cline{2-6}
       & \multirow{26}{*}{\textbf{Dec.}} & J1631+4048 & $-0.27\pm0.09$ &  49 & U \\
        & & J1502+0645 & $-0.46\pm0.11$ &  50 & U \\
        & & J1233+3101 & $-0.35\pm0.08$ &  61 & U \\
        & & J1524+3032 & $-0.33\pm0.09$ &  63 & U \\
        & & J1630+4711 & $-0.28\pm0.08$ &  64 & U \\
        & & J1206+2814 & $-0.38\pm0.07$ &  64 & U \\
        & & J1534+4658 & $-0.55\pm0.10$ &  66 & U \\
        & & J1102+0844 & $-0.41\pm0.08$ &  68 & U \\
        & & J0934+0306 & $-0.35\pm0.03$ &  71 & U \\
        & & J1443+4045 & $-0.57\pm0.06$ &  78 & U \\
        & & J1605+4834 & $-0.49\pm0.13$ &  79 & U \\
        & & J1125+2513 & $-0.49\pm0.08$ &  93 & U \\
        & & J0939+1710 & $-0.21\pm0.06$ &  94 & U \\
        & & J1458+4555 & $-0.46\pm0.10$ &  95 & U \\
        & & J1007+5007 & $-0.51\pm0.08$ &  99 & U \\
        & & J1334+1711 & $-0.51\pm0.10$ & 104 & U \\
        & & J1426+1955 & $-0.55\pm0.08$ & 110 & U \\
        & & J1005+4230 & $-0.68\pm0.07$ & 114 & U \\
        & & J1631+2953 & $-0.27\pm0.05$ & 125 & U \\
        & & J0929+4644 & $-0.22\pm0.05$ & 127 & U \\
        & & J0942+3451 & $-0.17\pm0.04$ & 155 & U \\
        & & J1321+0459 & $-0.47\pm0.05$ & 163 & U \\
        & & J1253+1227 & $-0.45\pm0.04$ & 221 & U \\
        & & J1252+1402 & $-0.33\pm0.04$ & 236 & U \\
        & & J1220+0641 & $-0.09\pm0.02$ & 392 & U \\
        & & J1218+3522 & $-0.12\pm0.02$ & 586 & U \\
       \bottomrule
    \end{tabular}
    \end{center}
  \tablecomments{RQQs whose 2019 A-configuration ($S_\mathrm{A}$) and 2010--2011 C-configuration ($S_\mathrm{C}$) flux densities differ by $>3\sigma$ (see \S\ref{subsec:angular_resolution}).  Sources are listed by morphology R (Resolved) or U (Unresolved) (see \S\ref{subsec:morphology}).  ``Inc." (increase) indicates $S_\mathrm{A}>S_\mathrm{C}$); ``Dec." (decrease) indicates $S_\mathrm{A}<S_\mathrm{C}$.}
\end{table}

It is not currently thought that starburst emission can contribute significant variability 
over the $\sim$1.3-kpc synthesized beam of these A-configuration observations \citep{Terlevich+1992,barvainis2005,BarcosMunoz+15} within the 8 years between the set of observations, although it is possible that a mixture of AGN and starburst activity are contributing to the total radio luminosity \citep{kimball2011,Condon+2013,kellerman16,MacFarlane+2021}. 
However, fractional increases in flux density of $\gtrsim100$\% in J0822$+$4553, J1000$+$1047, J1617$+$0638, and J1627$+$4736 suggest that a starburst would only be contributing slightly (if at all) to the observed radio, at least in those RQQs.  
Future joint investigations monitoring the optical/UV properties of RQQs in parallel with their observed radio variability would help shed light on the question as to whether the origin of this radio activity is intrinsic to the AGN \citep[e.g., due to enhancements of accretion activity;][]{KB20} or extrinsic \citep[e.g., from tidal disruption events;][]{Komossa2015}.  

An alternative possibility may be that newly formed jets in RQQs represent the beginning of evolution into RLQs \citep[e.g.,][]{Nyland+2020}, in which case there is no particular reason to expect a decrease in radio flux density of RQQs over decadal timescales.  The best way to investigate these possibilities would be to obtain repeated follow-up observations of the RQQ population at matching angular resolution, in order to measure the true variability structure function.

We discuss potential variability of the sample further in Section~\ref{section:variability}.

\section{Spectral Index and Variability Analysis}
\label{section:variability}

In this section, we incorporate ancillary data from past and ongoing radio surveys at other frequencies to determine spectral index properties across the full radio luminosity range of the \citet{kimball2011} parent sample.  Because the observations for the ancillary data were taken at different epochs over a couple of decades, this analysis must necessarily incorporate the consideration of potential variability. 

We reintroduce for this analysis the 27 RIQs and 14 RLQs from \citet{kimball2011} and \citet{kellerman16}, where we define ``radio-intermediate" as the luminosity range $10^{23}$--$10^{24}$\,W\,Hz$^{-1}$ between RQ and RL QSOs.  We only consider the unresolved core components (3\farcs5 scale) of the extended RIQs and RLQs in order to probe emission within their host galaxies; this consideration allows a more direct comparison to their RQ counterparts, which do not have large-scale emission extending beyond the host galaxy.  Furthermore, we note that \citet{Condon+2013} concluded that beaming does not significantly contribute to the overall luminosity function of these populations, owing partly to the fact that the most luminous extended sources are not compact or one-sided (as would be expected for beamed relativistic jets), and also that the majority of them do not appear compact in FIRST.  Although we did not include RIQs or RLQs in our 2019 A-configuration follow-up campaign, it is sufficient to use the lower-resolution (C-configuration) 2010--2011 observations of the entire QSO sample because the following analysis is based on the total radio emission from within the host galaxies, which is fully probed by the lower-resolution images.  The luminosity boundaries we use in defining RQ/RI/RL classes are reiterated in Table~\ref{tab:var_sample}.  

It is well known that the cores of RLQs exhibit variability on timescales of years, months, and even days \citep{barvainis2005, Hovatta+2007, Thygarajan+2011}, due presumably to the source of radio emission in such objects being an optically opaque, compact engine closely associated with an AGN \citep[e.g.,][]{Rees+1984}.  To see if the same effects were spread throughout quasars across the full range of radio loudness, \citet{barvainis2005} simultaneously studied samples of 8 RLQs, 11 RIQs, and 11 RQQs 
over $\sim2$ total years of observing with the legacy VLA. 
They concluded that similarities in the radio light curves between all three classes of objects---coupled with the presence of flat and inverted spectral indices ($\alpha>-0.5$) throughout the RQ sources---indicated that the underlying mechanism(s) driving the radio emission in the cores of RQQs may ultimately be similar to what drives radio emission in the cores of RLQs.  

In stark contrast with the spectral index findings of \citet{barvainis2005}, \citet{Condon+2013} found, through a statistical analysis of data from the NRAO VLA Sky Survey \citep[NVSS;][]{NVSS}, that the typical radio spectral index for low-redshift ($0.2<z<0.45$) RQQs is $\alpha\approx-0.7$ (steep spectrum), whereas for RLQs it is typically $\alpha\sim0$ (flat spectrum).  For synchrotron radiation, steep spectra are expected from optically thin emission---typical for star formation processes, or optically thin radio jets---whereas flat spectra are expected from optically thick emission, i.e., compact jets and jet cores \citep[e.g.,][]{Kell_Pauliny_compact}.  Differing spectral indices in RQQs vs.\ RLQs are thus consistent with the notion that different physical origins of radio emission dominate in the two types of sources, as concluded by \citet{kimball2011} and \citet{Condon+2013}.

In the following subsections, we more thoroughly describe the survey data used (\S~\ref{subsec:var_data}), present our spectral index measurements (\S~\ref{subsection:specindex}) along with a complementary time-domain analysis (\S~\ref{subsection:specindex_time}), then discuss a few individual sources with particularly interesting spectral properties (\S~\ref{subsec:var_intsources}).

\begin{table}[ht]
  \begin{center}
  \caption{Sample for Spectral Index, Time Domain Analysis}
  \label{tab:var_sample}

  \begin{tabular}{*{8}{p{.16\linewidth}}}
    \multicolumn{8}{c}{Panel A: Observational Details} \\
    \toprule
    \multicolumn{2}{l}{Observation} & 
    \multicolumn{2}{c}{$\nu_\mathrm{obs}$} & 
    \multicolumn{2}{c}{VLA Config.} & 
    \multicolumn{2}{r}{LAS} \\
    \multicolumn{2}{l}{} & 
    \multicolumn{2}{c}{(GHz)} & 
    \multicolumn{2}{c}{($\theta$)} & 
    \multicolumn{2}{l}{} \\
    \midrule
    \multicolumn{2}{l}{FIRST} & 
    \multicolumn{2}{c}{1.4} & 
    \multicolumn{2}{c}{B ($5''$)} & 
    \multicolumn{2}{r}{$60''$} \\
    \multicolumn{2}{l}{Kimball+2011} & 
    \multicolumn{2}{c}{6} & 
    \multicolumn{2}{c}{C ($3\farcs5$)} & 
    \multicolumn{2}{r}{$\geq120''$} \\
    \multicolumn{2}{l}{VLASS} & 
    \multicolumn{2}{c}{3} & 
    \multicolumn{2}{c}{B/BnA ($2\farcs5$)} & 
    \multicolumn{2}{r}{$30''$} \\
    \multicolumn{2}{l}{This Paper} & 
    \multicolumn{2}{c}{6} & 
    \multicolumn{2}{c}{A ($0\farcs33$)} & 
    \multicolumn{2}{r}{$\geq4\farcs5$} \\
    \bottomrule
  \end{tabular}
  \end{center}
  \tablecomments{Data are from FIRST \citep{first}, \citet{kimball2011}, VLASS \citep{vlass}, and the observations presented in this paper.  Columns 3 and 4 give each observation's corresponding VLA configuration (and fiducial angular resolution) and largest angular scale (LAS) detectable.\vspace{0.65cm}}
    
  \begin{center}
  \begin{tabular}{*{8}{p{.5cm}}}
    \multicolumn{8}{c}{Panel B: Source Classification} \\
    \toprule
    \multicolumn{2}{l}{Category} & 
    \multicolumn{2}{c}{$L_\mathrm{6GHz}$} &
    \multicolumn{2}{r}{N} \\
    \multicolumn{2}{l}{} & 
    \multicolumn{2}{c}{(W\,Hz$^{-1}$)} & 
    \multicolumn{2}{r}{}\\
    \midrule
    \multicolumn{2}{l}{Radio-loud} & 
    \multicolumn{2}{c}{$>10^{24}$} & 
    \multicolumn{2}{r}{14} \\
    \multicolumn{2}{l}{Radio-intermediate} & 
    \multicolumn{2}{c}{$10^{23}<L<10^{24}$} &
    \multicolumn{2}{r}{27} \\ 
    \multicolumn{2}{l}{Radio-quiet} & 
    \multicolumn{2}{c}{$<10^{23}$} & 
    \multicolumn{2}{r}{29} \\
    \bottomrule
  \end{tabular}
  \end{center}
  \tablecomments{Number counts and luminosity-based definitions of RL/RI/RQ QSOs \citep[as introduced in][]{kimball2011,kellerman16} presented in \S\ref{subsection:specindex}.}
\end{table}

\subsection{Survey data and subsample selection}
\label{subsec:var_data}


To investigate the spectral indices of our targets, we obtained comparison data at other radio frequencies from the FIRST catalog \citep{Helfand15} and from ``QuickLook" images from Epoch~1 of the VLA Sky Survey \citep[VLASS;][]{vlass}.  FIRST, performed with the legacy (pre-upgrade) VLA, was a 1.4-GHz survey with a 3$\sigma$ sensitivity of $\sim0.45$~mJy\,beam$^{-1}$ and resolution of $\sim5''$.  VLASS is a 3-GHz survey currently being undertaken with the upgraded VLA, with a 3$\sigma$ sensitivity of $\sim0.36$~mJy\,beam$^{-1}$ per Epoch and a resolution of $\sim2\farcs5$.

For counterparts from FIRST, we use the catalogued peak flux density when available.  For sources too faint to have a catalogued FIRST detection, but which appear to have an identifiable counterpart upon visual inspection, we determine a peak flux density using CASA's \emph{imfit} task, and add an additional 0.25\,mJy\,beam$^{-1}$ due to ``CLEAN bias" affecting the FIRST images \citep[see \S7.2 of][]{first}.  For VLASS, we use peak flux densities obtained from \emph{imfit} when we identify a visible counterpart.  When a counterpart was not identified in either survey, we use 3$\sigma$ sensitivity values as upper limits for our analysis; this is a typical S/N ratio used to estimate possible counterpart emission at known source positions.

A known issue for early VLASS data is that Epoch 1.1 and 1.2 QuickLook images yield flux densities that are systematically low by 15\% and 8\% respectively, with scatter up to $\sim8\%$\footnote{\url{https://library.nrao.edu/public/memos/vla/vlass/VLASS\_013.pdf}}.  We have accounted for these discrepancies by adjusting the measured flux densities accordingly; we have incorporated the scatter by including an additional 8\% error added in quadrature with the flux density errors reported by CASA's \emph{imfit} task.

\begin{figure}[t!]
    \epsscale{1.18}
    \plotone{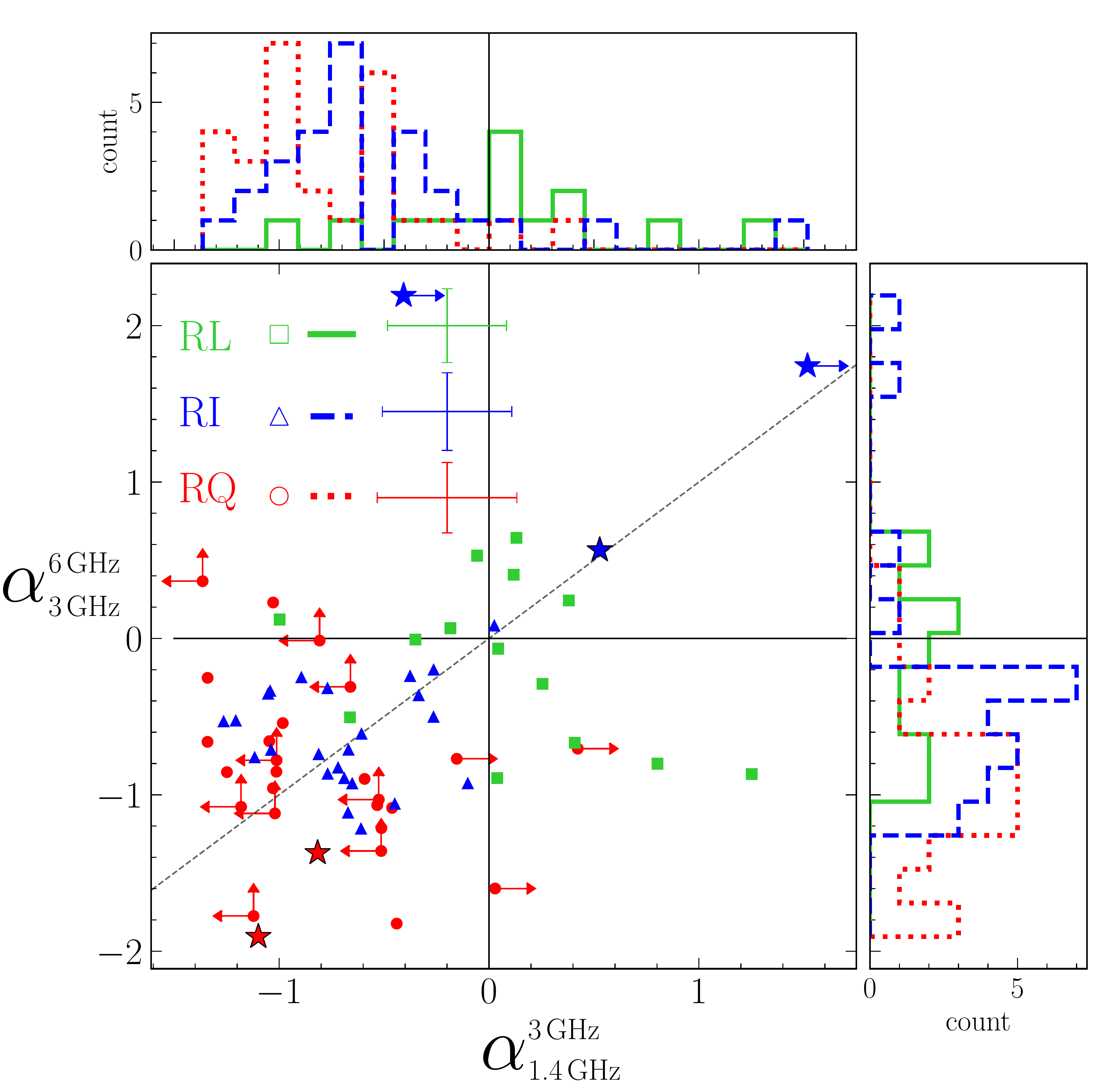}
    \caption{Spectral indices between 3 and 6~GHz compared to spectral indices between 1.4 and 3~GHz for the RL, RI, and RQ classes of QSOs.  The dashed line illustrates where the two spectral indices are equal, i.e., a constant power law across 1.4--6\,GHz.  Data at 6~GHz are from targeted VLA observations presented in \citet{kimball2011}, 3-GHz data are from VLASS, and 1.4-GHz data are from FIRST.  Several RQ and RI sources are not detected in FIRST or VLASS, as shown by the lower/upper-limit symbols (rightward-pointing for FIRST non-detections; upward- and leftward-pointing for VLASS non-detections).  Two RQQs (J1425$+$0803 and J1617$+$0638) are omitted from this figure as they are non-detections in both FIRST \textit{and} VLASS, so we cannot obtain meaningful estimates of spectral indices at this frequency range.  Median error bars for each radio loudness class are shown; the large errors primarily result from the high flux density uncertainties in the early QuickLook images from VLASS \citep[see discussion in \S\ref{subsec:var_data} and][]{vlass}.  Star symbols denote three RIQs and two RQQs that are highlighted in the discussion of \S\ref{subsec:var_intsources}.}
    \label{fig:fig10_alphaCS_vs_alphaLS}
\end{figure}

As the A-configuration observations presented in this paper are $\approx20\times$ deeper than FIRST and VLASS (and with many sources still being only marginally detected at $\mathrm{S/N}\approx3$--5), inclusion of the entire sample of 128 RQQs in this analysis would be dominated by upper limits of FIRST and VLASS flux densities.  Because our goal is to obtain meaningful estimates of spectral indices across all three radio loudness classes of QSOs (RQ/RI/RL), we limit our selection of QSOs for this analysis to those whose measured 6-GHz flux densities in C-configuration are $>200\,\mu\mathrm{Jy}$, which would correspond to a $>3\sigma_\mathrm{rms}$ detection in FIRST for a source with spectral index $\alpha=-0.7$ from 1.4--6~GHz (and no variability).  All RLQs and all RIQs are brighter than this value; of the 128 RQQs, 29 satisfy this criterion.

The surveys and sample selection for this analysis are summarized in Table~\ref{tab:var_sample}; the data for all sources in this analysis are provided in Table~\ref{tab:vartable} of Appendix~\ref{sec:app_var}.

\subsection{Spectral indices}
\label{subsection:specindex}

In Figure~\ref{fig:fig10_alphaCS_vs_alphaLS} we show the 1.4--3~GHz and 3--6~GHz spectral indices of the RL, RI, and RQ sources included in this analysis; note that calculation for these values implicitly assumes no variability between the different observing epochs.  The three radio loudness classes generally occupy separate regions of this parameter space; a two-sample Kolmogorov-Smirnov test \citep{K-S_Test} on the  RQ/RL populations yields a $p$-value of $6.5\times10^{-7}$, strongly indicating that the RQ/RL spectral indices follow two separate distributions.  The bulk of the RQQs have spectral indices $\alpha<0$ across the full frequency range; the RIQs largely overlap the RQ sources but with a tendency toward flatter spectral indices.
In contrast, the RLQs all manifest flat ($\alpha>-0.5$) or even inverted ($\alpha>0.5$) spectral index on at least one axis.  Note that the bottom-right quadrant is populated with sources that have a spectral peak between 1.4 and 6 GHz.

The behavior of the different radio loudness classes in Figure~\ref{fig:fig10_alphaCS_vs_alphaLS} is consistent with typically observed spectral indices for QSOs: RLQs exhibit flat-spectrum emission indicative of an optically thick, i.e., synchrotron-self-absorbed jet core \citep[][]{Cotton1980_SSA}, while RQQs show steeper spectra generally associated with optically thin synchrotron emission consistent with a star formation origin \citep[e.g.,][]{Condon+2013}, but also consistent with optically thin jets or shocks from winds.  The RIQs appear to have spectral indices in a range that bridges the RL and RQ classes.  Some sources lie far from the constant power law line in Figure~\ref{fig:fig10_alphaCS_vs_alphaLS}.  This spectral curvature may be explained by those sources containing both a significant steep-spectrum component related to star formation \textit{and} flat-spectrum component from the central AGN \citep{kimball2011,Condon+2013,kellerman16,MacFarlane+2021}.  The relative contributions of each component are frequency-dependent, meaning that a combination of the two will produce scatter from a constant power law across 1.4--6\,GHz in Figure~\ref{fig:fig10_alphaCS_vs_alphaLS}.  We note that, of the 70 sources in this sub-sample, 7/29 RQQs, 1/27 RIQs, and 5/14 RLQs are inconsistent with a constant power law beyond $1\sigma$ measurement errors.

Also evident are three obviously outlying RIQs; a possible 
explanation is that these three sources are variable, such that their radio emission changed between the different observing epochs of the 1.4, 3, and 6 GHz data.  It is interesting that these extreme outliers are found primarily in the RIQ class, and could suggest that radio-loudness is a transient phenomenon, with QSOs shifting between the two main radio-loudness classes \citep[e.g.,][]{Nyland+2020}.  The possible presence of variability, particularly in these outlier sources, is discussed below in \S\ref{subsec:var_intsources}.

\subsection{Time domain with spectral index analysis}
\label{subsection:specindex_time}

\begin{figure*}[t]
    \epsscale{1.05}
    \plottwo{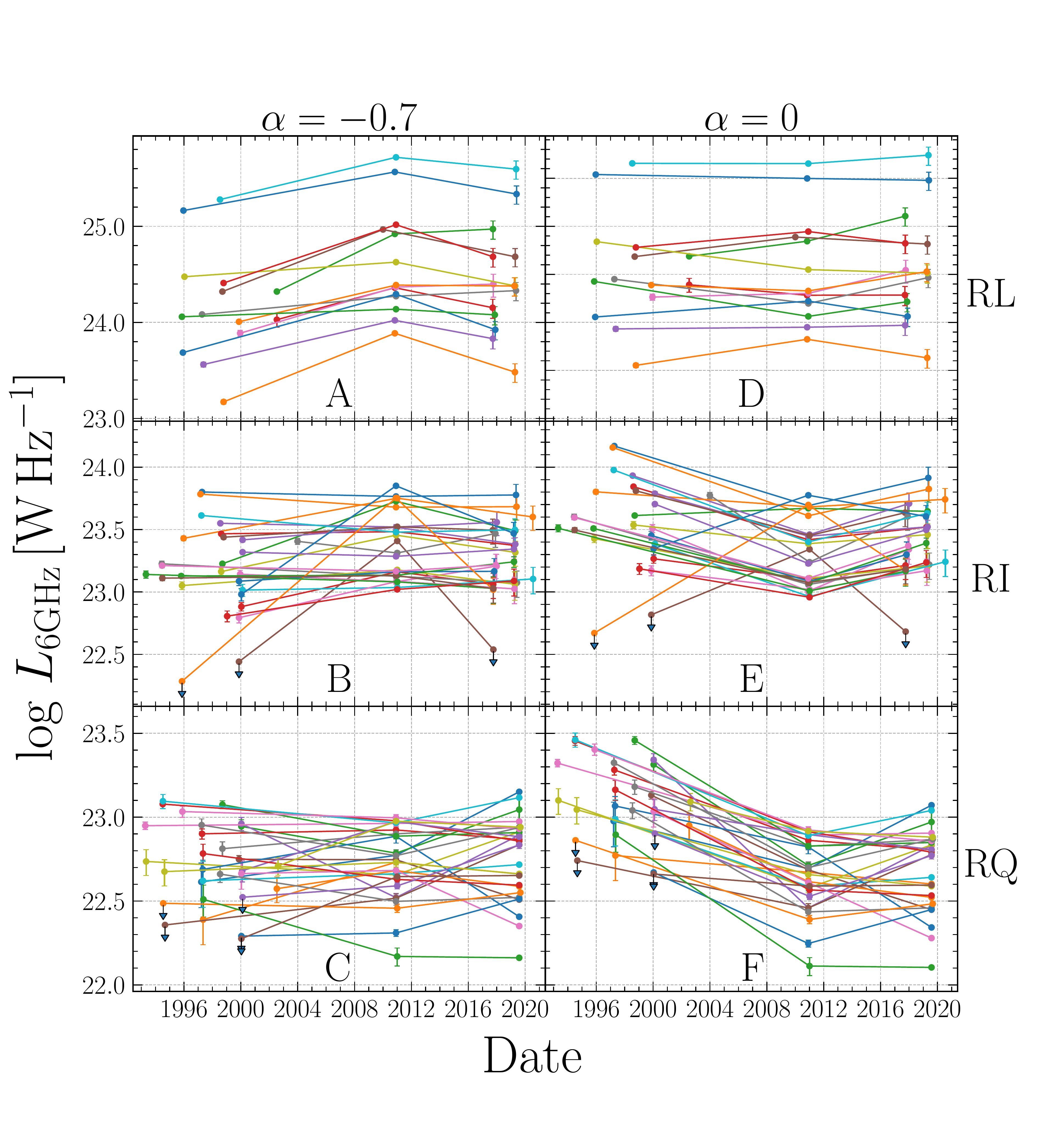}{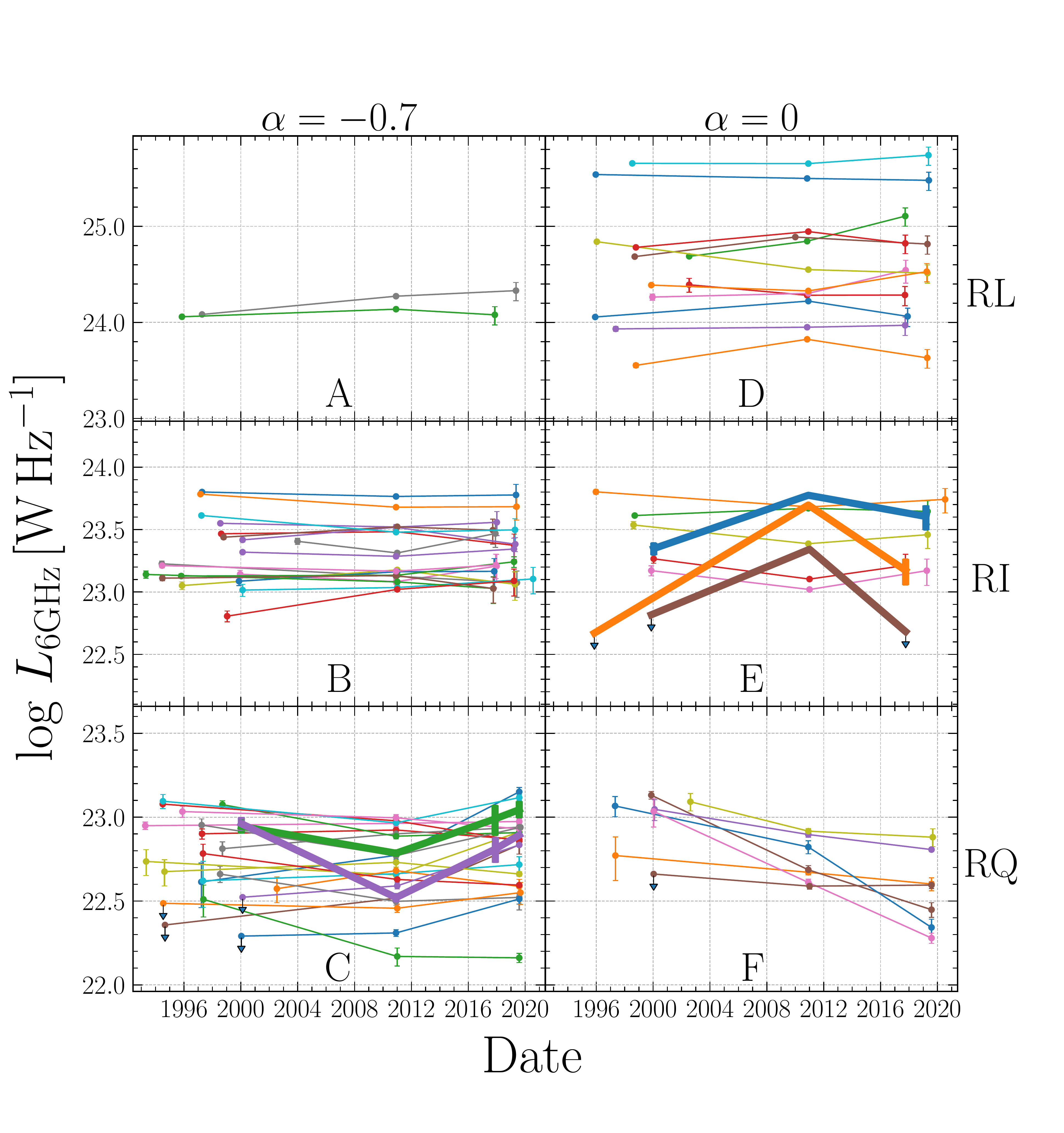}
    \caption{{\em Left panels:} Estimated 6-GHz luminosity over decades-long timescales of all QSOs considered in our time-domain analysis.  The top row shows the curves of the 14 RLQs examined, the middle row shows the 27 RIQs, and the bottom the 29 RQQs.  The left column shows the resultant curves assuming a steep spectral index ($\alpha=-0.7$) in converting FIRST and VLASS flux densities to our 6-GHz observing frequency, while the right-hand column shows the resultant curves assuming a flat spectral index ($\alpha=0.0$).  FIRST observations span 1993--2004; all 2010--2011 points reflect data from \citet{kellerman16}; for the RQQs, the 2019 data points represent flux densities from this paper; the 2017--2019 data points for the RIQs and RLQs reflect data from preliminary ``Quick Look" images from Epoch 1 of VLASS, and the 2019 data points for the RQQs are from the A-configuration observations in this paper. {\em Right panels:} Preferred results selected on their best-fit spectral index from the $\chi^2$ analysis. Sources not well-suited by either choice of spectral index are shown with thick lines (two in Panel~C and 3 in Panel~E), and discussed further in \S\ref{subsec:var_intsources}.
    \label{fig:fig11_lightcurves}}
\end{figure*}

The data we are using to investigate radio spectral indices of our targets are based on observations taken over multiple decades: FIRST observed the sky from 1993--2004; the \citet{kimball2011} C-configuration observations of our targets were performed in 2010--2011; VLASS Epoch~1 observations were obtained in 2017--2019; A-configuration observations of the RQQs (presented in this paper) were obtained in 2019.  We can therefore use these data to additionally probe variability within the sample over decade timescales.  Note that we employ the \citet{kellerman16} 6-GHz measurements for the 2010-2011 C-configuration data points, which for the handful of extended RIQs/RLQs include emission only from the core components.  This choice is reasonable as the extended features of RLQs/RIQs are not expected to significantly vary on decadal timescales \citep{Blandford+2019}, and it allows an appropriate comparison with the RQQs, which have no such large-scale emission.

We opt to convert all flux densities (FIRST at 1.4~GHz and VLASS at 3~GHz) to the 6-GHz frequency of our new observations presented here.  However, such a conversion requires assuming a specific value for radio spectral index.  As demonstrated in the previous section, the spectral indices vary across the sample, and in fact the possibility of variability of individual sources (which is supported by the spread seen in Figure~\ref{fig:fig8_fluxratio_2panel}) means that some calculated spectral indices may be incorrect.  Furthermore, flux density variation of a source could also coincide with spectral index variability; this would certainly be the case, for example, for a QSO whose radio emission is a combination of steep-spectrum star-formation processes along with a varying flat-spectrum jet component.

To convert the 1.4 GHz and 3 GHz data to 6 GHz, we choose to adopt a steep ($\alpha=-0.7$) or flat ($\alpha=0$) radio spectral index corresponding to the typical observed spectral slopes for optically thin and optically thick synchrotron emission, respectively.  Given the conflicting results in the literature regarding spectral indices in different-power QSOs \citep[e.g.,][]{barvainis2005, Condon+2013}, we recognize that employing one choice of $\alpha$ for all sources would yield dubious results.  We instead try both of the two values for every QSO, and compare the two results to determine which value of spectral index yields a more physically credible variability pattern for each source individually.  The distinction between the two values of $\alpha$ corresponds to a difference in 6-GHz flux density of roughly 0.37~dex for conversions from 1.4~GHz, and 0.14~dex for conversions from 3~GHz.  We note that, for any given source, it is possible that neither choice of spectral index is correct.

The resulting luminosity-vs-time values are shown in Figure~\ref{fig:fig11_lightcurves}, distinguished by class (RL, RI, RQ from top to bottom).  In both the left and right sets of panels, panels A--C correspond to the assumption of $\alpha=-0.7$ while panels D--F correspond to the assumption of $\alpha=0$.  The left-hand set of panels includes the two possible light curves (corresponding to different $\alpha$ value) for each QSO.  These left panels display some obviously unphysical patterns of luminosity variation.  For example, panel~A shows that when assuming $\alpha=-0.7$, nearly the entire population of RLQs displays a rising luminosity to 2010--2011 which then decreases to 2017--2019; conversely, panel~F shows that when assuming $\alpha=0$, the majority of RQQs display a sharp decrease in luminosity from the first to the second epoch.  We would not expect such coordinated behavior in QSOs that are spread across the sky and are clearly physically unassociated.  Therefore these left-hand panels clearly indicate that $\alpha=-0.7$ is not a reasonable spectral index to assume for all RLQs, while $\alpha=0$ is not a reasonable spectral index to assume for all RQQs.

The right-hand set of panels is effectively a subset of the left-hand-side that displays only the chosen ``best" value of $\alpha$ (of $0$ or $-0.7$) for each source.  To determine quantitatively which $\alpha$ value is the better match, we employ an assumption that the radio luminosity of each QSO should monotonically increase or decrease over the multiple epochs of observation, and therefore we attempt to fit each variation pattern to a power-law over time (with slope as a free variable).  We note that variability may produce a flux density variation pattern that does not follow this assumption; however, with just three data points, minimizing the degrees of freedom in this analysis prevents us from achieving overly spurious results.  
We use a reduced-$\chi^2$ goodness-of-fit test to determine which of the two values provides a better match to a power-law variability model.  We find that 22/29 ($76\pm8\%$) RQQs and 19/27 ($70\pm9\%$) RIQs are better represented with a steep spectral index ($\alpha=-0.7$), whereas 12/14 ($86\pm9\%$) of the RLQs are better represented with a flat spectral index ($\alpha=0$).  
The spectral index results for the different QSO classes are consistent with the results presented earlier in Figure~\ref{fig:fig10_alphaCS_vs_alphaLS}, but here we have added the opportunity to probe variability.  While a handful of sources show some evidence of variability, the majority are consistent with a non-variable light curve for the chosen choice of $\alpha$.  

An initially surprising observation from Figure~\ref{fig:fig11_lightcurves} is that the radio-loudness class showing the most discrepancies from flat light curves is the RQQ population.  In particular, panel F on the right seems to suggest that there is a population of flat-spectrum RQQs whose radio luminosity all decreased similarly over about two decades. However, we recall to the reader's attention the possibility of missing flux in the higher-resolution 6-GHz observations from 2019 compared to those from 2010-2011 (see discussion in \S\ref{subsec:angular_resolution}), and a symptom of such missing flux would indeed result in apparently decreasing light curves for $\alpha=0$, thus populating this panel~F.  We calculate that an increase of $\sim 23\%$ of radio emission in the 2019 observations for these 7 RQQs would alter their results in such a way as to match best to invariable luminosity with $\alpha=-0.7$, and suggest this value of 23\% as a rough estimate for the amount of undetected emission in the new observations for those 7 RQQs.  We find this conclusion to be more reasonable than the idea that we coincidentally observed a group of unrelated flat-spectrum RQQs that all decrease in luminosity over a similar time period, especially given the \emph{a priori} knowledge that RQQs tend to have steep radio spectra.

\begin{figure*}[ht!]
    \epsscale{1.0}
    \plotone{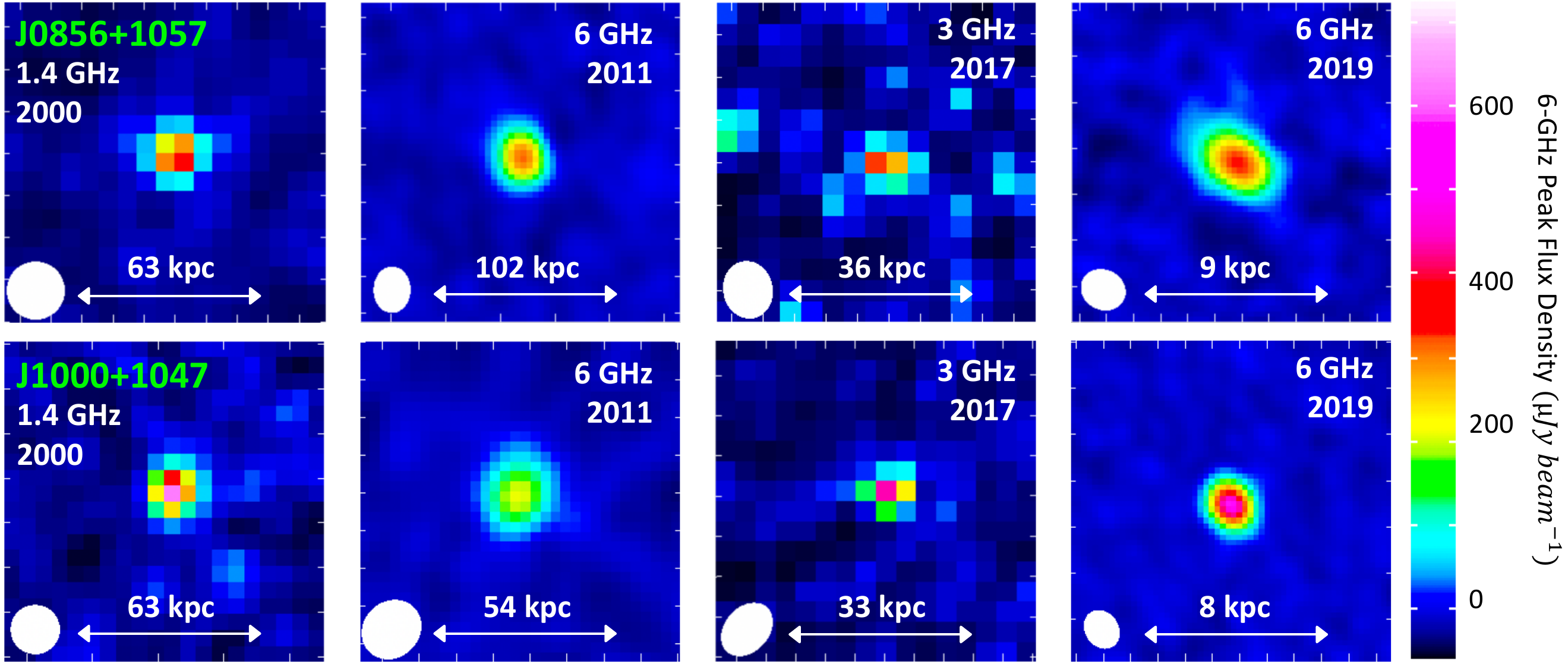}
    \caption{The two outlier RQ sources discussed in \S\ref{subsec:var_intsources}.  \textbf{Column 1} shows the FIRST image \citep{first,Helfand15}; \textbf{Column 2} shows the source in the VLA's C-configuration \citep{kimball2011,kellerman16}; \textbf{Column 3} displays the source's ``Quick Look" image from the VLA Sky Survey \citep{vlass}; \textbf{Column 4} contains the A-configuration images from the observations presented in this paper.  The resolution element (synthesized beam) is shown in the lower left corner for each image.  Note that images in the different columns were observed at different times \textit{and} frequencies.  The colorbar maps the source flux density \textit{at 6-GHz for each image} assuming $\alpha=-0.7$; the choice of spectral index is discussed in the text in \S\S\ref{subsection:specindex}--\ref{subsec:var_intsources}.  See Panel~A of Table~\ref{tab:var_sample} for a summary of each of the contributing surveys.
    \label{fig:fig12_RQ_triptych}}
\end{figure*}

\begin{figure}[hb!]
    \epsscale{1.17}
    \plotone{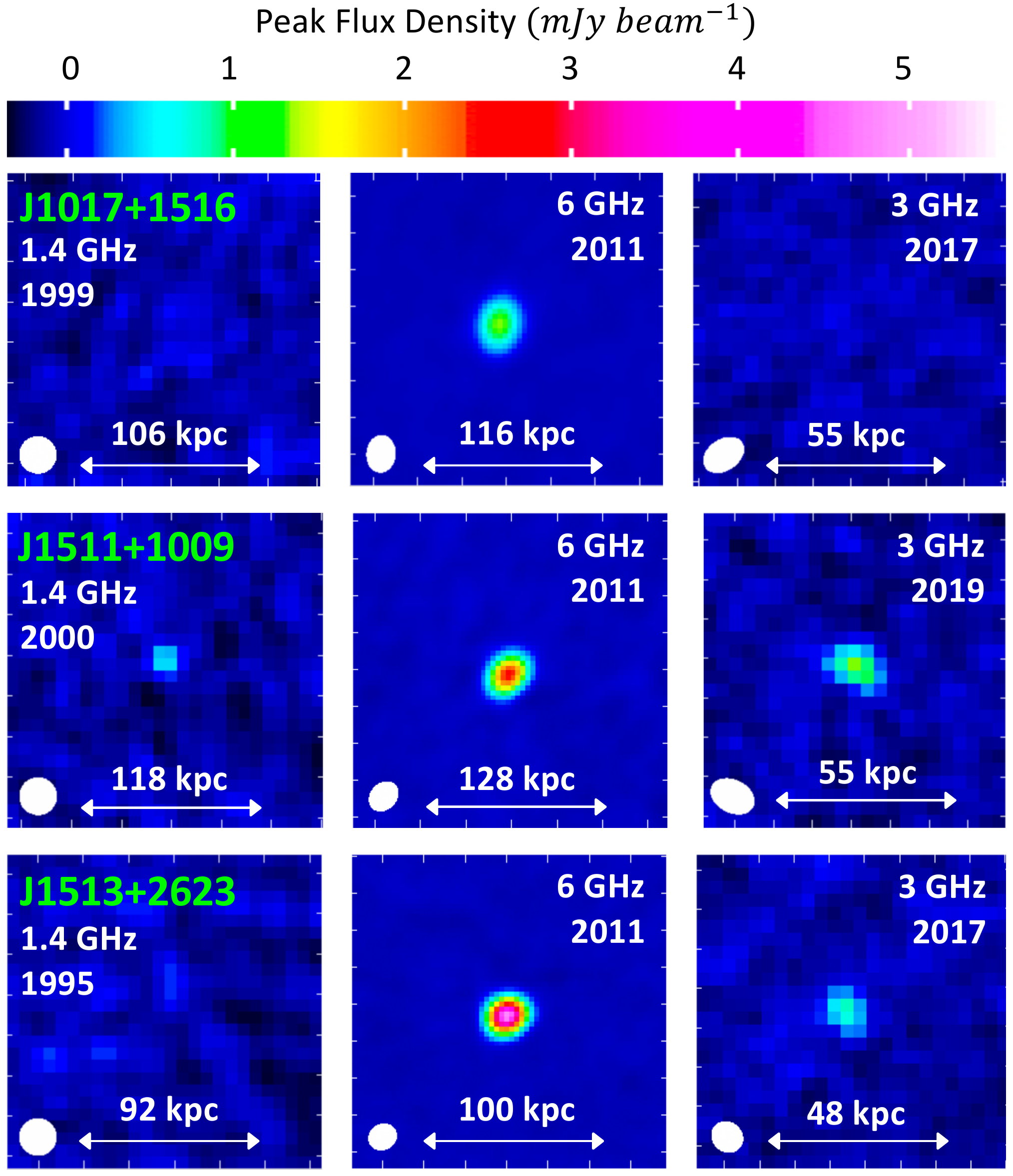}
    \caption{The three RIQs discussed in \S\ref{subsec:var_intsources}.  The panel layout follows that of the first three columns of Figure~\ref{fig:fig12_RQ_triptych}.  All images are on the same color scale.  Assumption of spectral index $\alpha=-0.7$ or 0 yields a conclusion that all three sources increased then decreased in flux density over a similar timescale; we suggest the likelier explanation is that these QSOs have radio spectra peaked at $>3$ GHz. 
    \label{fig:fig13_RI_triptych}}
\end{figure}

Finally, we call attention to two RQQs and three RIQs (the five sources marked as stars in Figure~\ref{fig:fig10_alphaCS_vs_alphaLS}, bold linestyles in the right-hand panels of Figure~\ref{fig:fig11_lightcurves}), which are ill-suited for either choice of spectral index resulting from this analysis.  The three RI sources are the outliers in Figure~\ref{fig:fig10_alphaCS_vs_alphaLS}; we discuss all five of these sources in more detail in the following subsection.

\subsection{Outlier Sources}
\label{subsec:var_intsources}

As just discussed, some of the QSOs studied in this section appear to be poorly represented by either choice of spectral index ($\alpha=0$ or $-0.7$). In Figure~\ref{fig:fig11_lightcurves}, the thick linestyle in the right set of panels highlights two RQQs (J0856$+$1057, marked in green in panel~C; J1000$+$1047, marked in purple) and three RIQs (J1017$+$1516, marked in brown in panel~E; J1511$+$1009, marked in orange; J1513$+$2623, marked in blue), each of whose evolution of flux density deviates strongly from the power-law variability assumption that appears to be sufficient for most QSOs.  These five QSOs are either extremely variable, and/or they have unusual radio spectral indices such that we erred in converting FIRST/VLASS flux densities to the fiducial 6-GHz observing frequency.  They are also the five sources marked as stars in Figure~\ref{fig:fig10_alphaCS_vs_alphaLS}.  We show images of the two RQQs in Figure~\ref{fig:fig12_RQ_triptych} and the three RIQs in Figure~\ref{fig:fig13_RI_triptych}.

\begin{figure}[h!]
    \epsscale{1.2}
    \plotone{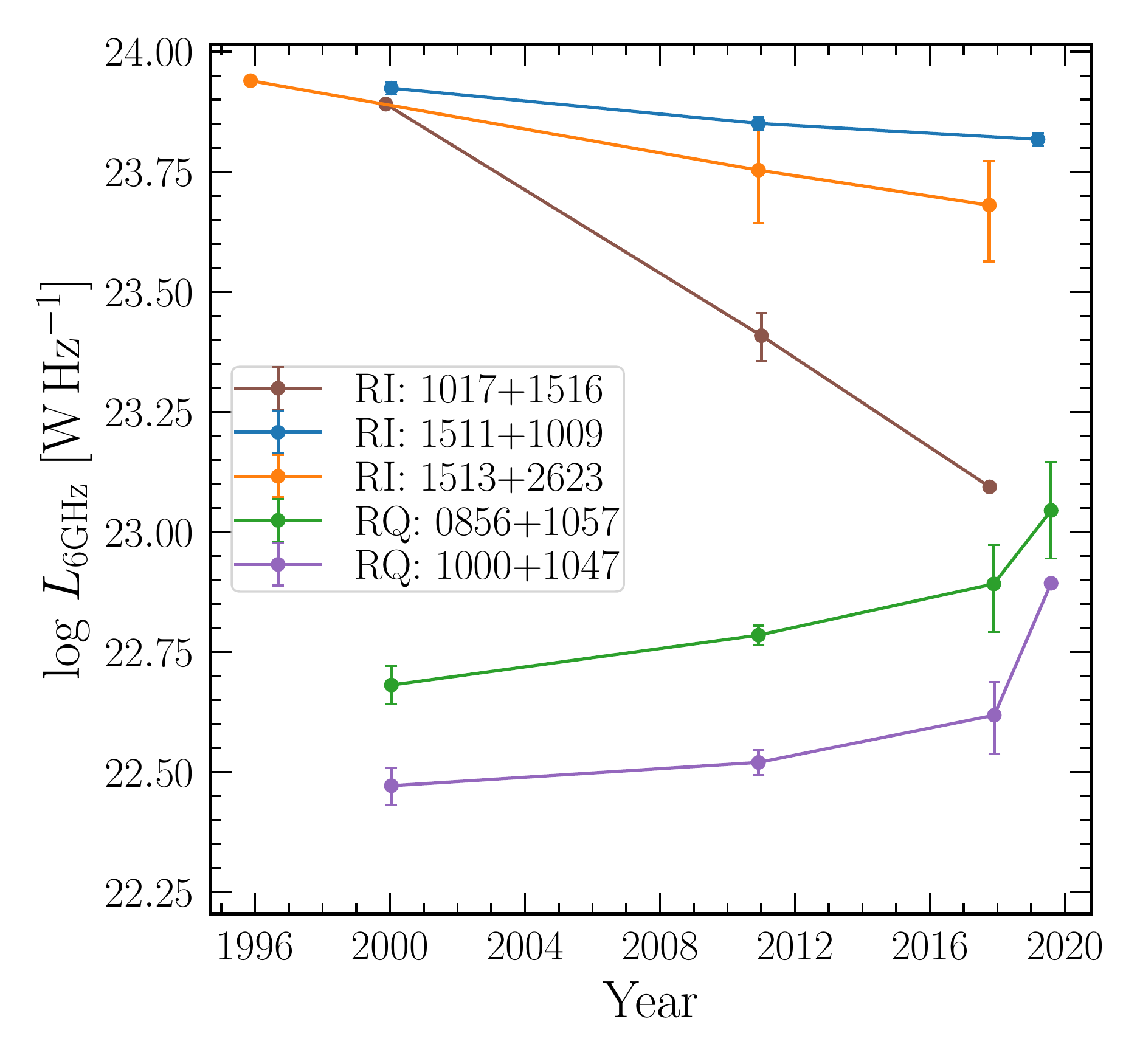}
    \caption{The five sources in this figure are those whose observations over multiple epochs clearly suggest that they are variable rather than having spectral index values of $\alpha\sim-0.7,0$ that are suitable for the bulk of the QSO population analyzed in this section.  Instead, we solved for $\alpha$ as a free-parameter while attempting to fit the light curves to a constant power-law.  The results shown here correspond to the ``best" $\alpha$ value for each source; see text of Section~\ref{subsec:var_intsources}.}
    \label{fig:fig14_alphafreeparam_5QSOs}
\end{figure}

The two RQQs---J1000+1047 and J0856+1057---demonstrate similar behavior in their light curves when assuming a spectral index of $\alpha=-0.7$: each shows a decrease in luminosity from 2000 (detected at 1.4~GHz in FIRST) to 2011 \citep[detected in 6~GHz by][]{kimball2011}, then an increase to 2019.  Note that each of these sources has a counterpart in VLASS, with corresponding values (from 2017) included in Figure~\ref{fig:fig11_lightcurves}.  Converting each one's 2017 VLASS 3-GHz observation to 6-GHz assuming $\alpha=-0.7$ yields a value that is consistent with a straightforward interpolation between the two epochs of 6-GHz observations (see Panel~C on the righthand-side of Figure~\ref{fig:fig11_lightcurves}).  For further investigation of these two sources, we performed spectral index analysis similar to that in \S\ref{subsec:var_data} but leaving $\alpha$ as a free parameter rather than requiring a value of 0 or $-0.7$.  The constant spectral index values that would lead to a power-law variation of luminosity for each target, over all four epochs of observations, are $\alpha=-1.6$ for J1000$+$1047 and $\alpha=-1.2$ for J0856+1057, and the corresponding light curves are shown in Figure~\ref{fig:fig14_alphafreeparam_5QSOs}.  Regardless of spectral index value, the fact that each source increased between the two epochs of 6-GHz observations proves that they are variable; the angular resolution difference could not cause such an increase.  Thus the behavior of these objects is similar to that of radio transients highlighted recently by \citet{Nyland+2020}, suggesting that these two RQQs could be additional examples of newly launched radio jets.  We note that J1000+1047 appears compact in our 0\farcs33-resolution observations, while J0856+1057 is slightly resolved, with some extended emission that is apparent above the 3.5-$\sigma$ level.

The three outlier RIQs show the apparent opposite behavior in Figure~\ref{fig:fig11_lightcurves} as the two outlier RQQs.  Assuming a flat ($\alpha=0$) spectral index, all three were faint in the 1990s (two were non-detections in FIRST), had brightened by 2011, then decreased again by the time of the VLASS observations in 2017--2019.  We suspect that rather than all three sources manifesting this similar variability pattern, it is instead likely that these RIQs possess inverted radio spectra ($\alpha>0$) from 1.4--6~GHz.  We performed a similar analysis as for the two outlier RQQs, wherein we leave $\alpha$ as a free parameter to identify the ``best" spectral index to yield a power-law variation in luminosity with time. The resulting best-fit spectral index values are $\alpha=2.0$ for J1017+1516; $\alpha=1.1$ for J1511+1009; and $\alpha=2.3$ for J1513+2623.  Such high values are common in the radio spectra of active stars \citep[due to the gyro-synchroton emission mechanisms present during Solar flares;][]{Ramaty1969,Melrose1980_bursts,Bastian+1998_bursts}, and may even be prevalent below the spectral peak of optically thick sources of emission at lower radio frequencies \citep[e.g.,][]{RL16}.  However, to our knowledge no research to-date has reported $\sim$GHz-regime power-law spectral indices as high as these for any QSOs.  We suggest the more plausible scenario is that these three RIQs are examples of Gigahertz-Peaked Spectrum (GPS) sources \citep{Odea1998}, which is a categorization describing compact QSOs with radio SEDs that peak in the few-GHz range; the peak would have to be above 3~GHz in order to be consistent with the observation that these sources were brighter at 6~GHz than at 3~GHz.  Values of $\alpha>2$ are commonly observed below the peaks of emission in GPS sources \citep[e.g.,][Fig.\ 4]{Sotnikova_GPSbelowPeak}.  It is also possible that these sources are variable in luminosity or spectral index or both.

\section{Summary}
\label{section:summary}

We have obtained new high-resolution VLA data for a low-redshift ($0.2<z<0.3$) radio-quiet QSO population (128 RQQs; $L_\mathrm{6\,GHz}<10^{23}$\,W\,Hz$^{-1}$) drawn from a homogeneous optically selected, volume-limited sample of 178 QSOs.  These 6-GHz observations yield $\sim0\farcs33$-resolution images ($\sim1.3$\,kpc at $z=0.25$), and a typical 1-$\sigma$ image sensitivity level of $\sim7\mu$Jy~beam$^{-1}$. 
Until now, the sub-galactic morphologies across a homogeneous volume-limited population of RQQs have been unknown.  Examination of the images yields the following.

\begin{itemize}
    \item The detected sources can be classified as ``Unresolved" (67 of 128) or ``Resolved" (55 of 128) based on their appearance in the $\sim0\farcs33$-resolution images at 6-GHz.  Six of the RQQs were undetected in these new observations.
    \item Extended sub-galactic emission (i.e., within the host galaxy but outside the region of peak radio emission) is visible in $\geq45\%$ of the population.
    \item A handful of the RQQs have sub-galactic radio morphology that seems to strongly indicate $\sim$kpc-scale jets or AGN-driven shocks as the power source of the radio emission.  In particular, J0935+4819 has a symmetric double-lobe morphology; several other sources manifest apparent linear emission extended outward from the core, or widespread flocculent emission that extends symmetrically outward from the core.
\end{itemize}

Through comparison of the new data with lower-resolution ($3\farcs5$) 6-GHz observations from the literature, we concluded the following.

\begin{itemize}
    \item Although not always directly visible in the images, extended sub-galactic emission is present within $\gtrapprox$70\% of RQQs in this population.
    \item The bulk of sub-galactic radio emission--- about 65\% on average--- is dominated by an unresolved core component coincident with the QSO's optical position.
    \item There is direct evidence of variability in the sample: six Resolved RQQs and four Unresolved RQQs increased in 6-GHz flux density from 2010 to 2019.  We expect that a similar number decreased in 6-GHz emission, but the latter cannot be definitively identified due to the change in angular resolution between the two sets of observations.
\end{itemize}

We additionally utilized 1.4-GHz FIRST data \citep{first,Helfand15} and ``Quick Look" images from Epoch~1 of the 3-GHz VLA Sky Survey \citep{vlass} to investigate the 1.4--3--6~GHz spectral indices for the subset of the full QSO sample (178 QSOs, $0.2<z<0.3$) that is brighter than 200\,$\mu$Jy\,beam$^{-1}$ at 6~GHz (3\farcs5 resolution), comprising 29 RQQs, 27 RIQs, and 14 RLQs (core-component only).  We make the following observations.

\begin{itemize}
    \item RQQs generally have steep spectral indices ($\alpha\approx-0.7$), while RLQs have spectral indices that are flat or even inverted ($\alpha\gtrsim0$).  These calculated spectral indices are consistent with other observations in the literature.
    \item RIQs have spectral indices that span between the RQ and RL populations.  This may indicate that RIQs are a transitional population.
    \item Three RIQs are clear outliers from the rest of the population; we conclude that they are likely to be GPS sources, which may indeed be undergoing this hypothesized transitional phase.
\end{itemize}

While we have achieved an important step forward in the analysis of origins of radio emission from RQQs, further observational data are needed in order to form more concrete conclusions.  For example, follow-up observations at this same frequency and resolution would probe possible variability--- both in radio power and morphology--- while observations at different frequencies would provide more concrete conclusions regarding the spectral slopes/curvature of individual sources. 
Far-infrared data could yield estimates of the star-formation contribution in individual sources, while X-ray data may do the same in terms of a potential coronal or jet component.
Much higher-resolution radio observations (achievable with Very Long Baseline Interferometry) would put limits on the brightness temperatures of individual features, and allow for direct separation between starburst/jet contributions in individual sources.  A more detailed analysis of optical/UV spectra of this sample may also help interpret the role that AGN-related mechanisms play in shaping the RQQ population.

\color{black}

\acknowledgements

We thank Pedro Beaklini, Kristina Nyland, Ken Kellermann, and James Condon for helpful comments that improved the manuscript.  We thank the referee for suggestions that improved the clarity of this paper.

Trevor McCaffrey is a student at the National Radio Astronomy Observatory. The National Radio Astronomy Observatory is a facility of the National Science Foundation operated under cooperative agreement by Associated Universities, Inc.  This project was partially funded by the NSF through the REU program at NRAO.

This research has made use of NASA’s Astrophysics Data System Bibliographic Services, Ned Wright's online Javascript Cosmology Calculator \citep{cosmocalc}, AstroPy \citep{astropy:2013, astropy:2018}, Matplotlib \citep{Hunter:2007}, NumPy \citep{numpy}, SciPy \citep{scipy}, Pandas \citep{reback2020pandas}, and the CIRADA cutout service at URL cutouts.cirada.ca, operated by the Canadian Initiative for Radio Astronomy Data Analysis (CIRADA). CIRADA is funded by a grant from the Canada Foundation for Innovation 2017 Innovation Fund (Project 35999), as well as by the Provinces of Ontario, British Columbia, Alberta, Manitoba and Quebec, in collaboration with the National Research Council of Canada, the US National Radio Astronomy Observatory and Australia’s Commonwealth Scientific and Industrial Research Organisation.

\bibliography{bibliography}{}

\begin{thebibliography}{}
\expandafter\ifx\csname natexlab\endcsname\relax\def\natexlab#1{#1}\fi
\providecommand{\url}[1]{\href{#1}{#1}}
\providecommand{\dodoi}[1]{doi:~\href{http://doi.org/#1}{\nolinkurl{#1}}}
\providecommand{\doeprint}[1]{\href{http://ascl.net/#1}{\nolinkurl{http://ascl.net/#1}}}
\providecommand{\doarXiv}[1]{\href{https://arxiv.org/abs/#1}{\nolinkurl{https://arxiv.org/abs/#1}}}

\bibitem[{{Abazajian} {et~al.}(2009){Abazajian}, {Adelman-McCarthy},
  {Ag{\"u}eros}, {Allam}, {Allende Prieto}, {An}, {Anderson}, {Anderson},
  {Annis}, {Bahcall}, {Bailer-Jones}, {Barentine}, {Bassett}, {Becker},
  {Beers}, {Bell}, {Belokurov}, {Berlind}, {Berman}, {Bernardi}, {Bickerton},
  {Bizyaev}, {Blakeslee}, {Blanton}, {Bochanski}, {Boroski}, {Brewington},
  {Brinchmann}, {Brinkmann}, {Brunner}, {Budav{\'a}ri}, {Carey}, {Carliles},
  {Carr}, {Castander}, {Cinabro}, {Connolly}, {Csabai}, {Cunha}, {Czarapata},
  {Davenport}, {de Haas}, {Dilday}, {Doi}, {Eisenstein}, {Evans}, {Evans},
  {Fan}, {Friedman}, {Frieman}, {Fukugita}, {G{\"a}nsicke}, {Gates},
  {Gillespie}, {Gilmore}, {Gonzalez}, {Gonzalez}, {Grebel}, {Gunn},
  {Gy{\"o}ry}, {Hall}, {Harding}, {Harris}, {Harvanek}, {Hawley}, {Hayes},
  {Heckman}, {Hendry}, {Hennessy}, {Hindsley}, {Hoblitt}, {Hogan}, {Hogg},
  {Holtzman}, {Hyde}, {Ichikawa}, {Ichikawa}, {Im}, {Ivezi{\'c}}, {Jester},
  {Jiang}, {Johnson}, {Jorgensen}, {Juri{\'c}}, {Kent}, {Kessler}, {Kleinman},
  {Knapp}, {Konishi}, {Kron}, {Krzesinski}, {Kuropatkin}, {Lampeitl},
  {Lebedeva}, {Lee}, {Lee}, {French Leger}, {L{\'e}pine}, {Li}, {Lima}, {Lin},
  {Long}, {Loomis}, {Loveday}, {Lupton}, {Magnier}, {Malanushenko},
  {Malanushenko}, {Mand elbaum}, {Margon}, {Marriner}, {Mart{\'\i}nez-Delgado},
  {Matsubara}, {McGehee}, {McKay}, {Meiksin}, {Morrison}, {Mullally}, {Munn},
  {Murphy}, {Nash}, {Nebot}, {Neilsen}, {Newberg}, {Newman}, {Nichol},
  {Nicinski}, {Nieto-Santisteban}, {Nitta}, {Okamura}, {Oravetz}, {Ostriker},
  {Owen}, {Padmanabhan}, {Pan}, {Park}, {Pauls}, {Peoples}, {Percival}, {Pier},
  {Pope}, {Pourbaix}, {Price}, {Purger}, {Quinn}, {Raddick}, {Re Fiorentin},
  {Richards}, {Richmond}, {Riess}, {Rix}, {Rockosi}, {Sako}, {Schlegel},
  {Schneider}, {Scholz}, {Schreiber}, {Schwope}, {Seljak}, {Sesar}, {Sheldon},
  {Shimasaku}, {Sibley}, {Simmons}, {Sivarani}, {Allyn Smith}, {Smith},
  {Smol{\v{c}}i{\'c}}, {Snedden}, {Stebbins}, {Steinmetz}, {Stoughton},
  {Strauss}, {SubbaRao}, {Suto}, {Szalay}, {Szapudi}, {Szkody}, {Tanaka},
  {Tegmark}, {Teodoro}, {Thakar}, {Tremonti}, {Tucker}, {Uomoto}, {Vanden
  Berk}, {Vandenberg}, {Vidrih}, {Vogeley}, {Voges}, {Vogt}, {Wadadekar},
  {Watters}, {Weinberg}, {West}, {White}, {Wilhite}, {Wonders}, {Yanny},
  {Yocum}, {York}, {Zehavi}, {Zibetti}, \& {Zucker}}]{sdss}
{Abazajian}, K.~N., {Adelman-McCarthy}, J.~K., {Ag{\"u}eros}, M.~A., {et~al.}
  2009, \apjs, 182, 543, \dodoi{10.1088/0067-0049/182/2/543}

\bibitem[{{An} \& {Baan}(2012)}]{anBaan}
{An}, T., \& {Baan}, W.~A. 2012, \apj, 760, 77,
  \dodoi{10.1088/0004-637X/760/1/77}

\bibitem[{{Anantharamaiah} {et~al.}(2000){Anantharamaiah}, {Viallefond},
  {Mohan}, {Goss}, \& {Zhao}}]{anantharamaiah2000}
{Anantharamaiah}, K.~R., {Viallefond}, F., {Mohan}, N.~R., {Goss}, W.~M., \&
  {Zhao}, J.~H. 2000, \apj, 537, 613, \dodoi{10.1086/309063}

\bibitem[{{Astropy Collaboration} {et~al.}(2013){Astropy Collaboration},
  {Robitaille}, {Tollerud}, {Greenfield}, {Droettboom}, {Bray}, {Aldcroft},
  {Davis}, {Ginsburg}, {Price-Whelan}, {Kerzendorf}, {Conley}, {Crighton},
  {Barbary}, {Muna}, {Ferguson}, {Grollier}, {Parikh}, {Nair}, {Unther},
  {Deil}, {Woillez}, {Conseil}, {Kramer}, {Turner}, {Singer}, {Fox}, {Weaver},
  {Zabalza}, {Edwards}, {Azalee Bostroem}, {Burke}, {Casey}, {Crawford},
  {Dencheva}, {Ely}, {Jenness}, {Labrie}, {Lim}, {Pierfederici}, {Pontzen},
  {Ptak}, {Refsdal}, {Servillat}, \& {Streicher}}]{astropy:2013}
{Astropy Collaboration}, {Robitaille}, T.~P., {Tollerud}, E.~J., {et~al.} 2013,
  \aap, 558, A33, \dodoi{10.1051/0004-6361/201322068}

\bibitem[{{Baldi} {et~al.}(2015){Baldi}, {Capetti}, \&
  {Giovannini}}]{Baldi+2015}
{Baldi}, R.~D., {Capetti}, A., \& {Giovannini}, G. 2015, \aap, 576, A38,
  \dodoi{10.1051/0004-6361/201425426}

\bibitem[{{Baldi} {et~al.}(2018){Baldi}, {Capetti}, \& {Massaro}}]{Baldi+2018}
{Baldi}, R.~D., {Capetti}, A., \& {Massaro}, F. 2018, \aap, 609, A1,
  \dodoi{10.1051/0004-6361/201731333}

\bibitem[{{Balokovi{\'c}} {et~al.}(2012){Balokovi{\'c}}, {Smol{\v c}i{\'c}},
  {Ivezi{\'c}}, {Zamorani}, {Schinnerer}, \& {Kelly}}]{Balokovic2012}
{Balokovi{\'c}}, M., {Smol{\v c}i{\'c}}, V., {Ivezi{\'c}}, {\v Z}., {et~al.}
  2012, \apj, 759, 30, \dodoi{10.1088/0004-637X/759/1/30}

\bibitem[{{Barcos-Mu{\~n}oz} {et~al.}(2015){Barcos-Mu{\~n}oz}, {Leroy},
  {Evans}, {Privon}, {Armus}, {Condon}, {Mazzarella}, {Meier}, {Momjian},
  {Murphy}, {Ott}, {Reichardt}, {Sakamoto}, {Sanders}, {Schinnerer},
  {Stierwalt}, {Surace}, {Thompson}, \& {Walter}}]{BarcosMunoz+15}
{Barcos-Mu{\~n}oz}, L., {Leroy}, A.~K., {Evans}, A.~S., {et~al.} 2015, \apj,
  799, 10, \dodoi{10.1088/0004-637X/799/1/10}

\bibitem[{{Barvainis} {et~al.}(2005){Barvainis}, {Leh{\'a}r}, {Birkinshaw},
  {Falcke}, \& {Blundell}}]{barvainis2005}
{Barvainis}, R., {Leh{\'a}r}, J., {Birkinshaw}, M., {Falcke}, H., \&
  {Blundell}, K.~M. 2005, \apj, 618, 108, \dodoi{10.1086/425859}

\bibitem[{{Bastian} {et~al.}(1998){Bastian}, {Benz}, \&
  {Gary}}]{Bastian+1998_bursts}
{Bastian}, T.~S., {Benz}, A.~O., \& {Gary}, D.~E. 1998, \araa, 36, 131,
  \dodoi{10.1146/annurev.astro.36.1.131}

\bibitem[{{Beck} {et~al.}(2016){Beck}, {Dobos}, {Budav{\'a}ri}, {Szalay}, \&
  {Csabai}}]{photoZ_sdss}
{Beck}, R., {Dobos}, L., {Budav{\'a}ri}, T., {Szalay}, A.~S., \& {Csabai}, I.
  2016, \mnras, 460, 1371, \dodoi{10.1093/mnras/stw1009}

\bibitem[{{Becker} {et~al.}(1995){Becker}, {White}, \& {Helfand}}]{first}
{Becker}, R.~H., {White}, R.~L., \& {Helfand}, D.~J. 1995, \apj, 450, 559,
  \dodoi{10.1086/176166}

\bibitem[{{Blandford} {et~al.}(2019){Blandford}, {Meier}, \&
  {Readhead}}]{Blandford+2019}
{Blandford}, R., {Meier}, D., \& {Readhead}, A. 2019, \araa, 57, 467,
  \dodoi{10.1146/annurev-astro-081817-051948}

\bibitem[{{Blundell} \& {Rawlings}(2001)}]{BR01}
{Blundell}, K.~M., \& {Rawlings}, S. 2001, \apjl, 562, L5,
  \dodoi{10.1086/337970}

\bibitem[{{Burbidge}(1967)}]{Burbidge67}
{Burbidge}, E.~M. 1967, \araa, 5, 399,
  \dodoi{10.1146/annurev.aa.05.090167.002151}

\bibitem[{{Burbidge} \& {Burbidge}(1967)}]{BB67_book}
{Burbidge}, G.~R., \& {Burbidge}, E.~M. 1967, {Quasi-stellar objects}

\bibitem[{{Chi} {et~al.}(2013){Chi}, {Barthel}, \& {Garrett}}]{Chi+13}
{Chi}, S., {Barthel}, P.~D., \& {Garrett}, M.~A. 2013, \aap, 550, A68,
  \dodoi{10.1051/0004-6361/201220783}

\bibitem[{{Colina} \& {Perez-Olea}(1993)}]{colina1993}
{Colina}, L., \& {Perez-Olea}, D. 1993, \apss, 205, 99,
  \dodoi{10.1007/BF00657963}

\bibitem[{{Condon}(1992)}]{condon1992}
{Condon}, J.~J. 1992, \araa, 30, 575,
  \dodoi{10.1146/annurev.aa.30.090192.003043}

\bibitem[{{Condon} {et~al.}(2002){Condon}, {Cotton}, \&
  {Broderick}}]{Condon+2002}
{Condon}, J.~J., {Cotton}, W.~D., \& {Broderick}, J.~J. 2002, \aj, 124, 675,
  \dodoi{10.1086/341650}

\bibitem[{{Condon} {et~al.}(1998){Condon}, {Cotton}, {Greisen}, {Yin},
  {Perley}, {Taylor}, \& {Broderick}}]{NVSS}
{Condon}, J.~J., {Cotton}, W.~D., {Greisen}, E.~W., {et~al.} 1998, \aj, 115,
  1693, \dodoi{10.1086/300337}

\bibitem[{{Condon} {et~al.}(1992){Condon}, {Huang}, {Yin}, \&
  {Thuan}}]{condon1992b}
{Condon}, J.~J., {Huang}, Z.~P., {Yin}, Q.~F., \& {Thuan}, T.~X. 1992, in
  Astronomical Society of the Pacific Conference Series, Vol.~31, Relationships
  Between Active Galactic Nuclei and Starburst Galaxies, ed. A.~V.
  {Filippenko}, 79

\bibitem[{{Condon} {et~al.}(2013){Condon}, {Kellermann}, {Kimball},
  {Ivezi{\'c}}, \& {Perley}}]{Condon+2013}
{Condon}, J.~J., {Kellermann}, K.~I., {Kimball}, A.~E., {Ivezi{\'c}}, {\v{Z}}.,
  \& {Perley}, R.~A. 2013, \apj, 768, 37, \dodoi{10.1088/0004-637X/768/1/37}

\bibitem[{{Cornwell} {et~al.}(2005){Cornwell}, {Golap}, \& {Bhatnagar}}]{wproj}
{Cornwell}, T.~J., {Golap}, K., \& {Bhatnagar}, S. 2005, in Astronomical
  Society of the Pacific Conference Series, Vol. 347, Astronomical Data
  Analysis Software and Systems XIV, ed. P.~{Shopbell}, M.~{Britton}, \&
  R.~{Ebert}, 86

\bibitem[{{Cotton} {et~al.}(1980){Cotton}, {Wittels}, {Shapiro}, {Marcaide},
  {Owen}, {Spangler}, {Rius}, {Angulo}, {Clark}, \& {Knight}}]{Cotton1980_SSA}
{Cotton}, W.~D., {Wittels}, J.~J., {Shapiro}, I.~I., {et~al.} 1980, \apjl, 238,
  L123, \dodoi{10.1086/183271}

\bibitem[{{Eddington}(1913)}]{EddingtonBias}
{Eddington}, A.~S. 1913, \mnras, 73, 359, \dodoi{10.1093/mnras/73.5.359}

\bibitem[{{Fanaroff} \& {Riley}(1974)}]{Fanaroff1974}
{Fanaroff}, B.~L., \& {Riley}, J.~M. 1974, \mnras, 167, 31P,
  \dodoi{10.1093/mnras/167.1.31P}

\bibitem[{{Harris} {et~al.}(2020){Harris}, {Jarrod Millman}, {van der Walt},
  {Gommers}, {Virtanen}, {Cournapeau}, {Wieser}, {Taylor}, {Berg}, {Smith},
  {Kern}, {Picus}, {Hoyer}, {van Kerkwijk}, {Brett}, {Haldane}, {Fern{\'a}ndez
  del R{\'\i}o}, {Wiebe}, {Peterson}, {G{\'e}rard-Marchant}, {Sheppard},
  {Reddy}, {Weckesser}, {Abbasi}, {Gohlke}, \& {Oliphant}}]{numpy}
{Harris}, C.~R., {Jarrod Millman}, K., {van der Walt}, S.~J., {et~al.} 2020,
  \nat, 585, 357, \dodoi{10.1038/s41586-020-2649-2}

\bibitem[{{Helfand} {et~al.}(2015){Helfand}, {White}, \& {Becker}}]{Helfand15}
{Helfand}, D.~J., {White}, R.~L., \& {Becker}, R.~H. 2015, \apj, 801, 26,
  \dodoi{10.1088/0004-637X/801/1/26}

\bibitem[{{Herrera Ruiz} {et~al.}(2016){Herrera Ruiz}, {Middelberg}, {Norris},
  \& {Maini}}]{HR16}
{Herrera Ruiz}, N., {Middelberg}, E., {Norris}, R.~P., \& {Maini}, A. 2016,
  \aap, 589, L2, \dodoi{10.1051/0004-6361/201628302}

\bibitem[{{Herrera Ruiz} {et~al.}(2017){Herrera Ruiz}, {Middelberg}, {Deller},
  {Norris}, {Best}, {Brisken}, {Schinnerer}, {Smol{\v{c}}i{\'c}}, {Delvecchio},
  {Momjian}, {Bomans}, {Scoville}, \& {Carilli}}]{HR17}
{Herrera Ruiz}, N., {Middelberg}, E., {Deller}, A., {et~al.} 2017, \aap, 607,
  A132, \dodoi{10.1051/0004-6361/201731163}

\bibitem[{{Hodge} {et~al.}(2011){Hodge}, {Becker}, {White}, {Richards}, \&
  {Zeimann}}]{hodge2011}
{Hodge}, J.~A., {Becker}, R.~H., {White}, R.~L., {Richards}, G.~T., \&
  {Zeimann}, G.~R. 2011, \aj, 142, 3, \dodoi{10.1088/0004-6256/142/1/3}

\bibitem[{{Hovatta} {et~al.}(2007){Hovatta}, {Tornikoski}, {Lainela}, {Lehto},
  {Valtaoja}, {Torniainen}, {Aller}, \& {Aller}}]{Hovatta+2007}
{Hovatta}, T., {Tornikoski}, M., {Lainela}, M., {et~al.} 2007, \aap, 469, 899,
  \dodoi{10.1051/0004-6361:20077529}

\bibitem[{Hunter(2007)}]{Hunter:2007}
Hunter, J.~D. 2007, Computing In Science \& Engineering, 9, 90,
  \dodoi{10.1109/MCSE.2007.55}

\bibitem[{{Jagannathan} {et~al.}(2018){Jagannathan}, {Bhatnagar}, {Brisken}, \&
  {Taylor}}]{aproj}
{Jagannathan}, P., {Bhatnagar}, S., {Brisken}, W., \& {Taylor}, A.~R. 2018,
  \aj, 155, 3, \dodoi{10.3847/1538-3881/aa989f}

\bibitem[{{Kellermann} {et~al.}(2016){Kellermann}, {Condon}, {Kimball},
  {Perley}, \& {Ivezi{\'c}}}]{kellerman16}
{Kellermann}, K.~I., {Condon}, J.~J., {Kimball}, A.~E., {Perley}, R.~A., \&
  {Ivezi{\'c}}, {\v{Z}}. 2016, \apj, 831, 168,
  \dodoi{10.3847/0004-637X/831/2/168}

\bibitem[{{Kellermann} \& {Pauliny-Toth}(1981)}]{Kell_Pauliny_compact}
{Kellermann}, K.~I., \& {Pauliny-Toth}, I.~I.~K. 1981, \araa, 19, 373,
  \dodoi{10.1146/annurev.aa.19.090181.002105}

\bibitem[{{Kellermann} {et~al.}(1989){Kellermann}, {Sramek}, {Schmidt},
  {Shaffer}, \& {Green}}]{Kellerman1989}
{Kellermann}, K.~I., {Sramek}, R., {Schmidt}, M., {Shaffer}, D.~B., \& {Green},
  R. 1989, \aj, 98, 1195, \dodoi{10.1086/115207}

\bibitem[{{Kimball} \& {Ivezi{\'c}}(2008)}]{kimballIvezic}
{Kimball}, A.~E., \& {Ivezi{\'c}}, {\v{Z}}. 2008, \aj, 136, 684,
  \dodoi{10.1088/0004-6256/136/2/684}

\bibitem[{{Kimball} {et~al.}(2011{\natexlab{a}}){Kimball}, {Ivezi{\'c}},
  {Wiita}, \& {Schneider}}]{Kimball+11b}
{Kimball}, A.~E., {Ivezi{\'c}}, {\v{Z}}., {Wiita}, P.~J., \& {Schneider}, D.~P.
  2011{\natexlab{a}}, \aj, 141, 182, \dodoi{10.1088/0004-6256/141/6/182}

\bibitem[{{Kimball} {et~al.}(2011{\natexlab{b}}){Kimball}, {Kellermann},
  {Condon}, {Ivezi{\'c}}, \& {Perley}}]{kimball2011}
{Kimball}, A.~E., {Kellermann}, K.~I., {Condon}, J.~J., {Ivezi{\'c}}, {\v{Z}}.,
  \& {Perley}, R.~A. 2011{\natexlab{b}}, \apjl, 739, L29,
  \dodoi{10.1088/2041-8205/739/1/L29}

\bibitem[{{Klindt} {et~al.}(2019){Klindt}, {Alexander}, {Rosario}, {Lusso}, \&
  {Fotopoulou}}]{Klindt+2019}
{Klindt}, L., {Alexander}, D.~M., {Rosario}, D.~J., {Lusso}, E., \&
  {Fotopoulou}, S. 2019, \mnras, 488, 3109, \dodoi{10.1093/mnras/stz1771}

\bibitem[{{Komossa}(2015)}]{Komossa2015}
{Komossa}, S. 2015, Journal of High Energy Astrophysics, 7, 148,
  \dodoi{10.1016/j.jheap.2015.04.006}

\bibitem[{{Kunert-Bajraszewska} {et~al.}(2020){Kunert-Bajraszewska},
  {Wo{\l}owska}, {Mooley}, {Kharb}, \& {Hallinan}}]{KB20}
{Kunert-Bajraszewska}, M., {Wo{\l}owska}, A., {Mooley}, K., {Kharb}, P., \&
  {Hallinan}, G. 2020, \apj, 897, 128, \dodoi{10.3847/1538-4357/ab9598}

\bibitem[{{Lacy} {et~al.}(2020){Lacy}, {Baum}, {Chandler}, {Chatterjee},
  {Clarke}, {Deustua}, {English}, {Farnes}, {Gaensler}, {Gugliucci},
  {Hallinan}, {Kent}, {Kimball}, {Law}, {Lazio}, {Marvil}, {Mao}, {Medlin},
  {Mooley}, {Murphy}, {Myers}, {Osten}, {Richards}, {Rosolowsky}, {Rudnick},
  {Schinzel}, {Sivakoff}, {Sjouwerman}, {Taylor}, {White}, {Wrobel},
  {Andernach}, {Beasley}, {Berger}, {Bhatnager}, {Birkinshaw}, {Bower},
  {Brandt}, {Brown}, {Burke-Spolaor}, {Butler}, {Comerford}, {Demorest}, {Fu},
  {Giacintucci}, {Golap}, {G{\"u}th}, {Hales}, {Hiriart}, {Hodge}, {Horesh},
  {Ivezi{\'c}}, {Jarvis}, {Kamble}, {Kassim}, {Liu}, {Loinard}, {Lyons},
  {Masters}, {Mezcua}, {Moellenbrock}, {Mroczkowski}, {Nyland},
  {O{\textquoteright}Dea}, {O{\textquoteright}Sullivan}, {Peters}, {Radford},
  {Rao}, {Robnett}, {Salcido}, {Shen}, {Sobotka}, {Witz}, {Vaccari}, {van
  Weeren}, {Vargas}, {Williams}, \& {Yoon}}]{vlass}
{Lacy}, M., {Baum}, S.~A., {Chandler}, C.~J., {et~al.} 2020, \pasp, 132,
  035001, \dodoi{10.1088/1538-3873/ab63eb}

\bibitem[{{Lerner}(2018)}]{Lerner2018}
{Lerner}, E.~J. 2018, \mnras, 477, 3185, \dodoi{10.1093/mnras/sty728}

\bibitem[{{Lonsdale} {et~al.}(2006){Lonsdale}, {Diamond}, {Thrall}, {Smith}, \&
  {Lonsdale}}]{lonsdale2006}
{Lonsdale}, C.~J., {Diamond}, P.~J., {Thrall}, H., {Smith}, H.~E., \&
  {Lonsdale}, C.~J. 2006, \apj, 647, 185, \dodoi{10.1086/505193}

\bibitem[{{Macfarlane} {et~al.}(2021){Macfarlane}, {Best}, {Sabater},
  {G{\"u}rkan}, {Jarvis}, {R{\"o}ttgering}, {Baldi}, {Calistro Rivera},
  {Duncan}, {Morabito}, {Prandoni}, \& {Retana-Montenegro}}]{MacFarlane+2021}
{Macfarlane}, C., {Best}, P.~N., {Sabater}, J., {et~al.} 2021, \mnras, 506,
  5888, \dodoi{10.1093/mnras/stab1998}

\bibitem[{{Maini} {et~al.}(2016){Maini}, {Prandoni}, {Norris}, {Giovannini}, \&
  {Spitler}}]{Maini+16}
{Maini}, A., {Prandoni}, I., {Norris}, R.~P., {Giovannini}, G., \& {Spitler},
  L.~R. 2016, \aap, 589, L3, \dodoi{10.1051/0004-6361/201628305}

\bibitem[{Massey(1951)}]{K-S_Test}
Massey, F.~J. 1951, Journal of the American Statistical Association, 46, 68.
\newblock \url{http://www.jstor.org/stable/2280095}

\bibitem[{{McMullin} {et~al.}(2007){McMullin}, {Waters}, {Schiebel}, {Young},
  \& {Golap}}]{casa}
{McMullin}, J.~P., {Waters}, B., {Schiebel}, D., {Young}, W., \& {Golap}, K.
  2007, in Astronomical Society of the Pacific Conference Series, Vol. 376,
  Astronomical Data Analysis Software and Systems XVI, ed. R.~A. {Shaw},
  F.~{Hill}, \& D.~J. {Bell}, 127

\bibitem[{{Melrose}(1980)}]{Melrose1980_bursts}
{Melrose}, D.~B. 1980, \ssr, 26, 3, \dodoi{10.1007/BF00212597}

\bibitem[{{Mohan} \& {Rafferty}(2015)}]{PyBDSF}
{Mohan}, N., \& {Rafferty}, D. 2015, {PyBDSF: Python Blob Detection and Source
  Finder}.
\newblock \doeprint{1502.007}

\bibitem[{{Morabito} {et~al.}(2022){Morabito}, {Jackson}, {Mooney}, {Sweijen},
  {Badole}, {Kukreti}, {Venkattu}, {Groeneveld}, {Kappes}, {Bonnassieux},
  {Drabent}, {Iacobelli}, {Croston}, {Best}, {Bondi}, {Callingham}, {Conway},
  {Deller}, {Hardcastle}, {McKean}, {Miley}, {Moldon}, {R{\"o}ttgering},
  {Tasse}, {Shimwell}, {van Weeren}, {Anderson}, {Asgekar}, {Avruch}, {van
  Bemmel}, {Bentum}, {Bonafede}, {Brouw}, {Butcher}, {Ciardi}, {Corstanje},
  {Coolen}, {Damstra}, {de Gasperin}, {Duscha}, {Eisl{\"o}ffel}, {Engels},
  {Falcke}, {Garrett}, {Griessmeier}, {Gunst}, {van Haarlem}, {Hoeft}, {van der
  Horst}, {J{\"u}tte}, {Kadler}, {Koopmans}, {Krankowski}, {Mann}, {Nelles},
  {Oonk}, {Orru}, {Paas}, {Pandey}, {Pizzo}, {Pandey-Pommier}, {Reich},
  {Rothkaehl}, {Ruiter}, {Schwarz}, {Shulevski}, {Soida}, {Tagger}, {Vocks},
  {Wijers}, {Wijnholds}, {Wucknitz}, {Zarka}, \& {Zucca}}]{Morabito+22}
{Morabito}, L.~K., {Jackson}, N.~J., {Mooney}, S., {et~al.} 2022, \aap, 658,
  A1, \dodoi{10.1051/0004-6361/202140649}

\bibitem[{{Nyland} {et~al.}(2020){Nyland}, {Dong}, {Patil}, {Lacy}, {Kimball},
  {Hallinan}, {Sarbadhicary}, {Polisensky}, {Kassim}, {Peters}, {Clarke},
  {Mukherjee}, {van Velzen}, \& {Baldassare}}]{Nyland+2020}
{Nyland}, K., {Dong}, D., {Patil}, P., {et~al.} 2020, arXiv e-prints,
  arXiv:2005.04734.
\newblock \doarXiv{2005.04734}

\bibitem[{{O'Dea}(1998)}]{Odea1998}
{O'Dea}, C.~P. 1998, \pasp, 110, 493, \dodoi{10.1086/316162}

\bibitem[{{O'Dea} \& {Saikia}(2021)}]{Odea+Saikia21}
{O'Dea}, C.~P., \& {Saikia}, D.~J. 2021, \aapr, 29, 3,
  \dodoi{10.1007/s00159-021-00131-w}

\bibitem[{{O'Dea} {et~al.}(2008){O'Dea}, {Baum}, {Privon}, {Noel-Storr},
  {Quillen}, {Zufelt}, {Park}, {Edge}, {Russell}, {Fabian}, {Donahue},
  {Sarazin}, {McNamara}, {Bregman}, \& {Egami}}]{Odea2008}
{O'Dea}, C.~P., {Baum}, S.~A., {Privon}, G., {et~al.} 2008, \apj, 681, 1035,
  \dodoi{10.1086/588212}

\bibitem[{{Oke} \& {Gunn}(1983)}]{ABmag}
{Oke}, J.~B., \& {Gunn}, J.~E. 1983, \apj, 266, 713, \dodoi{10.1086/160817}

\bibitem[{{Padovani} {et~al.}(2015){Padovani}, {Bonzini}, {Kellermann},
  {Miller}, {Mainieri}, \& {Tozzi}}]{Padovani+2015}
{Padovani}, P., {Bonzini}, M., {Kellermann}, K.~I., {et~al.} 2015, \mnras, 452,
  1263, \dodoi{10.1093/mnras/stv1375}

\bibitem[{{Padovani} {et~al.}(2011){Padovani}, {Miller}, {Kellermann},
  {Mainieri}, {Rosati}, \& {Tozzi}}]{padovani11}
{Padovani}, P., {Miller}, N., {Kellermann}, K.~I., {et~al.} 2011, \apj, 740,
  20, \dodoi{10.1088/0004-637X/740/1/20}

\bibitem[{pandas~development team(2020)}]{reback2020pandas}
pandas~development team, T. 2020, pandas-dev/pandas: Pandas, latest,  Zenodo,
  \dodoi{10.5281/zenodo.3509134}

\bibitem[{{Perley} \& {Butler}(2017)}]{perleyButlerFlux}
{Perley}, R.~A., \& {Butler}, B.~J. 2017, \apjs, 230, 7,
  \dodoi{10.3847/1538-4365/aa6df9}

\bibitem[{{Perley} {et~al.}(2011){Perley}, {Chandler}, {Butler}, \&
  {Wrobel}}]{EVLA}
{Perley}, R.~A., {Chandler}, C.~J., {Butler}, B.~J., \& {Wrobel}, J.~M. 2011,
  \apjl, 739, L1, \dodoi{10.1088/2041-8205/739/1/L1}

\bibitem[{{Price-Whelan} {et~al.}(2018){Price-Whelan}, {Sip{\H{o}}cz},
  {G{\"u}nther}, {Lim}, {Crawford}, {Conseil}, {Shupe}, {Craig}, {Dencheva},
  {Ginsburg}, {VanderPlas}, {Bradley}, {P{\'e}rez-Su{\'a}rez}, {de Val-Borro},
  {Paper Contributors}, {Aldcroft}, {Cruz}, {Robitaille}, {Tollerud},
  {Coordination Committee}, {Ardelean}, {Babej}, {Bach}, {Bachetti}, {Bakanov},
  {Bamford}, {Barentsen}, {Barmby}, {Baumbach}, {Berry}, {Biscani}, {Boquien},
  {Bostroem}, {Bouma}, {Brammer}, {Bray}, {Breytenbach}, {Buddelmeijer},
  {Burke}, {Calderone}, {Cano Rodr{\'\i}guez}, {Cara}, {Cardoso}, {Cheedella},
  {Copin}, {Corrales}, {Crichton}, {D{\textquoteright}Avella}, {Deil},
  {Depagne}, {Dietrich}, {Donath}, {Droettboom}, {Earl}, {Erben}, {Fabbro},
  {Ferreira}, {Finethy}, {Fox}, {Garrison}, {Gibbons}, {Goldstein}, {Gommers},
  {Greco}, {Greenfield}, {Groener}, {Grollier}, {Hagen}, {Hirst}, {Homeier},
  {Horton}, {Hosseinzadeh}, {Hu}, {Hunkeler}, {Ivezi{\'c}}, {Jain}, {Jenness},
  {Kanarek}, {Kendrew}, {Kern}, {Kerzendorf}, {Khvalko}, {King}, {Kirkby},
  {Kulkarni}, {Kumar}, {Lee}, {Lenz}, {Littlefair}, {Ma}, {Macleod},
  {Mastropietro}, {McCully}, {Montagnac}, {Morris}, {Mueller}, {Mumford},
  {Muna}, {Murphy}, {Nelson}, {Nguyen}, {Ninan}, {N{\"o}the}, {Ogaz}, {Oh},
  {Parejko}, {Parley}, {Pascual}, {Patil}, {Patil}, {Plunkett}, {Prochaska},
  {Rastogi}, {Reddy Janga}, {Sabater}, {Sakurikar}, {Seifert}, {Sherbert},
  {Sherwood-Taylor}, {Shih}, {Sick}, {Silbiger}, {Singanamalla}, {Singer},
  {Sladen}, {Sooley}, {Sornarajah}, {Streicher}, {Teuben}, {Thomas},
  {Tremblay}, {Turner}, {Terr{\'o}n}, {van Kerkwijk}, {de la Vega}, {Watkins},
  {Weaver}, {Whitmore}, {Woillez}, {Zabalza}, \& {Contributors}}]{astropy:2018}
{Price-Whelan}, A.~M., {Sip{\H{o}}cz}, B.~M., {G{\"u}nther}, H.~M., {et~al.}
  2018, \aj, 156, 123, \dodoi{10.3847/1538-3881/aabc4f}

\bibitem[{{Raginski} \& {Laor}(2016)}]{RL16}
{Raginski}, I., \& {Laor}, A. 2016, \mnras, 459, 2082,
  \dodoi{10.1093/mnras/stw772}

\bibitem[{{Ramaty}(1969)}]{Ramaty1969}
{Ramaty}, R. 1969, \apj, 158, 753, \dodoi{10.1086/150235}

\bibitem[{{Rau} \& {Cornwell}(2011)}]{mtmfs}
{Rau}, U., \& {Cornwell}, T.~J. 2011, \aap, 532, A71,
  \dodoi{10.1051/0004-6361/201117104}

\bibitem[{{Rees}(1984)}]{Rees+1984}
{Rees}, M.~J. 1984, \araa, 22, 471, \dodoi{10.1146/annurev.aa.22.090184.002351}

\bibitem[{{Richards} {et~al.}(2002){Richards}, {Fan}, {Newberg}, {Strauss},
  {Vanden Berk}, {Schneider}, {Yanny}, {Boucher}, {Burles}, {Frieman}, {Gunn},
  {Hall}, {Ivezi{\'c}}, {Kent}, {Loveday}, {Lupton}, {Rockosi}, {Schlegel},
  {Stoughton}, {SubbaRao}, \& {York}}]{Richards+2002a}
{Richards}, G.~T., {Fan}, X., {Newberg}, H.~J., {et~al.} 2002, \aj, 123, 2945,
  \dodoi{10.1086/340187}

\bibitem[{{Rosario} {et~al.}(2021){Rosario}, {Alexander}, {Moldon}, {Klindt},
  {Thomson}, {Morabito}, {Fawcett}, \& {Harrison}}]{Rosario+21}
{Rosario}, D.~J., {Alexander}, D.~M., {Moldon}, J., {et~al.} 2021, \mnras, 505,
  5283, \dodoi{10.1093/mnras/stab1653}

\bibitem[{{Sadler}(2016)}]{sadlerReview}
{Sadler}, E.~M. 2016, Astronomische Nachrichten, 337, 105,
  \dodoi{10.1002/asna.201512274}

\bibitem[{{Schlegel} {et~al.}(1998){Schlegel}, {Finkbeiner}, \&
  {Davis}}]{schlegel98}
{Schlegel}, D.~J., {Finkbeiner}, D.~P., \& {Davis}, M. 1998, \apj, 500, 525,
  \dodoi{10.1086/305772}

\bibitem[{{Schmidt}(1969)}]{Schmidt69}
{Schmidt}, M. 1969, \araa, 7, 527, \dodoi{10.1146/annurev.aa.07.090169.002523}

\bibitem[{{Schmidt} \& {Green}(1983)}]{SG83}
{Schmidt}, M., \& {Green}, R.~F. 1983, \apj, 269, 352, \dodoi{10.1086/161048}

\bibitem[{{Schneider} {et~al.}(2010){Schneider}, {Richards}, {Hall}, {Strauss},
  {Anderson}, {Boroson}, {Ross}, {Shen}, {Brandt}, {Fan}, {Inada}, {Jester},
  {Knapp}, {Krawczyk}, {Thakar}, {Vanden Berk}, {Voges}, {Yanny}, {York},
  {Bahcall}, {Bizyaev}, {Blanton}, {Brewington}, {Brinkmann}, {Eisenstein},
  {Frieman}, {Fukugita}, {Gray}, {Gunn}, {Hibon}, {Ivezi{\'c}}, {Kent}, {Kron},
  {Lee}, {Lupton}, {Malanushenko}, {Malanushenko}, {Oravetz}, {Pan}, {Pier},
  {Price}, {Saxe}, {Schlegel}, {Simmons}, {Snedden}, {SubbaRao}, {Szalay}, \&
  {Weinberg}}]{schneider10}
{Schneider}, D.~P., {Richards}, G.~T., {Hall}, P.~B., {et~al.} 2010, \aj, 139,
  2360, \dodoi{10.1088/0004-6256/139/6/2360}

\bibitem[{{Shimwell} {et~al.}(2019){Shimwell}, {Tasse}, {Hardcastle}, {Mechev},
  {Williams}, {Best}, {R{\"o}ttgering}, {Callingham}, {Dijkema}, {de Gasperin},
  {Hoang}, {Hugo}, {Mirmont}, {Oonk}, {Prandoni}, {Rafferty}, {Sabater},
  {Smirnov}, {van Weeren}, {White}, {Atemkeng}, {Bester}, {Bonnassieux},
  {Br{\"u}ggen}, {Brunetti}, {Chy{\.z}y}, {Cochrane}, {Conway}, {Croston},
  {Danezi}, {Duncan}, {Haverkorn}, {Heald}, {Iacobelli}, {Intema}, {Jackson},
  {Jamrozy}, {Jarvis}, {Lakhoo}, {Mevius}, {Miley}, {Morabito}, {Morganti},
  {Nisbet}, {Orr{\'u}}, {Perkins}, {Pizzo}, {Schrijvers}, {Smith}, {Vermeulen},
  {Wise}, {Alegre}, {Bacon}, {van Bemmel}, {Beswick}, {Bonafede}, {Botteon},
  {Bourke}, {Brienza}, {Calistro Rivera}, {Cassano}, {Clarke}, {Conselice},
  {Dettmar}, {Drabent}, {Dumba}, {Emig}, {En{\ss}lin}, {Ferrari}, {Garrett},
  {G{\'e}nova-Santos}, {Goyal}, {G{\"u}rkan}, {Hale}, {Harwood}, {Heesen},
  {Hoeft}, {Horellou}, {Jackson}, {Kokotanekov}, {Kondapally},
  {Kunert-Bajraszewska}, {Mahatma}, {Mahony}, {Mandal}, {McKean}, {Merloni},
  {Mingo}, {Miskolczi}, {Mooney}, {Nikiel-Wroczy{\'n}ski}, {O'Sullivan},
  {Quinn}, {Reich}, {Roskowi{\'n}ski}, {Rowlinson}, {Savini}, {Saxena},
  {Schwarz}, {Shulevski}, {Sridhar}, {Stacey}, {Urquhart}, {van der Wiel},
  {Varenius}, {Webster}, \& {Wilber}}]{LoTSS1}
{Shimwell}, T.~W., {Tasse}, C., {Hardcastle}, M.~J., {et~al.} 2019, \aap, 622,
  A1, \dodoi{10.1051/0004-6361/201833559}

\bibitem[{{Shimwell} {et~al.}(2022){Shimwell}, {Hardcastle}, {Tasse}, {Best},
  {R{\"o}ttgering}, {Williams}, {Botteon}, {Drabent}, {Mechev}, {Shulevski},
  {van Weeren}, {Bester}, {Br{\"u}ggen}, {Brunetti}, {Callingham}, {Chy{\.z}y},
  {Conway}, {Dijkema}, {Duncan}, {de Gasperin}, {Hale}, {Haverkorn}, {Hugo},
  {Jackson}, {Mevius}, {Miley}, {Morabito}, {Morganti}, {Offringa}, {Oonk},
  {Rafferty}, {Sabater}, {Smith}, {Schwarz}, {Smirnov}, {O'Sullivan},
  {Vedantham}, {White}, {Albert}, {Alegre}, {Asabere}, {Bacon}, {Bonafede},
  {Bonnassieux}, {Brienza}, {Bilicki}, {Bonato}, {Calistro Rivera}, {Cassano},
  {Cochrane}, {Croston}, {Cuciti}, {Dallacasa}, {Danezi}, {Dettmar}, {Di
  Gennaro}, {Edler}, {En{\ss}lin}, {Emig}, {Franzen}, {Garc{\'\i}a-Vergara},
  {Grange}, {G{\"u}rkan}, {Hajduk}, {Heald}, {Heesen}, {Hoang}, {Hoeft},
  {Horellou}, {Iacobelli}, {Jamrozy}, {Jeli{\'c}}, {Kondapally}, {Kukreti},
  {Kunert-Bajraszewska}, {Magliocchetti}, {Mahatma}, {Ma{\l}ek}, {Mandal},
  {Massaro}, {Meyer-Zhao}, {Mingo}, {Mostert}, {Nair}, {Nakoneczny},
  {Nikiel-Wroczy{\'n}ski}, {Orr{\'u}}, {Pajdosz-{\'S}mierciak}, {Pasini},
  {Prandoni}, {van Piggelen}, {Rajpurohit}, {Retana-Montenegro}, {Riseley},
  {Rowlinson}, {Saxena}, {Schrijvers}, {Sweijen}, {Siewert}, {Timmerman},
  {Vaccari}, {Vink}, {West}, {Wo{\l}owska}, {Zhang}, \&
  {Zheng}}]{Shimwell+22_LoTSS2}
{Shimwell}, T.~W., {Hardcastle}, M.~J., {Tasse}, C., {et~al.} 2022, \aap, 659,
  A1, \dodoi{10.1051/0004-6361/202142484}

\bibitem[{{Smol{\v{c}}i{\'c}} {et~al.}(2017){Smol{\v{c}}i{\'c}}, {Novak},
  {Bondi}, {Ciliegi}, {Mooley}, {Schinnerer}, {Zamorani}, {Navarrete},
  {Bourke}, {Karim}, {Vardoulaki}, {Leslie}, {Delhaize}, {Carilli}, {Myers},
  {Baran}, {Delvecchio}, {Miettinen}, {Banfield}, {Balokovi{\'c}}, {Bertoldi},
  {Capak}, {Frail}, {Hallinan}, {Hao}, {Herrera Ruiz}, {Horesh}, {Ilbert},
  {Intema}, {Jeli{\'c}}, {Kl{\"o}ckner}, {Krpan}, {Kulkarni}, {McCracken},
  {Laigle}, {Middleberg}, {Murphy}, {Sargent}, {Scoville}, \&
  {Sheth}}]{Smolcic2017_radioCOSMOS}
{Smol{\v{c}}i{\'c}}, V., {Novak}, M., {Bondi}, M., {et~al.} 2017, \aap, 602,
  A1, \dodoi{10.1051/0004-6361/201628704}

\bibitem[{{Sotnikova} {et~al.}(2019){Sotnikova}, {Mufakharov}, {Majorova},
  {Mingaliev}, {Udovitskii}, {Bursov}, \& {Semenova}}]{Sotnikova_GPSbelowPeak}
{Sotnikova}, Y.~V., {Mufakharov}, T.~V., {Majorova}, E.~K., {et~al.} 2019,
  Astrophysical Bulletin, 74, 348, \dodoi{10.1134/S1990341319040023}

\bibitem[{{Sweijen} {et~al.}(2022){Sweijen}, {van Weeren}, {R{\"o}ttgering},
  {Morabito}, {Jackson}, {Offringa}, {van der Tol}, {Veenboer}, {Oonk}, {Best},
  {Bondi}, {Shimwell}, {Tasse}, \& {Thomson}}]{Sweijen+22}
{Sweijen}, F., {van Weeren}, R.~J., {R{\"o}ttgering}, H.~J.~A., {et~al.} 2022,
  Nature Astronomy, 6, 350, \dodoi{10.1038/s41550-021-01573-z}

\bibitem[{{Terashima} \& {Wilson}(2003)}]{TW03_Rx}
{Terashima}, Y., \& {Wilson}, A.~S. 2003, \apj, 583, 145,
  \dodoi{10.1086/345339}

\bibitem[{{Terlevich} {et~al.}(1992){Terlevich}, {Tenorio-Tagle}, {Franco}, \&
  {Melnick}}]{Terlevich+1992}
{Terlevich}, R., {Tenorio-Tagle}, G., {Franco}, J., \& {Melnick}, J. 1992,
  \mnras, 255, 713, \dodoi{10.1093/mnras/255.4.713}

\bibitem[{{Thyagarajan} {et~al.}(2011){Thyagarajan}, {Helfand}, {White}, \&
  {Becker}}]{Thygarajan+2011}
{Thyagarajan}, N., {Helfand}, D.~J., {White}, R.~L., \& {Becker}, R.~H. 2011,
  \apj, 742, 49, \dodoi{10.1088/0004-637X/742/1/49}

\bibitem[{{Ulvestad} {et~al.}(2005){Ulvestad}, {Antonucci}, \&
  {Barvainis}}]{ulvestad2005}
{Ulvestad}, J.~S., {Antonucci}, R. R.~J., \& {Barvainis}, R. 2005, \apj, 621,
  123, \dodoi{10.1086/427426}

\bibitem[{{van Haarlem} {et~al.}(2013){van Haarlem}, {Wise}, {Gunst}, {Heald},
  {McKean}, {Hessels}, {de Bruyn}, {Nijboer}, {Swinbank}, {Fallows},
  {Brentjens}, {Nelles}, {Beck}, {Falcke}, {Fender}, {H{\"o}randel},
  {Koopmans}, {Mann}, {Miley}, {R{\"o}ttgering}, {Stappers}, {Wijers},
  {Zaroubi}, {van den Akker}, {Alexov}, {Anderson}, {Anderson}, {van Ardenne},
  {Arts}, {Asgekar}, {Avruch}, {Batejat}, {B{\"a}hren}, {Bell}, {Bell}, {van
  Bemmel}, {Bennema}, {Bentum}, {Bernardi}, {Best}, {B{\^\i}rzan}, {Bonafede},
  {Boonstra}, {Braun}, {Bregman}, {Breitling}, {van de Brink}, {Broderick},
  {Broekema}, {Brouw}, {Br{\"u}ggen}, {Butcher}, {van Cappellen}, {Ciardi},
  {Coenen}, {Conway}, {Coolen}, {Corstanje}, {Damstra}, {Davies}, {Deller},
  {Dettmar}, {van Diepen}, {Dijkstra}, {Donker}, {Doorduin}, {Dromer}, {Drost},
  {van Duin}, {Eisl{\"o}ffel}, {van Enst}, {Ferrari}, {Frieswijk}, {Gankema},
  {Garrett}, {de Gasperin}, {Gerbers}, {de Geus}, {Grie{\ss}meier}, {Grit},
  {Gruppen}, {Hamaker}, {Hassall}, {Hoeft}, {Holties}, {Horneffer}, {van der
  Horst}, {van Houwelingen}, {Huijgen}, {Iacobelli}, {Intema}, {Jackson},
  {Jelic}, {de Jong}, {Juette}, {Kant}, {Karastergiou}, {Koers}, {Kollen},
  {Kondratiev}, {Kooistra}, {Koopman}, {Koster}, {Kuniyoshi}, {Kramer},
  {Kuper}, {Lambropoulos}, {Law}, {van Leeuwen}, {Lemaitre}, {Loose}, {Maat},
  {Macario}, {Markoff}, {Masters}, {McFadden}, {McKay-Bukowski}, {Meijering},
  {Meulman}, {Mevius}, {Middelberg}, {Millenaar}, {Miller-Jones}, {Mohan},
  {Mol}, {Morawietz}, {Morganti}, {Mulcahy}, {Mulder}, {Munk}, {Nieuwenhuis},
  {van Nieuwpoort}, {Noordam}, {Norden}, {Noutsos}, {Offringa}, {Olofsson},
  {Omar}, {Orr{\'u}}, {Overeem}, {Paas}, {Pandey-Pommier}, {Pandey}, {Pizzo},
  {Polatidis}, {Rafferty}, {Rawlings}, {Reich}, {de Reijer}, {Reitsma},
  {Renting}, {Riemers}, {Rol}, {Romein}, {Roosjen}, {Ruiter}, {Scaife}, {van
  der Schaaf}, {Scheers}, {Schellart}, {Schoenmakers}, {Schoonderbeek},
  {Serylak}, {Shulevski}, {Sluman}, {Smirnov}, {Sobey}, {Spreeuw}, {Steinmetz},
  {Sterks}, {Stiepel}, {Stuurwold}, {Tagger}, {Tang}, {Tasse}, {Thomas},
  {Thoudam}, {Toribio}, {van der Tol}, {Usov}, {van Veelen}, {van der Veen},
  {ter Veen}, {Verbiest}, {Vermeulen}, {Vermaas}, {Vocks}, {Vogt}, {de Vos},
  {van der Wal}, {van Weeren}, {Weggemans}, {Weltevrede}, {White}, {Wijnholds},
  {Wilhelmsson}, {Wucknitz}, {Yatawatta}, {Zarka}, {Zensus}, \& {van
  Zwieten}}]{LOFAR}
{van Haarlem}, M.~P., {Wise}, M.~W., {Gunst}, A.~W., {et~al.} 2013, \aap, 556,
  A2, \dodoi{10.1051/0004-6361/201220873}

\bibitem[{{Virtanen} {et~al.}(2020){Virtanen}, {Gommers}, {Oliphant},
  {Haberland}, {Reddy}, {Cournapeau}, {Burovski}, {Peterson}, {Weckesser},
  {Bright}, {van der Walt}, {Brett}, {Wilson}, {Millman}, {Mayorov}, {Nelson},
  {Jones}, {Kern}, {Larson}, {Carey}, {Polat}, {Feng}, {Moore}, {VanderPlas},
  {Laxalde}, {Perktold}, {Cimrman}, {Henriksen}, {Quintero}, {Harris},
  {Archibald}, {Ribeiro}, {Pedregosa}, {van Mulbregt}, \& {SciPy 1. 0
  Contributors}}]{scipy}
{Virtanen}, P., {Gommers}, R., {Oliphant}, T.~E., {et~al.} 2020, Nature
  Methods, 17, 261, \dodoi{10.1038/s41592-019-0686-2}

\bibitem[{{Wright}(2006)}]{cosmocalc}
{Wright}, E.~L. 2006, \pasp, 118, 1711, \dodoi{10.1086/510102}

\bibitem[{{Zakamska} \& {Greene}(2014)}]{ZG14}
{Zakamska}, N.~L., \& {Greene}, J.~E. 2014, \mnras, 442, 784,
  \dodoi{10.1093/mnras/stu842}

\bibitem[{{Zakamska} {et~al.}(2016){Zakamska}, {Lampayan}, {Petric}, {Dicken},
  {Greene}, {Heckman}, {Hickox}, {Ho}, {Krolik}, {Nesvadba}, {Strauss},
  {Geach}, {Oguri}, \& {Strateva}}]{Zakamska+16}
{Zakamska}, N.~L., {Lampayan}, K., {Petric}, A., {et~al.} 2016, \mnras, 455,
  4191, \dodoi{10.1093/mnras/stv2571}

\end{thebibliography}
\bibliographystyle{aasjournal}

\appendix
\section{Images and notes for individual sources}
\label{sec:app_images}
\begin{figure}[p]
    \vspace*{1cm}
    \makebox[\linewidth]{
        \includegraphics[width=1.0\linewidth]{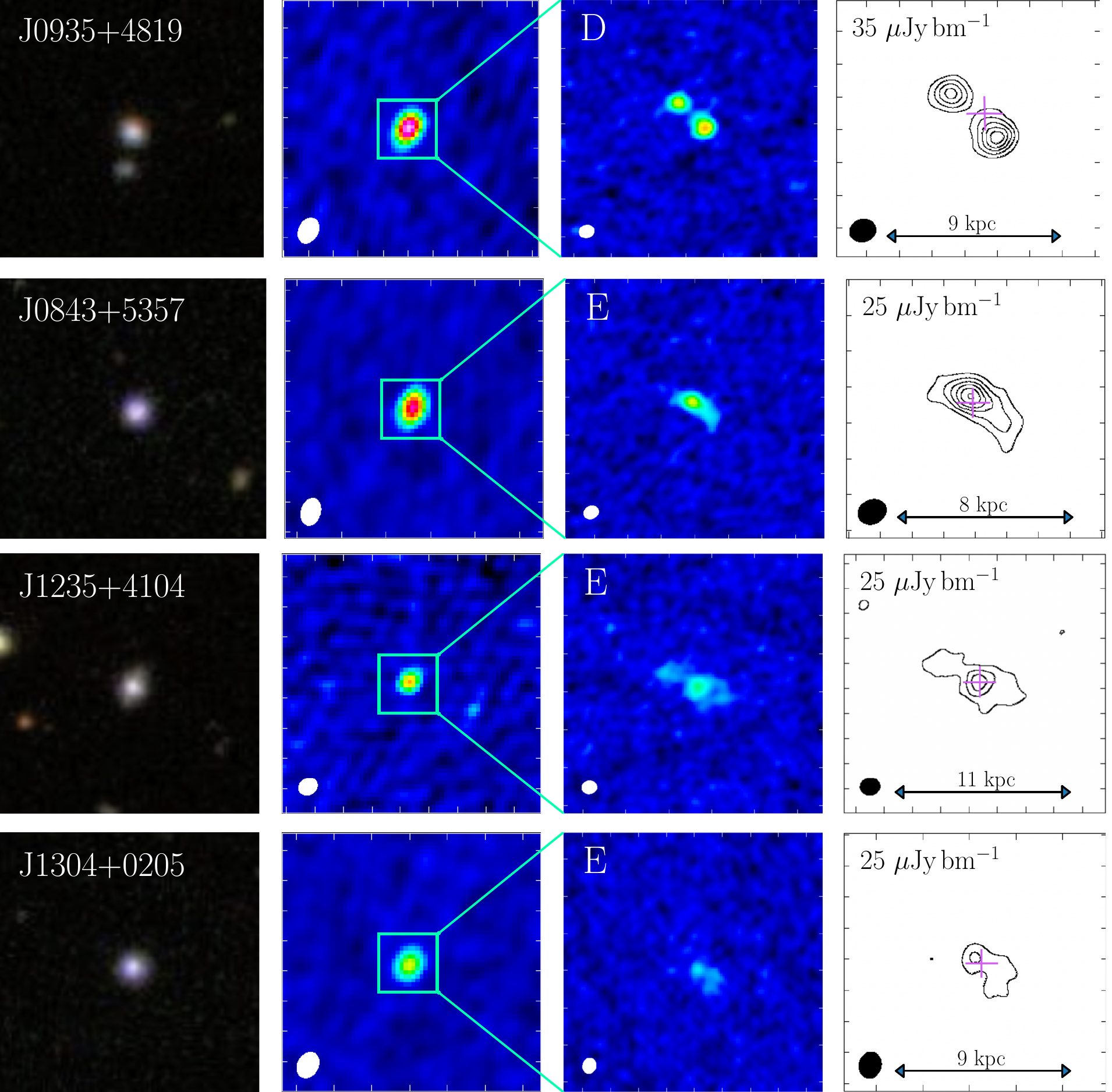}
    }
    \caption{Images of resolved sources with complex radio morphologies in our sample.  The panel layout follows that of Figure~\ref{fig:fig3_image_inset_example}.  See \S\ref{section:morphologies} for a discussion on morphology classification.  Individual notes on each of these sources are available in Appendix~\ref{notes}.}
    \label{fig:images}
\end{figure}

In Figure~\ref{fig:images}, we present images of the Resolved (Extended, Multi-component, Slightly-resolved) sources.  Base-level contours correspond to signal-to-noise detections of 3.5--4 (see caption of  Figure~\ref{fig:fig3_image_inset_example}). Per the analysis in Section~\ref{subsec:imageconstraints}, we estimate that the expectation value of the number of unassociated faint background sources in the Multi-Component class is $\sim0.125$.

\renewcommand{\thefigure}{\arabic{figure} ({\em Continued})}
\addtocounter{figure}{-1}

\begin{figure}[p]
    \vspace*{2cm}
    \makebox[\linewidth]{
        \includegraphics[width=1.0\linewidth]{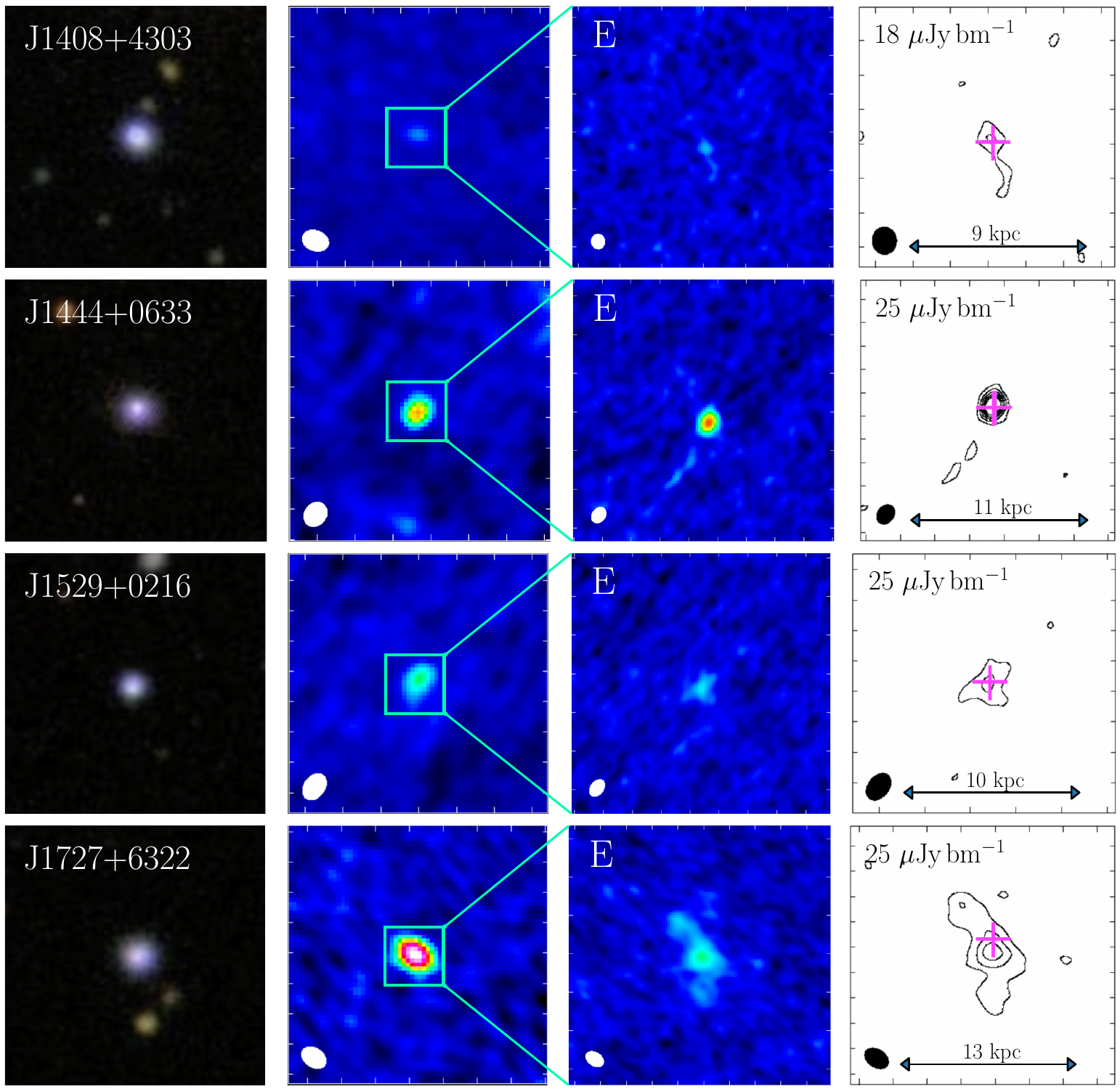}
    }
    \caption{}
    \label{fig:image_inset_grid_p2}
\end{figure}

\renewcommand{\thefigure}{\arabic{figure}}

\renewcommand{\thefigure}{\arabic{figure} ({\em Continued})}
\addtocounter{figure}{-1}

\begin{figure}[p]
    \vspace*{2cm}
    \makebox[\linewidth]{
        \includegraphics[width=1.0\linewidth]{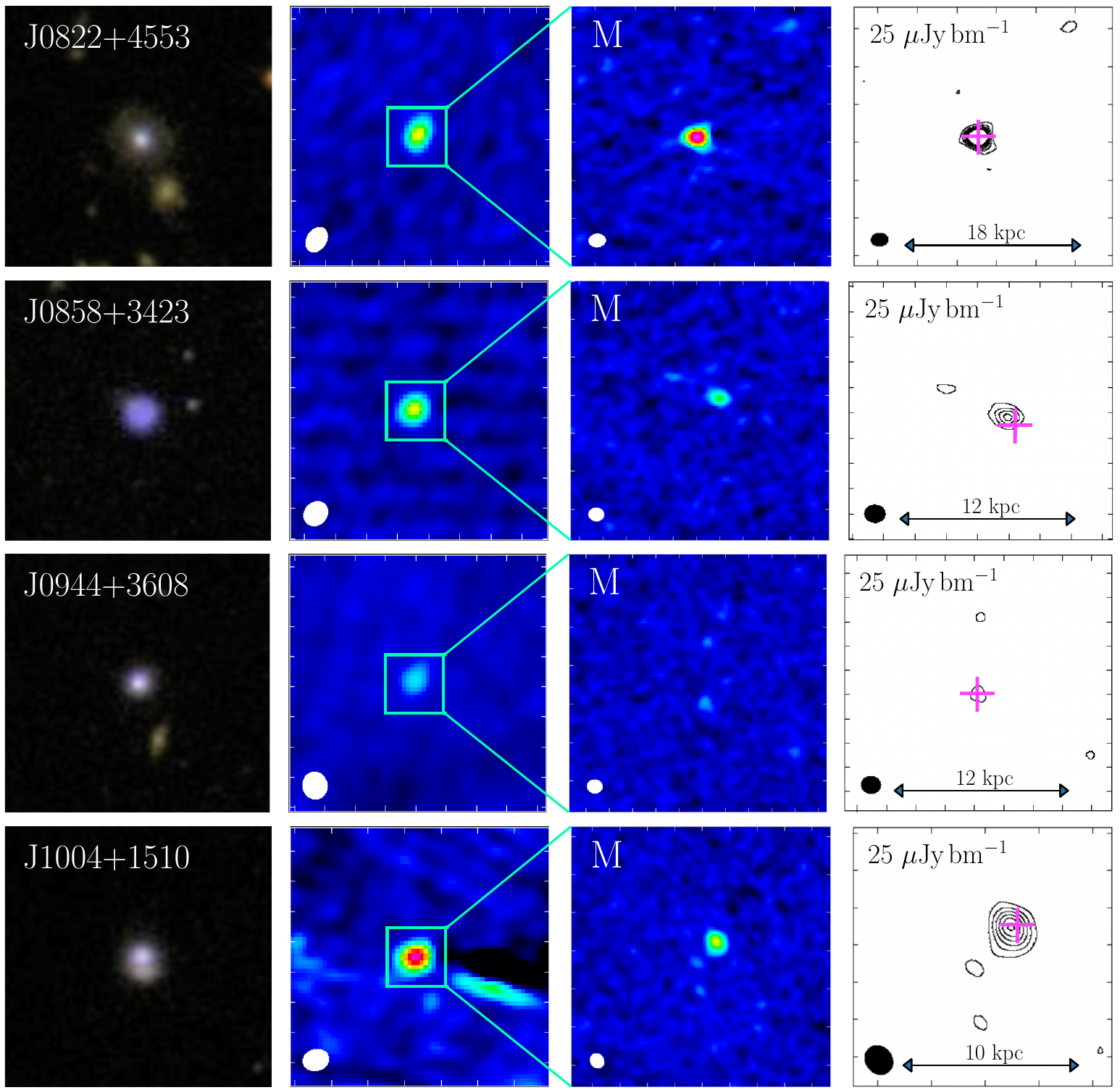}
    }
    \caption{}
    \label{fig:image_inset_grid_p3}
\end{figure}

\renewcommand{\thefigure}{\arabic{figure}}

\renewcommand{\thefigure}{\arabic{figure} ({\em Continued})}
\addtocounter{figure}{-1}

\begin{figure}[p]
    \vspace*{2cm}
    \makebox[\linewidth]{
        \includegraphics[width=1.0\linewidth]{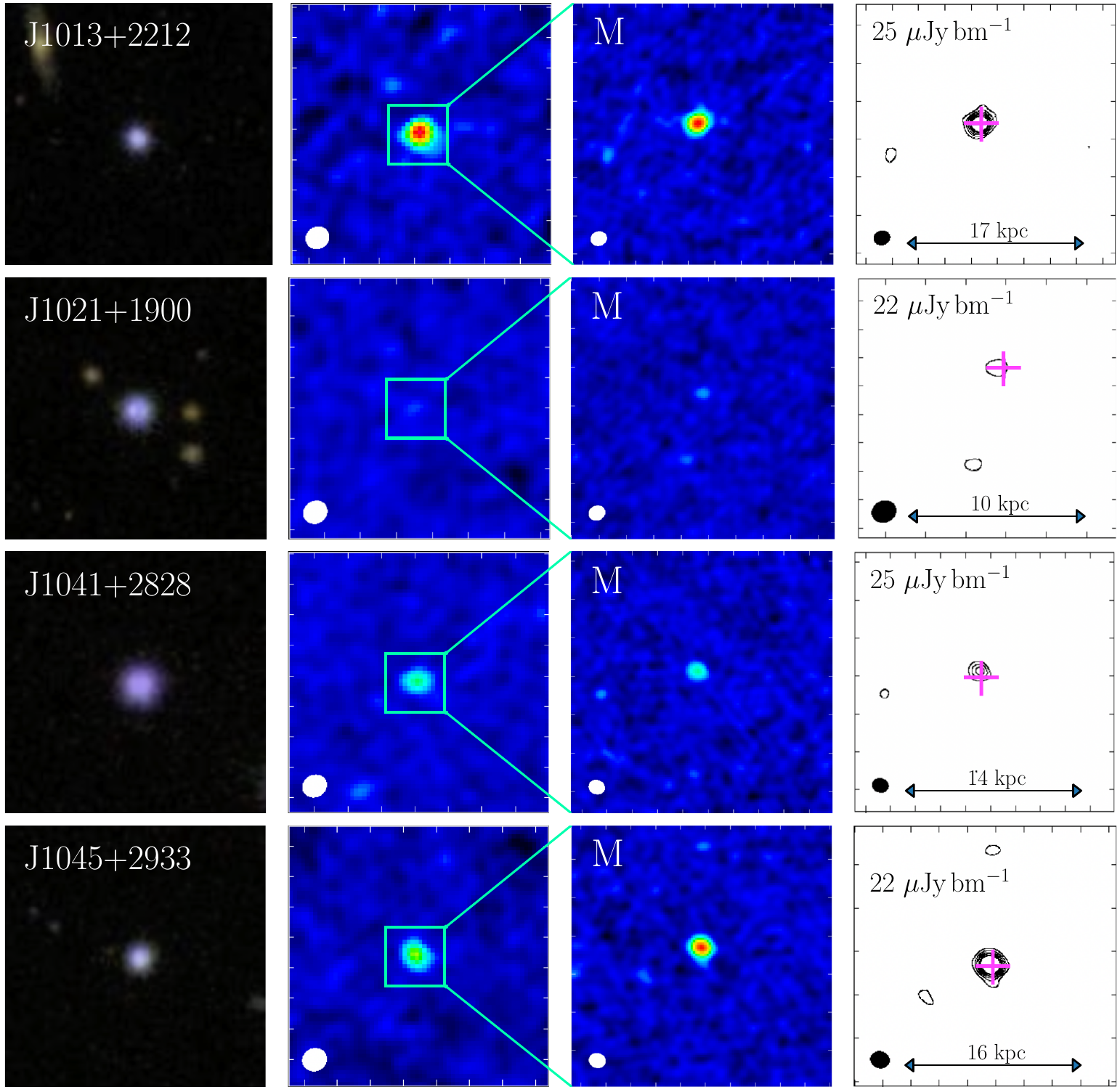}
    }
    \caption{}
    \label{fig:image_inset_grid_p4}
\end{figure}

\renewcommand{\thefigure}{\arabic{figure}}

\renewcommand{\thefigure}{\arabic{figure} ({\em Continued})}
\addtocounter{figure}{-1}

\begin{figure}[p]
    \vspace*{2cm}
    \makebox[\linewidth]{
        \includegraphics[width=1.0\linewidth]{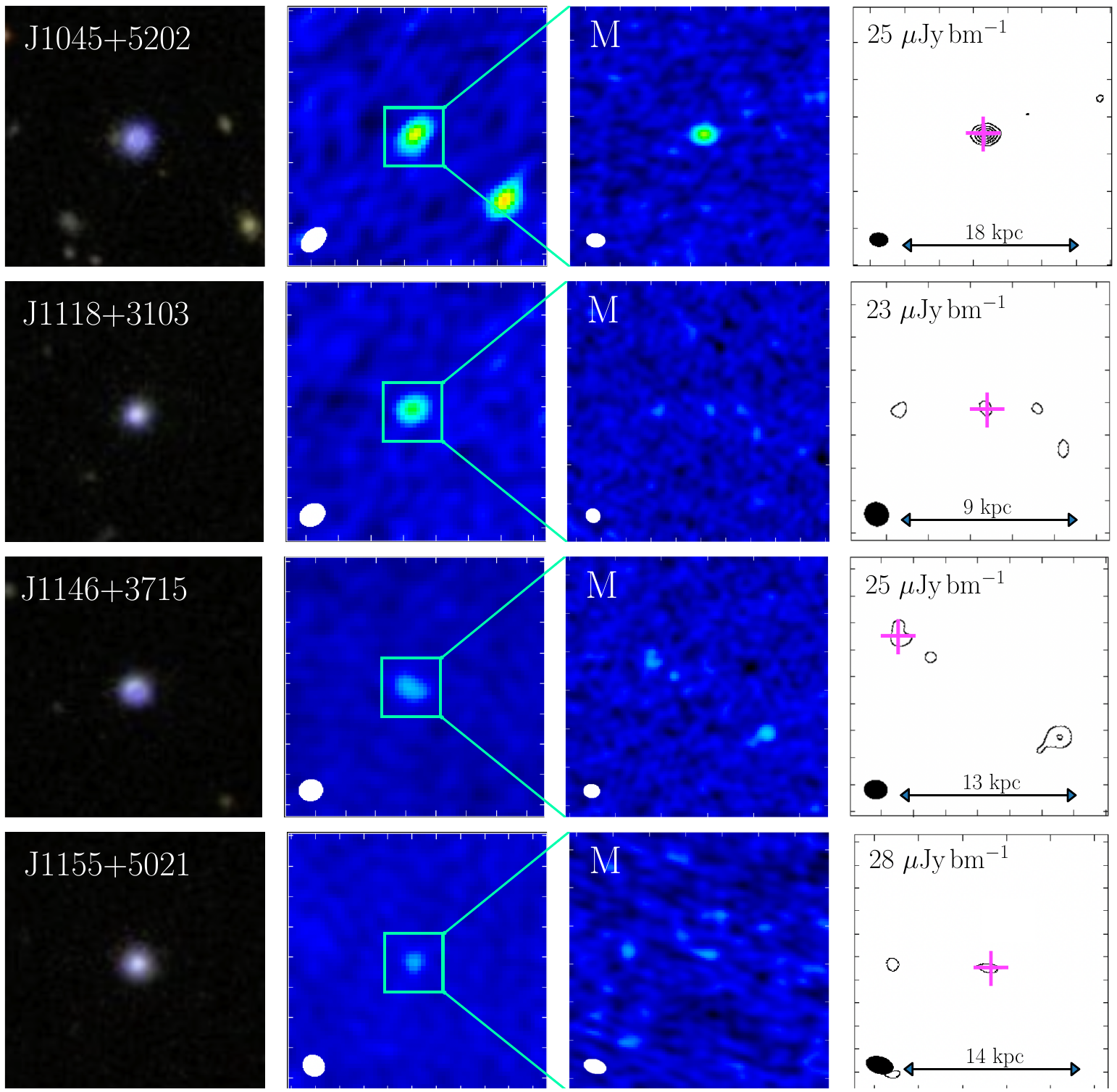}
    }
    \caption{}
    \label{fig:image_inset_grid_p5}
\end{figure}
\renewcommand{\thefigure}{\arabic{figure} ({\em Continued})}
\addtocounter{figure}{-1}

\begin{figure}[p]
    \vspace*{2cm}
    \makebox[\linewidth]{
        \includegraphics[width=1.0\linewidth]{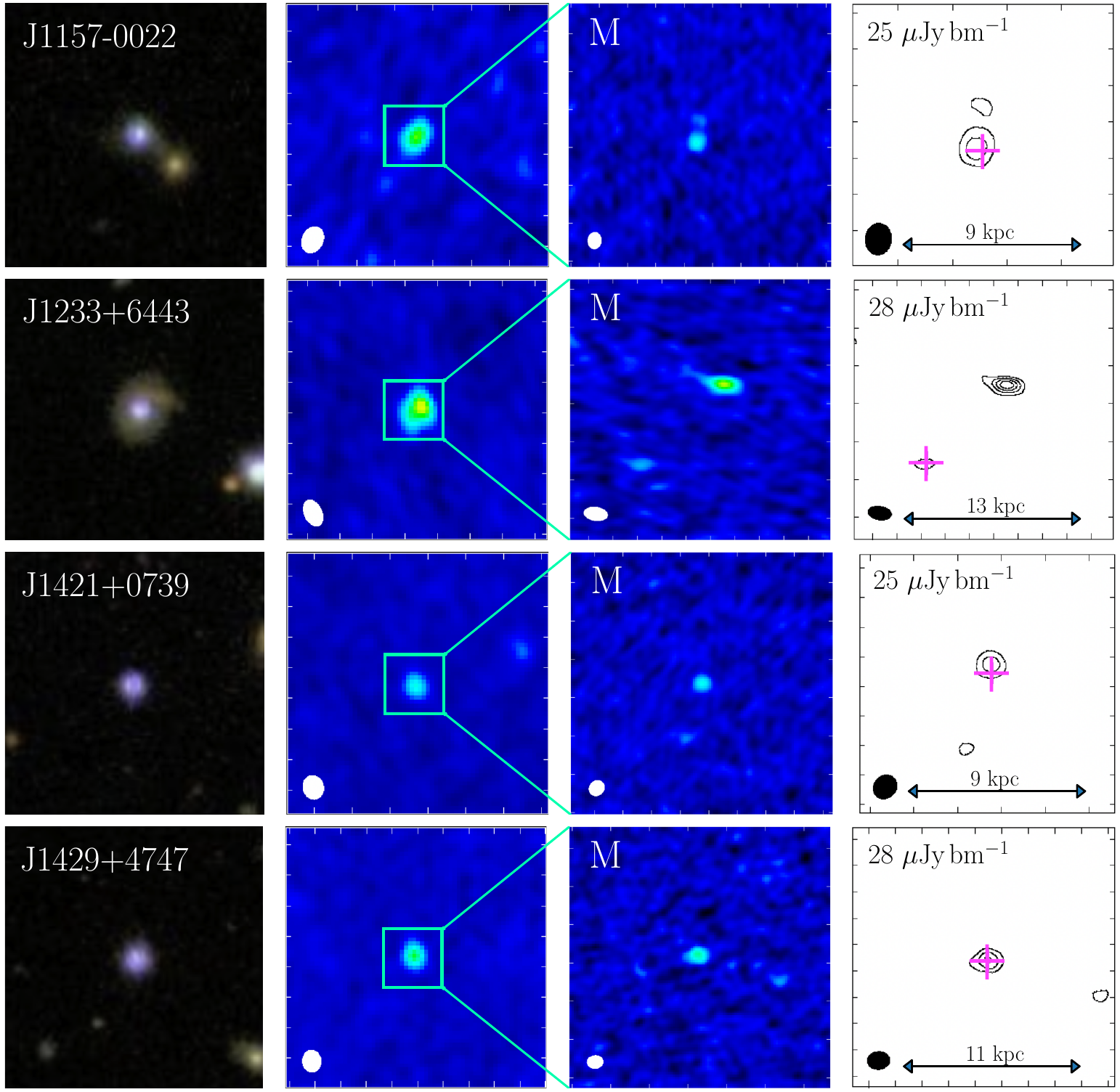}
    }
    \caption{}
    \label{fig:image_inset_grid_p5}
\end{figure}

\renewcommand{\thefigure}{\arabic{figure}}
\begin{figure}[p]
    \vspace*{2cm}
    \makebox[\linewidth]{
        \includegraphics[width=1.0\linewidth]{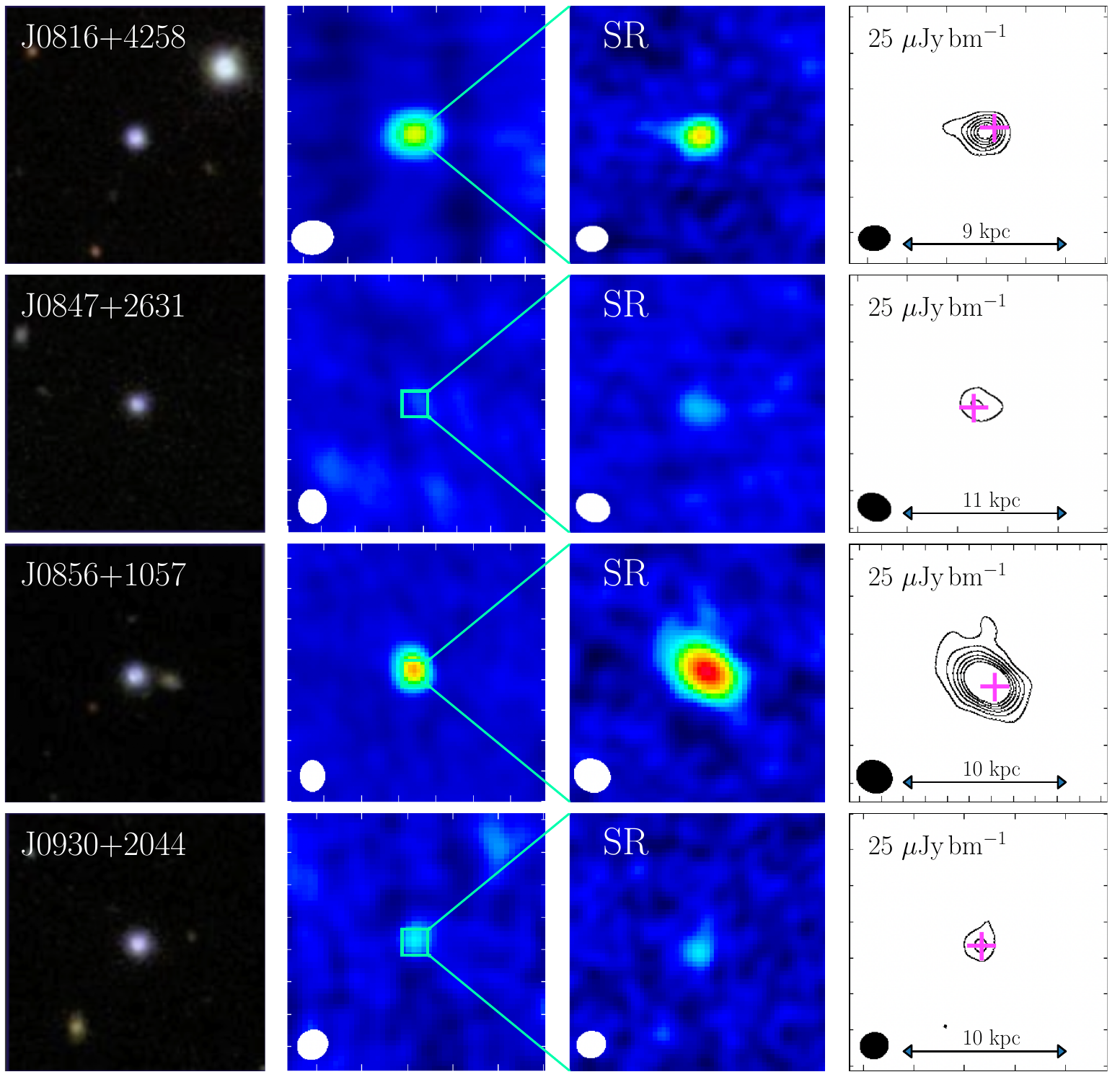}
    }
    \caption{Sources classified as ``Slightly Resolved".  The panel layouts follow that of Figure~\ref{fig:fig3_image_inset_example}.  See \S\ref{section:morphologies} for a discussion on morphology classification.}
    \label{fig:SRsources1}
\end{figure}

\renewcommand{\thefigure}{\arabic{figure} ({\em Continued})}
\addtocounter{figure}{-1}

\begin{figure}[p]
    \vspace*{2cm}
    \makebox[\linewidth]{
        \includegraphics[width=1.0\linewidth]{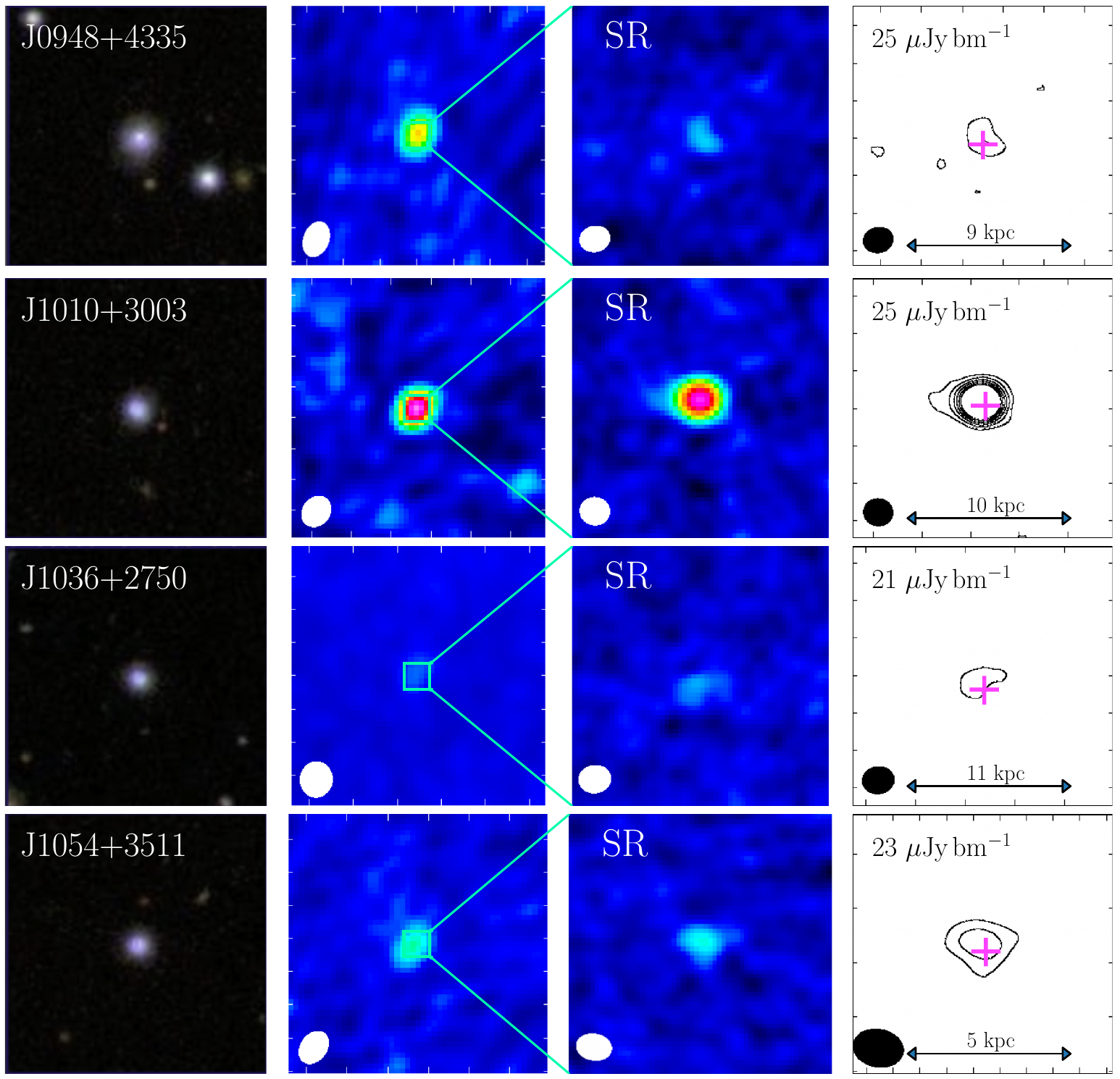}
    }
    \caption{}
    \label{fig:SRsources2}
\end{figure}

\renewcommand{\thefigure}{\arabic{figure} ({\em Continued})}
\addtocounter{figure}{-1}

\begin{figure}[p]
    \vspace*{2cm}
    \makebox[\linewidth]{
        \includegraphics[width=1.0\linewidth]{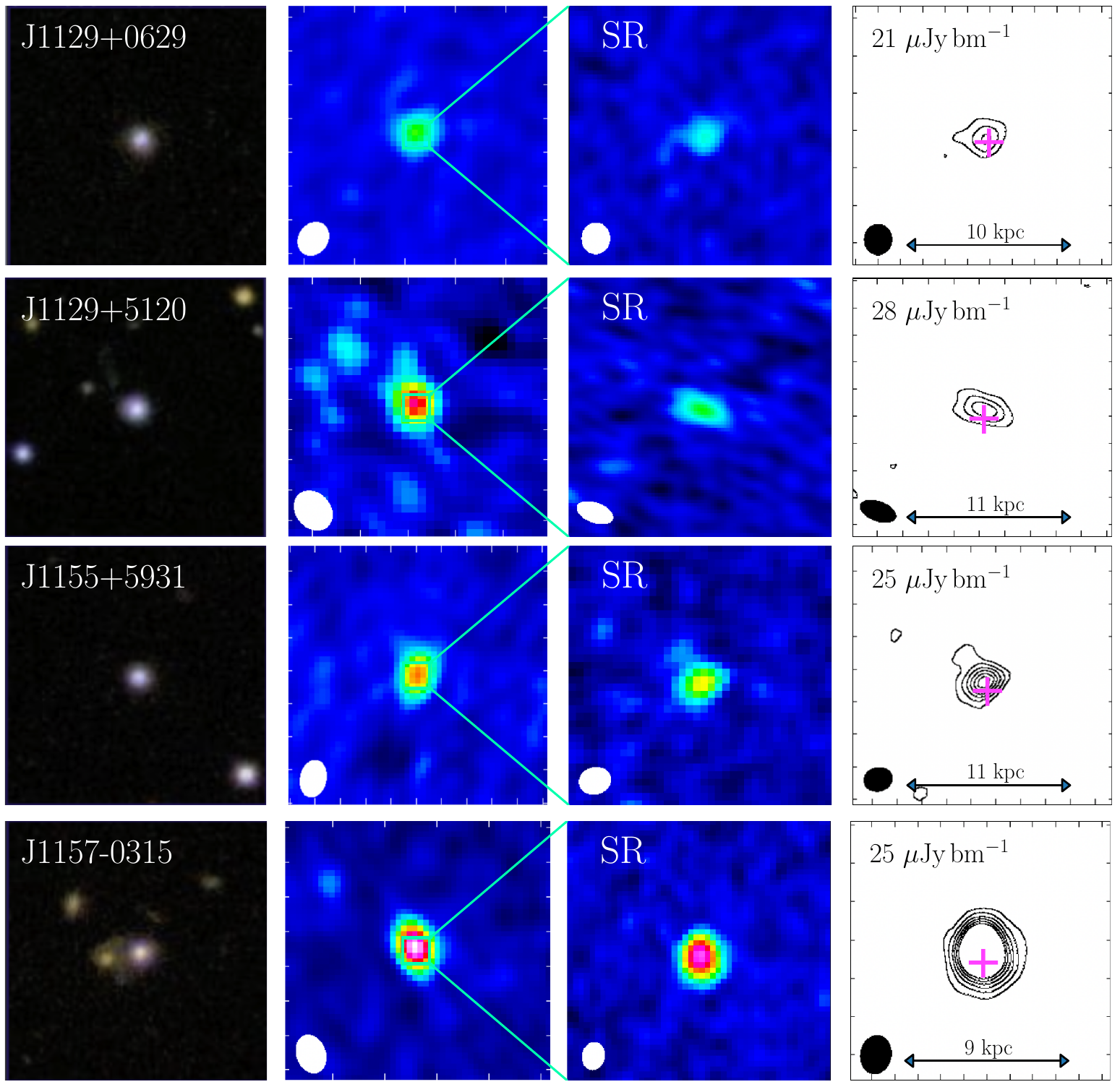}
    }
    \caption{}
    \label{fig:SRsources3}
\end{figure}

\renewcommand{\thefigure}{\arabic{figure} ({\em Continued})}
\addtocounter{figure}{-1}

\begin{figure}[p]
    \vspace*{2cm}
    \makebox[\linewidth]{
        \includegraphics[width=1.0\linewidth]{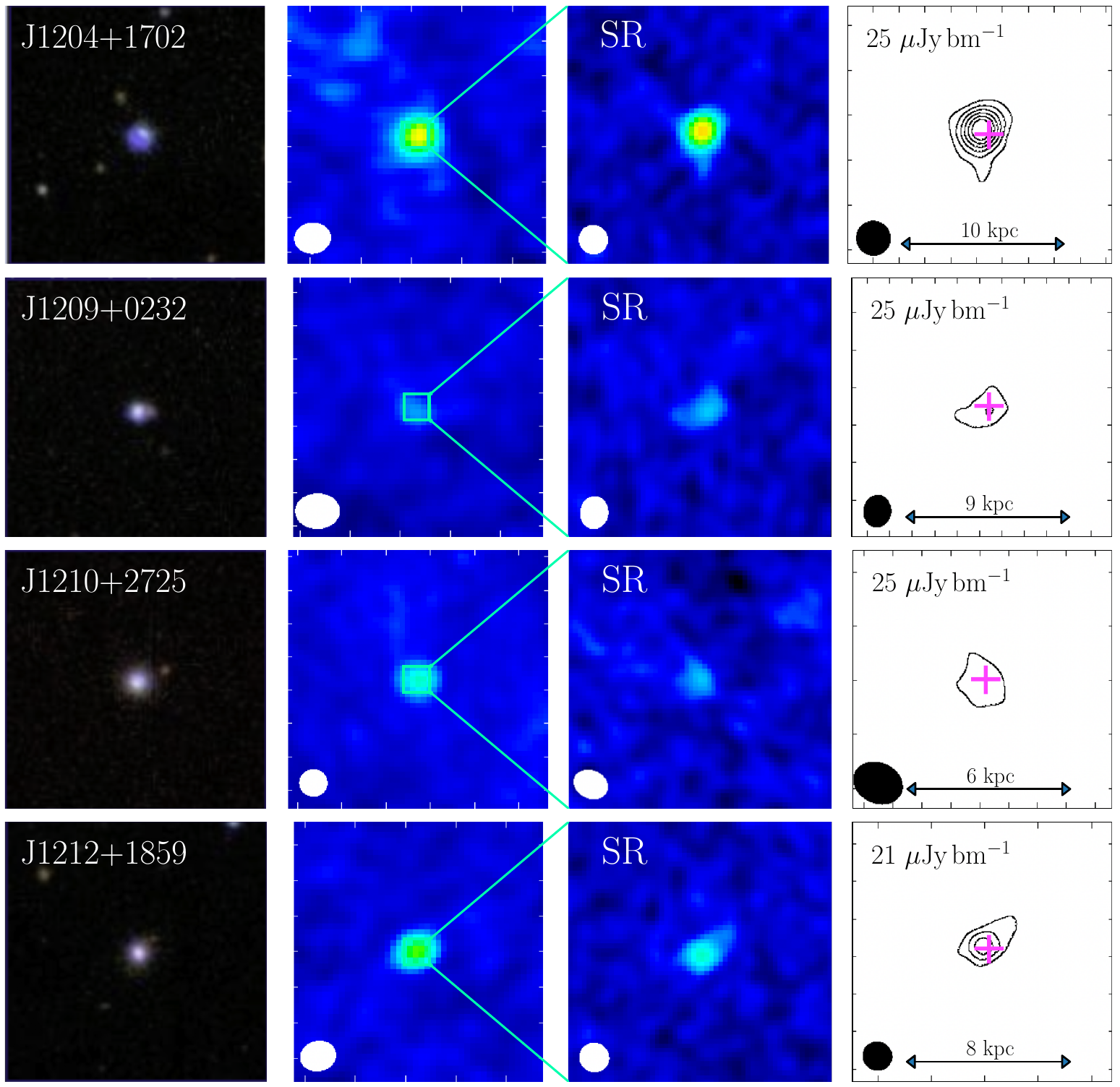}
    }
    \caption{}
    \label{fig:SRsources4}
\end{figure}

\renewcommand{\thefigure}{\arabic{figure} ({\em Continued})}
\addtocounter{figure}{-1}

\begin{figure}[p]
    \vspace*{2cm}
    \makebox[\linewidth]{
        \includegraphics[width=1.0\linewidth]{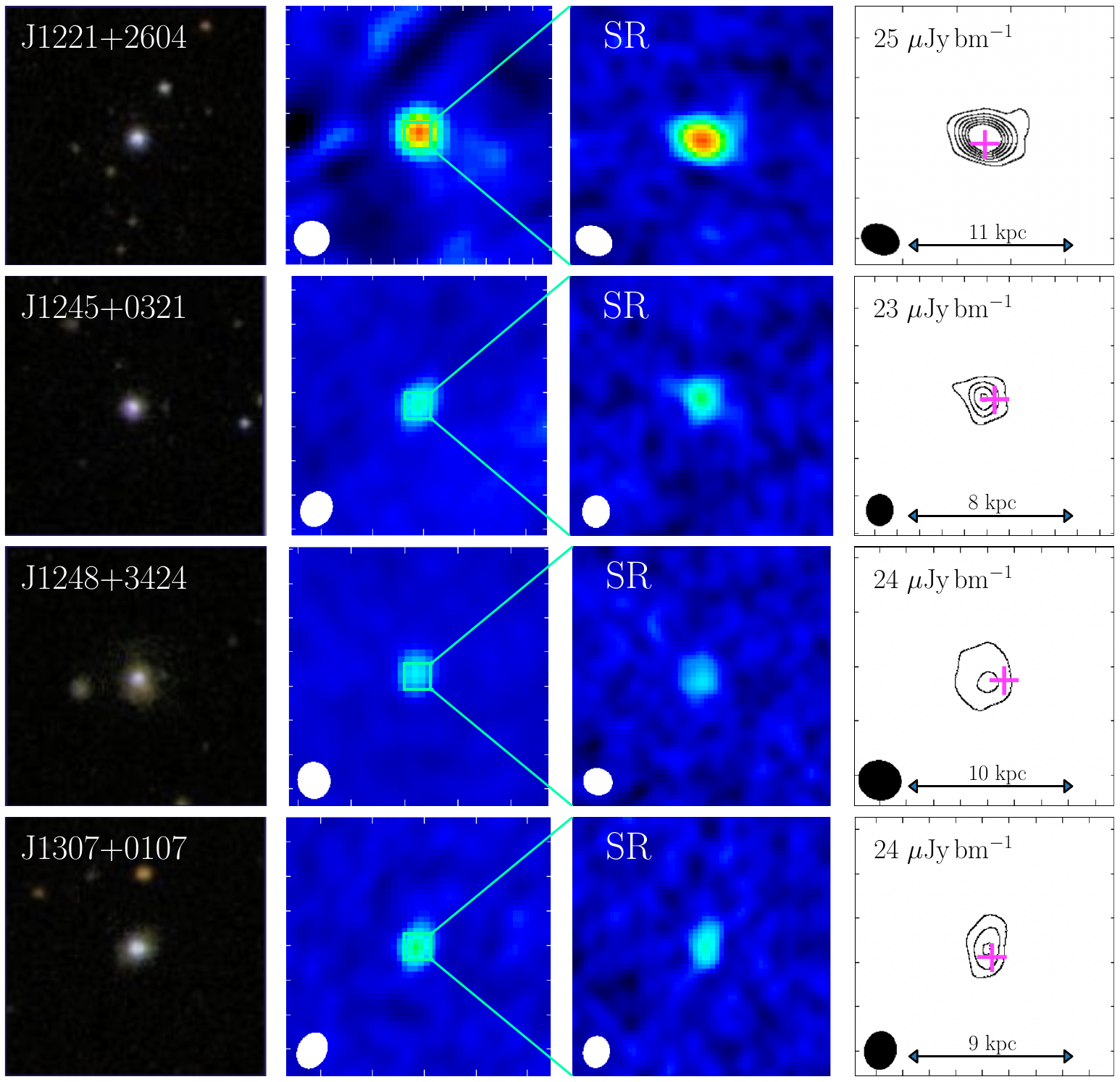}
    }
    \caption{}
    \label{fig:SRsources5}
\end{figure}

\renewcommand{\thefigure}{\arabic{figure} ({\em Continued})}
\addtocounter{figure}{-1}

\begin{figure}[p]
    \vspace*{2cm}
    \makebox[\linewidth]{
        \includegraphics[width=1.0\linewidth]{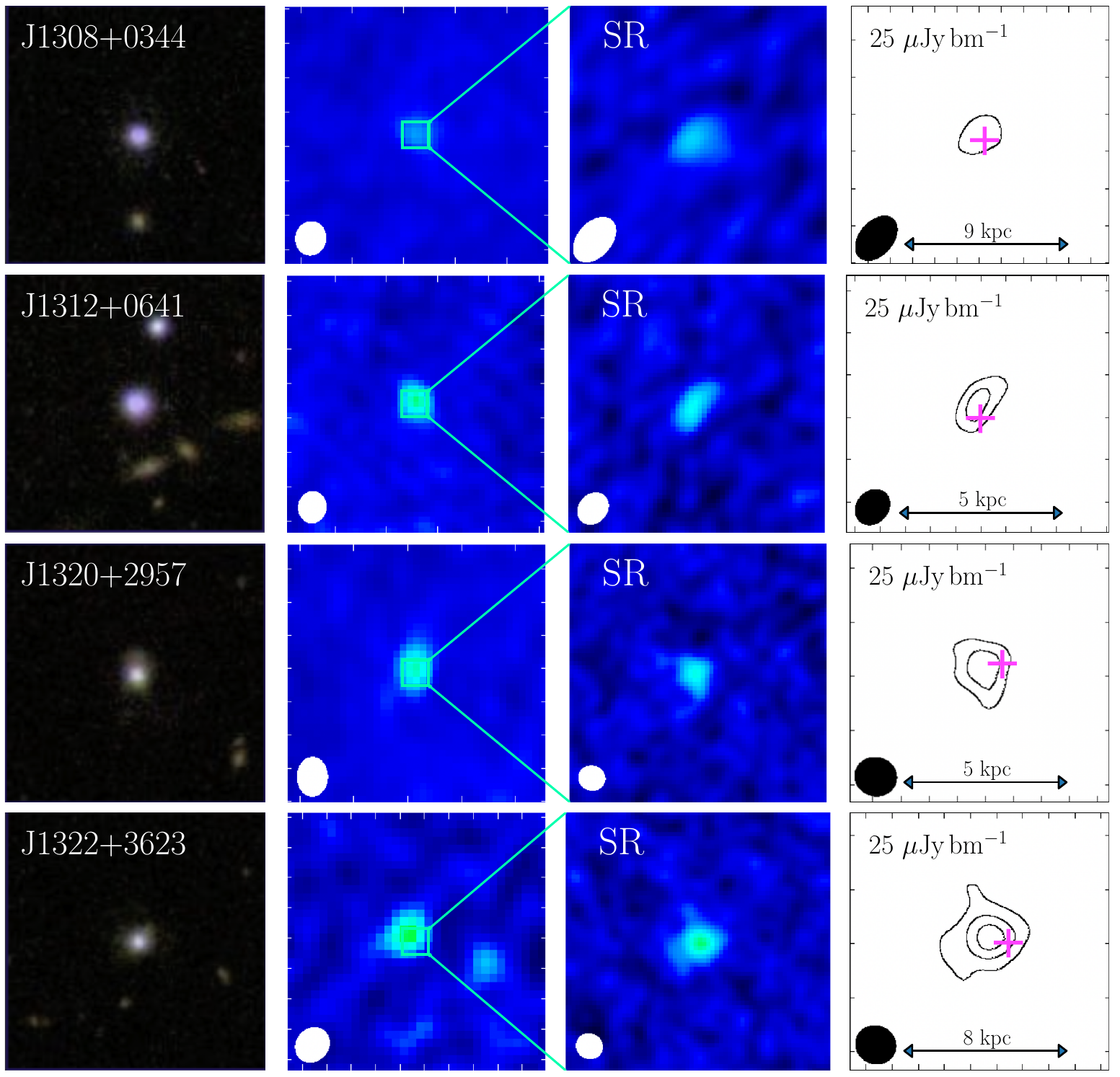}
    }
    \caption{}
    \label{fig:SRsources6}
\end{figure}

\renewcommand{\thefigure}{\arabic{figure} ({\em Continued})}
\addtocounter{figure}{-1}

\begin{figure}[p]
    \vspace*{2cm}
    \makebox[\linewidth]{
        \includegraphics[width=1.0\linewidth]{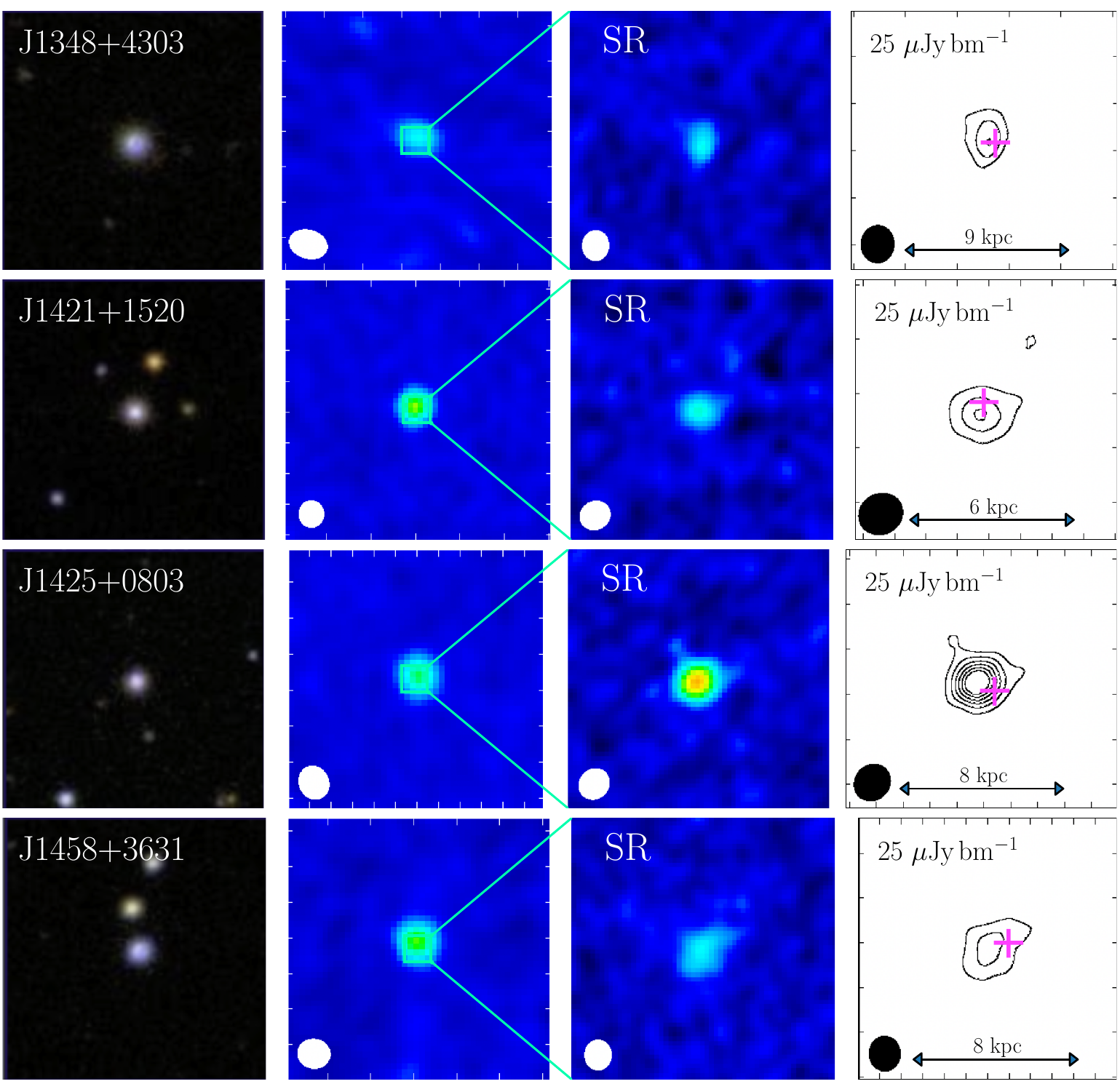}
    }
    \caption{}
    \label{fig:SRsources7}
\end{figure}

\renewcommand{\thefigure}{\arabic{figure} ({\em Continued})}
\addtocounter{figure}{-1}

\begin{figure}[p]
    \vspace*{2cm}
    \makebox[\linewidth]{
        \includegraphics[width=1.0\linewidth]{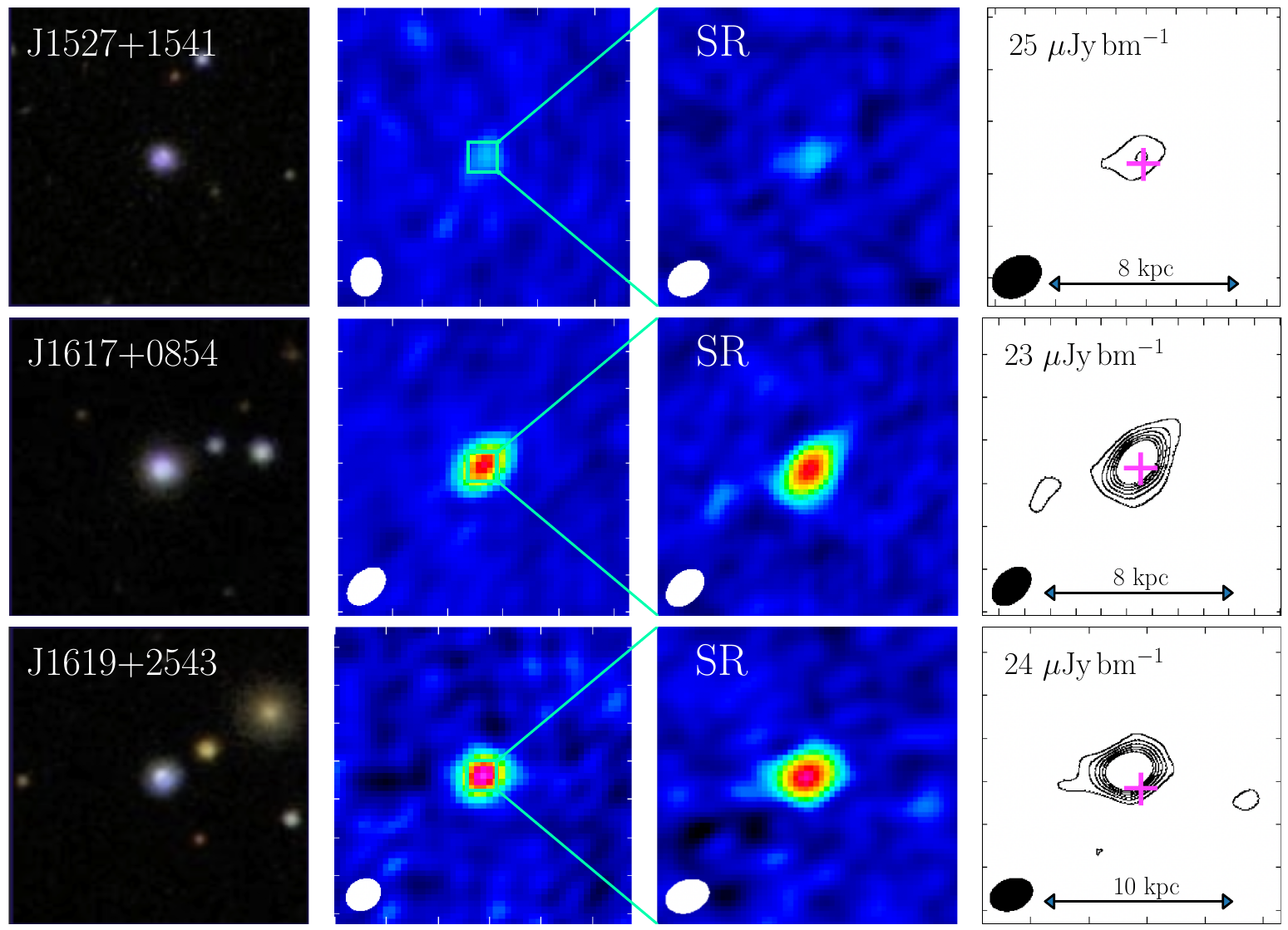}
    }
    \caption{}
    \label{fig:SRsources8}
\end{figure}

\renewcommand{\thefigure}{\arabic{figure}}

\label{section:notes}
\label{notes}
We separate notes into two main subsections to highlight sources with especially interesting radio morphologies and sources with large increases/decreases in flux density from 2010--2011 observations separately; a third subsection discusses images with remaining image artifacts or other issues.

\subsection{Sources with Interesting Morphological Features}

\subsubsection{J0843+5357}
A clear asymmetric, broad extended structure stretches $\sim4$\,kpc southeast and traverses roughly half the distance in the opposite direction.  Both extensions appear to abruptly terminate and curve inward southwest.


\subsubsection{J0935+4819}
A well-resolved double-lobed source (see top panel of the first page of Figure~\ref{fig:images}) that must be powered predominantly by AGN activity.  The southwest lobe has an extension toward the northeast that coincides with the SDSS optical position, suggesting radio emission from a core that is blended with the southwest lobe.  The linear extent of the two lobes is 1\farcs5, corresponding to a physical extent of $\gtrsim6$~kpc with unknown line-of-sight orientation.  With radio spectral index of $\alpha\sim-0.7$ from 1.4 to 6~GHz, this appears to be the first example of a Compact Steep Spectrum source in a radio-\emph{quiet} QSO.

\subsubsection{J0944+3608}
A $\sim50\,\mu$Jy core slightly resolved in two directions that also has two $\sim30\,\mu$Jy compact components oriented in the directions that the core exhibits resolution $\sim8\,\mathrm{kpc}$ from the optical position of the QSO.


\subsubsection{J1004+1510}
Note that the structure that we see in the C-configuration image appears to be a calibration error, with a corresponding negative feature just above it; the only real structure is the unresolved core.  Source has an interesting three-component morphology that stands alone from the rest of the sample (see the top panel of the third part of Figure~\ref{fig:images}).  There is a fairly strong $\sim150\,\mu$Jy compact source coinciding with the SDSS optical location, while two other $\sim40\,\mu$Jy compact radio sources produce the added-component radio emission.  The two sources offset $\sim1''$ from the core are both on the southwest side of the bright compact emission.  Source's total flux density has decreased $\sim$50\% from previous $3\farcs5$ observation.

\subsubsection{J1045+2933}
A slightly resolved $\sim280\,\mu$Jy component lies at the exact optical location of the QSO while there is another fainter $\sim40\,\mu$Jy compact component $\sim10$\,kpc north of the core component---the opposite direction from which the central source exhibits its resolution.  


\subsubsection{J1102+0844}
There is no clear point of emission overlapping with the optical location of the QSO, which we typically see in most of these detected targets---however, there is a faint 22\,$\mu$Jy source offset $\approx0\farcs81$ from the SDSS optical location, which we take to be associated with our main target.  

\subsubsection{J1118+3103}
The $\sim30\,\mu$Jy core may be symmetrically emitting a pair of equally bright ($\sim20\,\mu$Jy) lobes $\sim6$\,kpc from the detected core; there may be another slightly elongated component curved away from the rightward offset component.  

\subsubsection{J1125+2513}
The previous $\sim3\farcs5$ image of this source hinted at this RQQ being slightly resolved in the southwest direction and we have confirmed that suspicion by resolving it into two clear compact components:  a core $\sim30\,\mu$Jy source overlapping with the optical position of the QSO, and another source of comparable flux density $\sim16$\,kpc southwest of the core.  Inspection of the SDSS image around the area of this target reveals no clear presence of background sources, so the extended nature of this source is most certainly real and associated with the QSO.


\subsubsection{J1146+3715}
A faint, odd V-shaped radio source is at the exact location of the SDSS QSO and two other compact components are present along a line southeast of the QSO component (see the bottom panel of part 4 of Figure~\ref{fig:images} for the image).  One of the offset sources is of comparable flux density ($\sim35\,\mu\mathrm{Jy}$) to the V-shaped source and  just $\sim3$\,kpc away, while the more distant ($\sim15\,\mathrm{kpc}$) component is much brighter ($\sim70\,\mu\mathrm{Jy}$) and appears to be slightly resolved in the opposite direction from which the V-shaped source is oriented. 


\subsubsection{J1155+5021}
There is a faint $\sim35\,\mu\mathrm{Jy}$ source coincident with the QSO's optical position that exhibits some resolution beyond the synthesized beam.  The elongated component extends $\gtrsim2\,\mathrm{kpc}$ west, and another compact radio source of comparable flux density resides $\sim15\,\mathrm{kpc}$ from the component coinciding with the QSO in the same direction.

\subsubsection{1210+0154}
A two-component source where the main component coincides with the QSO's SDSS position as well as another second compact source extending $\sim19\,\mathrm{kpc}$ from the first.  This is the farthest distance from the QSO that we detect radio emission in any of the RQQs studied herein.


\subsubsection{J1233+6443}
Source potentially appeared to be slightly resolved in C-configuration observations, but has now been clearly resolved into two components:  one faint $\sim40\,\mu\mathrm{Jy}$ compact radio source at the exact optical location of the QSO, and one much brighter $\sim160\,\mu\mathrm{Jy}$ component offset $\sim12\,\mathrm{kpc}$ northeast from the fainter component.  Further assessment of the SDSS optical image at the location of this source is ambiguous as to discerning whether or not the brighter source could potentially be another separate galaxy interacting/interfering with our target, or if it is the edge-brightened extended QSO that it appears to be (see the second panel of the final part of Figure~\ref{fig:images} for the produced VLA image of this source).

\subsubsection{J1235+4104}

A source that is clearly extended at our $0\farcs33$ resolution.  A well-defined $\sim135\,\mu\mathrm{Jy}$ core coincides with the optical position of the QSO while the asymmetric extended emission contributes $\sim220\,\mu\mathrm{Jy}$ to the source's total radio luminosity (see the third panel of the first part of Figure~\ref{fig:images} for image).  Interestingly, the source (with total 6 GHz flux density $S_{\rm 6GHz}=406\pm36\,\mu\mathrm{Jy}$) does not appear even as a faint marginal detection in FIRST.  Even if the core were to completely ``shut off", and the extended emission was all that was contributing at the time of the FIRST observation, assuming a steep spectral index $\alpha=-0.7$ \citep[typical for synchrotron emisson;][]{condon1992}, this implies a 1.4 GHz spectral flux density of $S_{\rm 1.4GHz}=609\pm75\,\mu\mathrm{Jy}$---roughly 35\% above the typical 450-$\mu\mathrm{Jy}$ 3-$\sigma_{\rm rms}$ FIRST detection threshold.  This result indicates that the high resolution image of J1235+4104 either reveals the presence of nascent jets associated with a variable core, or the extended emission has a flat spectrum.  The 6-GHz observation of this source by \citet{kimball2011} ($S_{\rm 6GHz}=266\pm16.1\,\mu\mathrm{Jy}$) perhaps supports the scenario of the extended emission being due to some sort of enrichment of AGN activity---whether it be the emergence of young jets \citep{Nyland+2020} or an enhancement of accretion activity \citep{KB20}.

\subsubsection{J1304+0205}
Source exhibits clear extended emission---apparently in only one direction (see the lower panel of the first part of Figure~\ref{fig:images}).  The width of the extended component appears to increase with distance from the core.


\subsubsection{J1408+4303}
Source has a faint $\sim30\,\mu$Jy core component as well as some apparent narrow extended emission oriented in only one direction, bringing its total flux density up to $52.9\pm8.6\,\mu$Jy, agreeable with its C-configuration flux density of 52.3\,$\mu$Jy.

\subsubsection{J1444+0633}
The relatively strong ($255.8\pm7.2\,\mu\mathrm{Jy}$) core may have an associated narrow strip of emission detected in A-configuration contributing $\sim100\,\mu\mathrm{Jy}$ to the source's total flux density.

\subsubsection{J1458+4555}
This is the only RQQ from \citet{kellerman16} that was observed to appear resolved on $\sim$14\,kpc scales; our A-configuration observation clearly resolves the extended radio structure into two compact points of emission of roughly equal flux density ($\sim35\,\mu$Jy).  Investigation of the SDSS optical image reveals an overlapping source with the potential added radio component, so this is probably not a core-lobe structure---rather, it reinforces the sentiment of all RQQs in this sample being unresolved on host galaxy scales as even the one RQQ that may have had extended emission was really only producing compact radio emission.


\subsubsection{J1529+0216}
There is clear emission extending beyond the $\sim0\farcs33$ synthesized beam in what appears to be in four different directions, forming something of an X-shaped radio morphology (see the second panel of part two of Figure~\ref{fig:images}).  


\subsubsection{J1727+6322}
Source has strong, diffuse extended emission emerging from opposite sides of the core, reaching distances of $\sim3.5\,\mathrm{kpc}$ in each of the north and south directions (see the third panel of the second part of Figure~\ref{fig:images}).

\subsection{Notably Variable Sources}

\subsubsection{J0822+4553}
Source has a strong $\sim430\,\mu$Jy radio core coinciding with its optical position and another $\sim100\,\mu$Jy slightly elongated component $\sim16\,\mathrm{kpc}$ from the core.  This source's flux density was only $222\pm11.2\,\mu$Jy in \citet{kimball2011}, indicating significant variability ($+140\%$) over decadal timescales.  Each of these characteristics are strong indicators of an AGN being a dominant contributor to the radio in this QSO.

\subsubsection{J1000+1047}
Source remains compact on A-configuration scales and shows strong signs of variability; flux density has increased from $\sim223\,\mu$Jy at $\sim3\farcs5$ to $\sim558\,\mu$Jy at $\sim0\farcs33$.

\subsubsection{J1129+5120}
Our observation reveals a lone slightly resolved source at the optical location of this QSO that has decreased $\sim67\%$ from  $501\pm44.6\,\mu\mathrm{Jy}$ at $3\farcs5$ resolution to $161\pm21\,\mu\mathrm{Jy}$ at $\sim0\farcs33$.

\subsubsection{J1617+0638}
Source remains unresolved on A-configuration scales, but shows strong variability in the positive direction---nearing fractional flux density changes of 80\%, increasing from $\sim265\,\mu\mathrm{Jy}$ at $3\farcs5$ to $\sim465\,\mu\mathrm{Jy}$ at $0\farcs33$.

\subsubsection{J1627+4736}
Source stays unresolved at our higher resolution and doubles in flux density from $\sim300\,\mu\mathrm{Jy}$ at $3\farcs5$ to $\sim600\,\mu\mathrm{Jy}$ at $0\farcs33$.

\subsection{Sources with Image Defects}
\label{section:notes_bad}

\subsubsection{J0934+0306}
A bright $\sim80\,\mathrm{mJy}$ FIRST source at 093430.71+030545.43 $\sim1\farcm5$ from the target significantly propagates its individual errors through to our central target.  Although we detect faint $\sim50\,\mu\mathrm{Jy}$ radio emission at the QSO's optical position, such errors make it difficult to untangle any possible extended morphologies present, so we take it as being a simple point source.

\subsubsection{J1034+6053}
Image has both a complex jetted source at 103436.05+605252.60 offset $\sim 1'$ from our source, as well as a strong $\sim170\,\mathrm{mJy}$ source at 103351.43+605107.32 along the edge of the VLA's primary beam.  A confusing extended structure overlaps with our central target, making accurate determination of the source's morphology difficult.  However, this source was one of the few RQQs not detected by \citet{kimball2011} (3-$\sigma_{\rm rms}$ upper limit of $20.7\,\mu\mathrm{Jy}$), and the total flux of the core component alone sums to $27.6\pm7.3\,\mu\mathrm{Jy}$.  The total flux density of the core component plus the possible extended structure on the other hand sums to $\sim220\,\mu\mathrm{Jy}$.  While variability does play a role in shaping our results, extended jetted/lobed sources are not capable are being variable to this degree over these distances \citep{Blandford+2019}, so we take J1034+6053 to be limited to the $27.6\,\mu\mathrm{Jy}$ confined to its core component.

\subsubsection{J1443+4045}
Image is severely dynamic-range-limited, with a powerful 260mJy source at 144259.34+404429.12 offset $\sim 1'$ from the pointing center, degrading the image sensitivity by a factor of $\approx5$.

\section{Multi-frequency Data}
\label{sec:app_var}

Here we provide the multi-frequency data used throughout \S\ref{section:variability}; a full machine-readable version of this table is available in the online journal.  Data stored in Table~\ref{tab:vartable} are listed as follows:

\begin{itemize}
    \item \textbf{Column 1:} Full name of the QSO in SDSSJ format
    \item \textbf{Column 2:} Redshift from the SDSS-DR7 catalog \citep{schneider10}
    \item \textbf{Column 3:} Measured total 1.4-GHz flux density of each source as it appeared in FIRST \citep{first}.
    \item \textbf{Column 4:} Date of source's observation in FIRST in YYYY-MM-DD format.
    \item \textbf{Column 5:} Total 6-GHz flux density measured by \citet{kellerman16} in the VLA's C configuration ($\sim3\farcs5$ at a 6-GHz reference frequency).
    \item \textbf{Column 6:} Date of source's C-configuration VLA observation in YYYY-MM format \citep{kimball2011}. 
    \item \textbf{Column 7:} Measured total 3-GHz flux density in Epoch 1 of VLASS \citep{vlass}.  
    \item \textbf{Column 8:} Date of source's observation in VLASS in YYYY-MM-DD format; Epoch 1.1 observations span 2017 September--2018 February while Epoch 1.2 was undergone from March through July of 2019.  We discuss the reliability and resulting corrections of VLASS flux densities in~\S\ref{subsec:var_data}.
    \item \textbf{Column 9:}  Measured total 6-GHz flux density of source in the VLA's A configuration ($\sim0\farcs33$ at a 6-GHz reference frequency).
    \item \textbf{Column 10:} Date of source's A-configuration VLA observation in YYYY-MM format.
    \item \textbf{Column 11:} Source's radio-loud/-intermediate/-quiet classification based on 6-GHz luminosities measured by \citet{kellerman16}.  See Section~\ref{section:targets} for how ``radio-quietness" is defined.
    \item \textbf{Column 12:} Source's spectral index determination obtained through the time-domain analysis explained in \S~\ref{subsection:specindex_time}.
\end{itemize}

\begin{longrotatetable}
\begin{deluxetable*}{cccccccccccc}
\label{tab:vartable}
\tablewidth{700pt}
\tabletypesize{\scriptsize}
\tablehead{
\colhead{(1)} & \colhead{(2)} & \colhead{(3)} & \colhead{(4)} & \colhead{(5)} & \colhead{(6)} & \colhead{(7)} & \colhead{(8)} & 
\colhead{(9)} & \colhead{(10)} & \colhead{(11)} & \colhead{(12)} \\
\colhead{Name} & \colhead{$z$} & 
\colhead{$S_{\rm 1.4GHz, FIRST}$} & \colhead{DATE$_{\rm FIRST}$} & \colhead{$S_{\rm 6GHz, VLAC}$} & \colhead{DATE$_{\rm VLAC}$} & 
\colhead{$S_{\rm 3GHz, VLASS}$} & \colhead{DATE$_{\rm VLASS}$} &  \colhead{$S_{\rm 6GHz, VLAA}$} & \colhead{DATE$_{\rm VLAA}$} & 
\colhead{CLASS} & \colhead{$\alpha$}  \\ 
\colhead{(SDSSJ)} & \colhead{} & \colhead{(mJy)} & \colhead{(Y-M-D)} & \colhead{(mJy)} & \colhead{(Y-M)} & \colhead{(mJy)} & \colhead{(Y-M-D)} & 
\colhead{(mJy)} & \colhead{(Y-M)} & \colhead{} & \colhead{} 
}
\startdata
122011.88+020342.2 & 0.240 & $323.80\pm1.50$ & 1998-07-15 & $322.00\pm9.66$ & 2011-01-17 & $393.98\pm84.91$ & 2019-05-14 & \nodata & \nodata & RL & 0.0 \\
083353.88+422401.8 & 0.249 & $229.57\pm0.42$ & 1995-12-19 & $209.00\pm6.27$ & 2010-11-15 & $199.80\pm43.04$ & 2019-05-24 & \nodata & \nodata & RL & 0.0 \\
154743.53+205216.6 & 0.265 & $35.20\pm0.19$ & 1998-10-15 & $51.41\pm1.56$ & 2010-11-30 & $38.76\pm8.52$ & 2017-09-26 & \nodata & \nodata & RL & 0.0 \\
095407.02+212235.9 & 0.295 & $22.58\pm0.41$ & 1998-09-15 & $36.08\pm0.79$ & 2010-12-1 & $30.51\pm6.57$ & 2019-04-19 & \nodata & \nodata & RL & 0.0 \\
085632.99+595746.8 & 0.283 & $24.79\pm0.18$ & 2002-07-15 & $35.61\pm1.07$ & 2010-11-30 & $65.00\pm14.02$ & 2017-09-25 & \nodata & \nodata & RL & 0.0 \\
140336.43+174136.1 & 0.222 & $20.57\pm0.16$ & 1999-11-15 & $17.89\pm0.54$ & 2011-1-3 & $28.41\pm6.13$ & 2019-04-01 & \nodata & \nodata & RL & 0.0 \\
112952.99+221520.0 & 0.291 & $33.22\pm0.19$ & 1996-01-15 & $17.01\pm0.51$ & 2010-12-17 & $15.65\pm3.37$ & 2019-04-19 & \nodata & \nodata & RL & 0.0 \\
100726.10+124856.2 & 0.241 & $13.03\pm0.96$ & 1999-12-15 & $14.24\pm0.43$ & 2010-12-15 & $24.81\pm6.65$ & 2017-10-06 & \nodata & \nodata & RL & 0.0 \\
122539.55+245836.3 & 0.268 & $6.48\pm0.18$ & 1995-12-05 & $9.49\pm0.29$ & 2011-1-2 & $6.58\pm1.42$ & 2017-11-24 & \nodata & \nodata & RL & 0.0 \\
092837.98+602521.0 & 0.295 & $11.50\pm1.90$ & 2002-07-15 & $8.94\pm0.27$ & 2010-10-15 & $8.99\pm2.03$ & 2017-09-18 & \nodata & \nodata & RL & 0.0 \\
111121.71+482045.9 & 0.281 & $14.55\pm0.28$ & 1997-04-10 & $8.13\pm0.25$ & 2010-12-16 & $15.10\pm3.26$ & 2019-05-10 & \nodata & \nodata & RL & $-0.7$ \\
145331.47+264946.7 & 0.279 & $13.98\pm0.17$ & 1995-11-10 & $6.05\pm0.25$ & 2011-1-4 & $8.59\pm1.86$ & 2017-11-12 & \nodata & \nodata & RL & $-0.7$ \\
084347.84+203752.4 & 0.227 & $2.87\pm0.14$ & 1998-10-15 & $5.36\pm0.16$ & 2010-11-15 & $3.43\pm0.75$ & 2019-04-13 & \nodata & \nodata & RL & 0.0 \\
093200.07+553347.4 & 0.266 & $4.95\pm0.29$ & 1997-05-15 & $5.15\pm0.16$ & 2010-10-31 & $5.39\pm1.18$ & 2017-09-23 & \nodata & \nodata & RL & 0.0 \\
151351.50+262358.1 & 0.203 & $<0.47$ & 1995-11-13 & $5.04\pm0.15$ & 2010-12-4 & $1.51\pm0.36$ & 2017-10-02 & \nodata & \nodata & RI & 0.0 \\
091702.11+212337.5 & 0.202 & $4.19\pm0.13$ & 1998-09-15 & $4.79\pm0.14$ & 2011-1-15 & $4.53\pm0.99$ & 2019-04-22 & \nodata & \nodata & RI & 0.0 \\
151132.53+100953.1 & 0.282 & $1.14\pm0.13$ & 2000-01-15 & $3.05\pm0.09$ & 2010-12-3 & $2.06\pm0.46$ & 2019-03-12 & \nodata & \nodata & RI & 0.0 \\
125807.45+232921.6 & 0.258 & $3.91\pm0.16$ & 1995-12-23 & $2.96\pm0.09$ & 2010-11-1 & $3.40\pm0.75$ & 2020-07-15 & \nodata & \nodata & RI & 0.0 \\
075403.60+481428.0 & 0.276 & $7.91\pm0.16$ & 1997-04-10 & $2.63\pm0.08$ & 2011-1-15 & $4.40\pm0.95$ & 2019-05-10 & \nodata & \nodata & RI & $-0.7$ \\
083658.90+442602.2 & 0.254 & $9.15\pm0.12$ & 1997-02-28 & $2.60\pm0.08$ & 2011-1-15 & $4.25\pm0.93$ & 2019-05-24 & \nodata & \nodata & RI & $-0.7$ \\
094603.94+013923.6 & 0.220 & $7.33\pm0.15$ & 1998-07-24 & $2.47\pm0.08$ & 2011-1-15 & $4.38\pm0.96$ & 2018-01-02 & \nodata & \nodata & RI & $-0.7$ \\
134113.93-005315.0 & 0.237 & $5.13\pm0.16$ & 1998-08-15 & $1.93\pm0.06$ & 2010-11-2 & $2.43\pm0.56$ & 2019-03-24 & \nodata & \nodata & RI & $-0.7$ \\
140407.17+213321.6 & 0.261 & $3.89\pm0.15$ & 1998-10-15 & $1.70\pm0.05$ & 2010-11-4 & $2.60\pm0.59$ & 2017-09-21 & \nodata & \nodata & RI & $-0.7$ \\
161217.97+073145.4 & 0.207 & $4.92\pm0.11$ & 2000-02-15 & $1.65\pm0.07$ & 2010-12-16 & $3.06\pm0.68$ & 2019-03-18 & \nodata & \nodata & RI & $-0.7$ \\
134206.56+050523.8 & 0.266 & $3.55\pm0.14$ & 2000-02-15 & $1.62\pm0.08$ & 2010-11-3 & $1.92\pm0.25$ & 2019-04-22 & \nodata & \nodata & RI & $-0.7$ \\
144930.49-004746.3 & 0.253 & $2.21\pm0.15$ & 1998-08-15 & $1.56\pm0.05$ & 2010-12-1 & $1.85\pm0.42$ & 2019-04-23 & \nodata & \nodata & RI & 0.0 \\
101719.02+151620.8 & 0.245 & $<0.45$ & 1999-11-15 & $1.51\pm0.05$ & 2010-10-15 & $<0.33$ & 2017-10-06 & \nodata & \nodata & RI & 0.0 \\
171013.42+334402.5 & 0.208 & $3.80\pm0.15$ & 1994-06-19 & $1.24\pm0.04$ & 2010-12-19 & $2.25\pm0.50$ & 2017-12-10 & \nodata & \nodata & RI & $-0.7$ \\
145005.13+463521.3 & 0.293 & $4.49\pm0.13$ & 1997-03-25 & $1.19\pm0.04$ & 2010-12-2 & $2.02\pm0.46$ & 2019-04-14 & \nodata & \nodata & RI & $-0.7$ \\
162607.24+335915.3 & 0.205 & $3.11\pm0.14$ & 1994-06-25 & $1.18\pm0.04$ & 2010-12-18 & $1.51\pm0.35$ & 2017-10-06 & \nodata & \nodata & RI & $-0.7$ \\
094215.12+090015.8 & 0.213 & $1.69\pm0.13$ & 2000-01-15 & $1.16\pm0.04$ & 2011-1-15 & $1.50\pm0.34$ & 2017-10-02 & \nodata & \nodata & RI & 0.0 \\
133636.65+420934.1 & 0.223 & $2.69\pm0.10$ & 1995-10-23 & $0.99\pm0.03$ & 2010-11-1 & $2.05\pm0.20$ & 2019-03-19 & \nodata & \nodata & RI & $-0.7$ \\
144012.74+615633.0 & 0.275 & $3.21\pm0.18$ & 2003-12-28 & $0.94\pm0.03$ & 2010-11-6 & $2.17\pm0.49$ & 2017-11-27 & \nodata & \nodata & RI & $-0.7$ \\
113109.24+263207.8 & 0.244 & $1.85\pm0.13$ & 1995-11-13 & $0.89\pm0.03$ & 2010-10-18 & $1.11\pm0.28$ & 2019-04-12 & \nodata & \nodata & RI & $-0.7$ \\
110704.52+320630.0 & 0.243 & $2.79\pm0.14$ & 1994-06-11 & $0.80\pm0.03$ & 2010-10-17 & $1.16\pm0.28$ & 2019-06-04 & \nodata & \nodata & RI & $-0.7$ \\
141116.72+194440.0 & 0.238 & $1.08\pm0.10$ & 1999-11-15 & $0.76\pm0.03$ & 2010-11-5 & $1.08\pm0.26$ & 2019-04-01 & \nodata & \nodata & RI & 0.0 \\
154307.77+193751.7 & 0.229 & $1.21\pm0.12$ & 1999-01-15 & $0.72\pm0.03$ & 2010-12-15 & $1.36\pm0.33$ & 2019-03-21 & \nodata & \nodata & RI & $-0.7$ \\
115523.74+150756.9 & 0.287 & $1.40\pm0.11$ & 1999-11-15 & $0.60\pm0.02$ & 2010-10-20 & $0.98\pm0.26$ & 2017-10-28 & \nodata & \nodata & RI & $-0.7$ \\
151936.14+283827.6 & 0.269 & $1.83\pm0.12$ & 1993-04-29 & $0.58\pm0.02$ & 2010-12-5 & $0.83\pm0.20$ & 2017-10-03 & \nodata & \nodata & RI & $-0.7$ \\
114954.98+044812.8 & 0.270 & $1.36\pm0.15$ & 2000-02-15 & $0.52\pm0.02$ & 2010-10-19 & $0.98\pm0.24$ & 2020-07-22 & \nodata & \nodata & RI & $-0.7$ \\
105607.79+134443.6 & 0.284 & $1.63\pm0.12$ & 1999-12-15 & $0.51\pm0.02$ & 2010-10-16 & $1.11\pm0.27$ & 2017-12-28 & \nodata & \nodata & RI & $-0.7$ \\
172711.80+632242.1 & 0.217 & $1.09\pm0.13$ & 2002-08-15 & $0.73\pm0.03$ & 2010-12-13 & $0.62\pm0.17$ & 2017-09-11 & $0.67\pm0.08$ & 2019-09-20 & RQ & 0.0 \\
093509.48+481910.2 & 0.224 & $1.58\pm0.11$ & 1997-04-10 & $0.60\pm0.03$ & 2010-12-8 & $0.87\pm0.19$ & 2019-04-18 & $0.53\pm0.05$ & 2019-08-13 & RQ & $-0.7$ \\
115753.20-031537.1 & 0.215 & $2.58\pm0.14$ & 1998-09-16 & $0.60\pm0.02$ & 2010-12-3 & $1.09\pm0.21$ & 2018-01-12 & $0.64\pm0.02$ & 2019-08-20 & RQ & $-0.7$ \\
121832.68+352255.8 & 0.240 & $2.04\pm0.12$ & 1994-07-05 & $0.59\pm0.02$ & 2010-12-2 & $1.06\pm0.20$ & 2017-12-02 & $0.44\pm0.01$ & 2019-08-04 & RQ & $-0.7$ \\
112941.94+512050.5 & 0.234 & $0.88\pm0.12$ & 1997-04-29 & $0.50\pm0.04$ & 2010-12-14 & $<0.39$ & 2019-05-10 & $0.17\pm0.02$ & 2019-08-04 & RQ & 0.0 \\
101000.68+300321.5 & 0.256 & $1.31\pm0.07$ & 1993-04-13 & $0.49\pm0.02$ & 2010-12-6 & $1.03\pm0.20$ & 2019-04-22 & $0.50\pm0.02$ & 2019-08-04 & RQ & $-0.7$ \\
161940.56+254323.0 & 0.269 & $1.43\pm0.11$ & 1995-11-22 & $0.47\pm0.02$ & 2010-12-4 & $0.75\pm0.18$ & 2017-11-24 & $0.41\pm0.02$ & 2019-08-27 & RQ & $-0.7$ \\
100438.82+151057.3 & 0.219 & $1.17\pm0.06$ & 1999-11-15 & $0.42\pm0.03$ & 2010-12-11 & $0.50\pm0.16$ & 2017-10-06 & $0.24\pm0.02$ & 2019-08-04 & RQ & 0.0 \\
084313.41+535718.8 & 0.218 & $0.52\pm0.15$ & 1997-05-07 & $0.41\pm0.01$ & 2010-12-26 & $<0.45$ & 2017-09-26 & $0.35\pm0.03$ & 2019-08-13 & RQ & 0.0 \\
122018.43+064119.6 & 0.286 & $0.55\pm0.08$ & 2000-02-07 & $0.39\pm0.01$ & 2010-12-15 & $<0.49$ & 2019-05-01 & $0.32\pm0.01$ & 2019-08-20 & RQ & 0.0 \\
135326.12+362049.4 & 0.285 & $1.45\pm0.14$ & 1994-07-10 & $0.39\pm0.03$ & 2010-12-1 & $0.62\pm0.17$ & 2017-10-14 & $0.55\pm0.02$ & 2019-08-17 & RQ & $-0.7$ \\
161723.67+085414.7 & 0.206 & $<0.45$ & 2000-01-15 & $0.38\pm0.02$ & 2010-12-28 & $0.62\pm0.16$ & 2019-03-18 & $0.39\pm0.02$ & 2019-08-20 & RQ & 0.0 \\
101325.43+221229.4 & 0.274 & $0.83\pm0.08$ & 1998-09-15 & $0.36\pm0.02$ & 2010-12-9 & $0.68\pm0.17$ & 2019-04-18 & $0.40\pm0.02$ & 2019-08-17 & RQ & $-0.7$ \\
155620.23+521520.0 & 0.227 & $1.17\pm0.16$ & 1997-05-04 & $0.30\pm0.02$ & 2010-12-10 & $0.58\pm0.17$ & 2017-11-11 & $0.27\pm0.01$ & 2019-08-13 & RQ & $-0.7$ \\
162750.54+473623.5 & 0.262 & $1.26\pm0.11$ & 1997-03-31 & $0.30\pm0.02$ & 2010-12-7 & $1.06\pm0.20$ & 2019-05-04 & $0.44\pm0.01$ & 2019-09-20 & RQ & $-0.7$ \\
115558.97+593129.2 & 0.241 & $0.63\pm0.11$ & 2002-7-15 & $0.29\pm0.01$ & 2010-12-22 & $0.62\pm0.17$ & 2017-11-27 & $0.24\pm0.02$ & 2019-08-04 & RQ & $-0.7$ \\
085640.78+105755.8 & 0.274 & $1.35\pm0.14$ & 2000-01-15 & $0.28\pm0.01$ & 2010-12 & $0.72\pm0.18$ & 2017-11-21 & $0.51\pm0.10$ & 2019-08 & RQ & $-0.7$ \\
123532.83+410445.1 & 0.212 & $<0.51$ & 1994-09-02 & $0.27\pm0.02$ & 2010-12-27 & $0.45\pm0.15$ & 2019-04-15 & $0.55\pm0.07$ & 2019-08-13 & RQ & $-0.7$ \\
161711.42+063833.4 & 0.229 & $<0.63$ & 2000-02-15 & $0.27\pm0.01$ & 2010-12-23 & $<0.65$ & 2019-03-18 & $0.47\pm0.01$ & 2019-08-20 & RQ & $-0.7$ \\
130456.91+395111.4 & 0.271 & $0.62\pm0.11$ & 1994-08-19 & $0.25\pm0.02$ & 2010-12-16 & $0.59\pm0.17$ & 2017-10-28 & $0.22\pm0.01$ & 2019-08-17 & RQ & $-0.7$ \\
125236.15+140213.9 & 0.266 & $0.63\pm0.12$ & 2000-01-15 & $0.24\pm0.01$ & 2010-12-17 & $<0.41$ & 2019-04-25 & $0.11\pm0.01$ & 2019-08-17 & RQ & 0.0 \\
100033.88+104723.7 & 0.226 & $1.78\pm0.16$ & 2000-01-15 & $0.23\pm0.01$ & 2010-12-5 & $0.87\pm0.19$ & 2017-11-26 & $0.55\pm0.01$ & 2019-08-04 & RQ & $-0.7$ \\
082205.24+455349.1 & 0.300 & $0.43\pm0.13$ & 1997-03-22 & $0.22\pm0.01$ & 2010-12-20 & $<0.41$ & 2019-05-24 & $0.53\pm0.03$ & 2019-08-13 & RQ & $-0.7$ \\
130416.99+020537.0 & 0.229 & $0.87\pm0.10$ & 1998-07-18 & $0.21\pm0.01$ & 2010-12-18 & $<0.45$ & 2019-04-21 & $0.23\pm0.04$ & 2019-08-20 & RQ & $-0.7$ \\
104541.76+520235.5 & 0.284 & $0.49\pm0.15$ & 1997-05-04 & $0.20\pm0.01$ & 2010-12-19 & $<0.50$ & 2017-09-26 & $0.22\pm0.03$ & 2019-08-04 & RQ & $-0.7$ \\
104528.30+293344.7 & 0.293 & $0.60\pm0.10$ & 1993-05-04 & $0.18\pm0.01$ & 2010-12-12 & $<0.39$ & 2019-04-13 & $0.32\pm0.01$ & 2019-08-04 & RQ & $-0.7$ \\
145824.46+363119.5 & 0.246 & $<0.49$ & 1994-07-17 & $0.17\pm0.01$ & 2010-12-25 & $0.51\pm0.16$ & 2017-11-10 & $0.21\pm0.03$ & 2019-09-20 & RQ & $-0.7$ \\
142522.37+080327.1 & 0.230 & $<0.37$ & 2000-01-15 & $0.14\pm0.01$ & 2010-12-29 & $<0.47$ & 2019-05-08 & $0.22\pm0.01$ & 2019-08-17 & RQ & $-0.7$ \\
153539.25+564406.5 & 0.207 & $0.76\pm0.16$ & 1997-05-15 & $0.13\pm0.02$ & 2010-12-24 & $<0.43$ & 2017-10-28 & $0.12\pm0.01$ & 2019-08-13 & RQ & $-0.7$ \\
\enddata
\end{deluxetable*}
\end{longrotatetable}

\end{document}